\documentclass[aps,prc,twocolumn,showpacs,amssymb,amsmath,amsfonts,floatfix,nofootinbib]{revtex4-1}
\usepackage{graphicx}

\makeatletter
\def\p@subsection{}
\def\p@subsubsection{}
\makeatother

\usepackage{natbib}
\usepackage{amsmath}
\usepackage[usenames]{color}
\usepackage{epsfig}
\usepackage{epstopdf}
\usepackage{amssymb,amsmath}
\usepackage{amsmath}
\usepackage{subfig}
\usepackage{epstopdf}
\usepackage{sidecap}
\usepackage{cases}
\usepackage{enumitem}
\usepackage{bm}
\usepackage{overpic}
\usepackage{upgreek}
\usepackage{morefloats}

\definecolor{grey}{rgb}{0.9,0.9,0.9}
\definecolor{black}{rgb}{0,0,0}

\newcommand{\be}{\begin{eqnarray}}
\newcommand{\ee}{\end{eqnarray}}
\newcommand{\bc}{\begin{center}}
\newcommand{\ec}{\end{center}}
\newcommand{\beq}{\begin{eqnarray}}
\newcommand{\eea}{\end{eqnarray}}

\begin{document}

\title{Towards an understanding of discrete ambiguities in truncated partial wave analyses}


\author{  Y. Wunderlich\,$^1$, \\
 A. \v{S}varc\,$^2$, R. L. Workman$^3$ , L. Tiator$\,^4$ and R. Beck$\,^1$
  }

\affiliation{$\,^1$ Helmholtz-Institut f\"{u}r Strahlen- und Kernphysik der Universit\"{a}t Bonn, Nussallee 14-16, 53115 Bonn, Germany }
\affiliation{$\,^2$ Rudjer Bo\v{s}kovi\'{c} Institute, Bijeni\v{c}ka cesta 54,
                 P.O. Box 180, 10002 Zagreb, Croatia}
\affiliation{$\,^3$ Data Analysis Center at the Institute for Nuclear Studies, Department of Physics, The George Washington University, Washington, D.C. 20052, USA}
\affiliation{$\,^4$ Institut f\"{u}r Kernphysik, Universit\"{a}t Mainz, D-55099 Mainz, Germany}

\vspace{5cm}
\date{\today}

\begin{abstract}

It is well known that the observables in a single-channel scattering 
problem remain invariant once the amplitude is multiplied by an overall 
energy- and angle-dependent phase. This invariance is called the continuum 
ambiguity and acts on the infinite partial wave set. It has also long been 
known that, in the case of a truncated partial wave set, another 
invariance exists, originating from the replacement of the roots of  
partial wave amplitudes with their complex conjugate values. This discrete 
ambiguity is also known as the Omelaenko-Gersten-type ambiguity. 
In this paper, we show that for scalar particles, discrete ambiguities 
are just a subset of continuum ambiguities with a specific phase and 
thus mix partial waves, as the continuum ambiguity does. We present the 
main features of both, continuum and discrete ambiguities, and describe 
a numerical method which establishes the relevant phase connection.

\end{abstract}

\pacs{PACS numbers: 13.60.Le, 14.20.Gk, 11.80.Et }
\maketitle


\allowdisplaybreaks

\section{Introduction} \label{sec:Introduction}

\begin{figure}[h]
\begin{overpic}[width=0.48\textwidth]{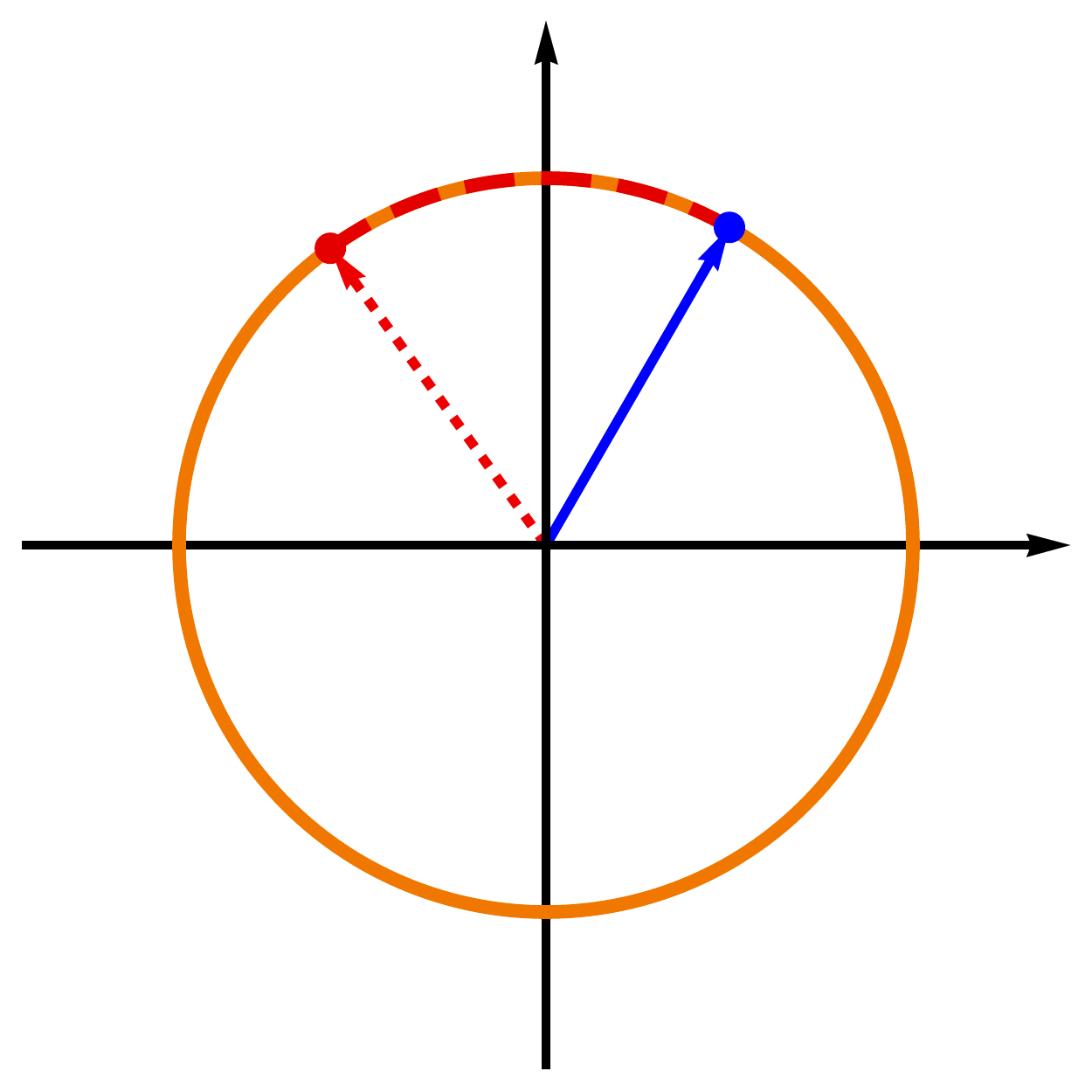}
  \put(90,43){\begin{Large}$\mathrm{Re}$\end{Large}}
  \put(52.5,93.5){\begin{Large}$\mathrm{Im}$\end{Large}}
  \put(53.5,53.5){\begin{Large} \textcolor{blue}{$A(W,\theta)$} \end{Large}}
  \put(18.5,53.5){\begin{Large} \textcolor{red}{$\tilde{A}(W,\theta)$} \end{Large}}
 \end{overpic}
\caption{The geometrical picture of the general continuum ambiguity (\ref{eq:ContAmbTrafoDefinition}) is depicted here. The differential cross section (\ref{eq:DiffCSDefinition}) constrains the true solution for the amplitude $A(W,\theta)$ to be located on a circle of radius $+ \sqrt{\sigma_{0}}$ in the complex plane. A rotation of the amplitude $A(W,\theta)$ does not alter the cross section.}
\label{fig:ContinuumAmbiguityGeometry}
\end{figure}
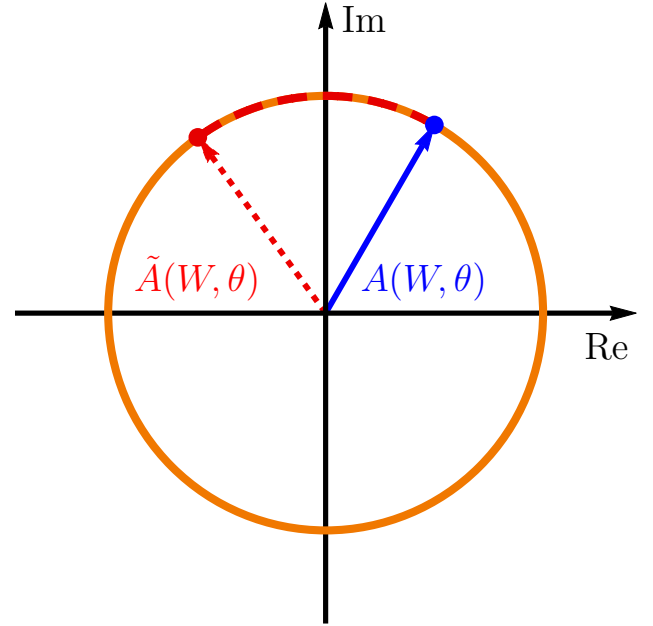

In this work, we will consider the simple case of a $2 \rightarrow 2$ reaction amplitude $A(W,\theta)$ for scalar particles. We make this choice for illustrative and pedagogical purposes. The amplitude has the conventional partial wave expansion
\begin{equation}
 A \left(W, \theta \right) = \sum_{\ell = 0}^{\infty} (2 \ell + 1) A_{\ell} (W) P_{\ell} (\cos \theta) \mathrm{.} \label{eq:BasicInfinitePWExpansion}
\end{equation}
The extraction of partial waves from data shall be studied, with data given in case of the scalar reaction just by the differential cross section, which is defined as the modulus squared of $A(W,\theta)$ (ignoring explicit phase-space factors)
\begin{equation}
 \sigma_{0} \left(W, \theta \right) = \left| A \left(W, \theta \right) \right|^{2} \mathrm{.} \label{eq:DiffCSDefinition}
\end{equation}
Taking the positive branch of the square-root on both sides of this equation, it is seen that at each point $(W,\theta)$ in phase-space, the cross section confines the amplitude to a circle in the complex plane: $\left| A (W,\theta) \right| = + \sqrt{\sigma_{0}}$. Figure \ref{fig:ContinuumAmbiguityGeometry} shows a depiction of this fact. \newline
From the geometrical depiction as well as from the mathematical form of (\ref{eq:DiffCSDefinition}), it is quickly seen that the cross section remains unchanged under a rotation of the amplitude $A(W,\theta)$ by a phase, which is generally allowed to depend on energy and angle (see Figure \ref{fig:ContinuumAmbiguityGeometry}):
\begin{equation}
 A(W,\theta) \rightarrow \tilde{A} (W,\theta) := e^{i \Phi(W,\theta)} A(W,\theta) \mathrm{.} \label{eq:ContAmbTrafoDefinition}
\end{equation}
The invariance under such transformations has long been well known and is generally referred
to as the {\it continuum ambiguity} \cite{BowcockBurkhardt}.

\clearpage

\begin{figure}[htb]
\includegraphics[width=0.315\textwidth]{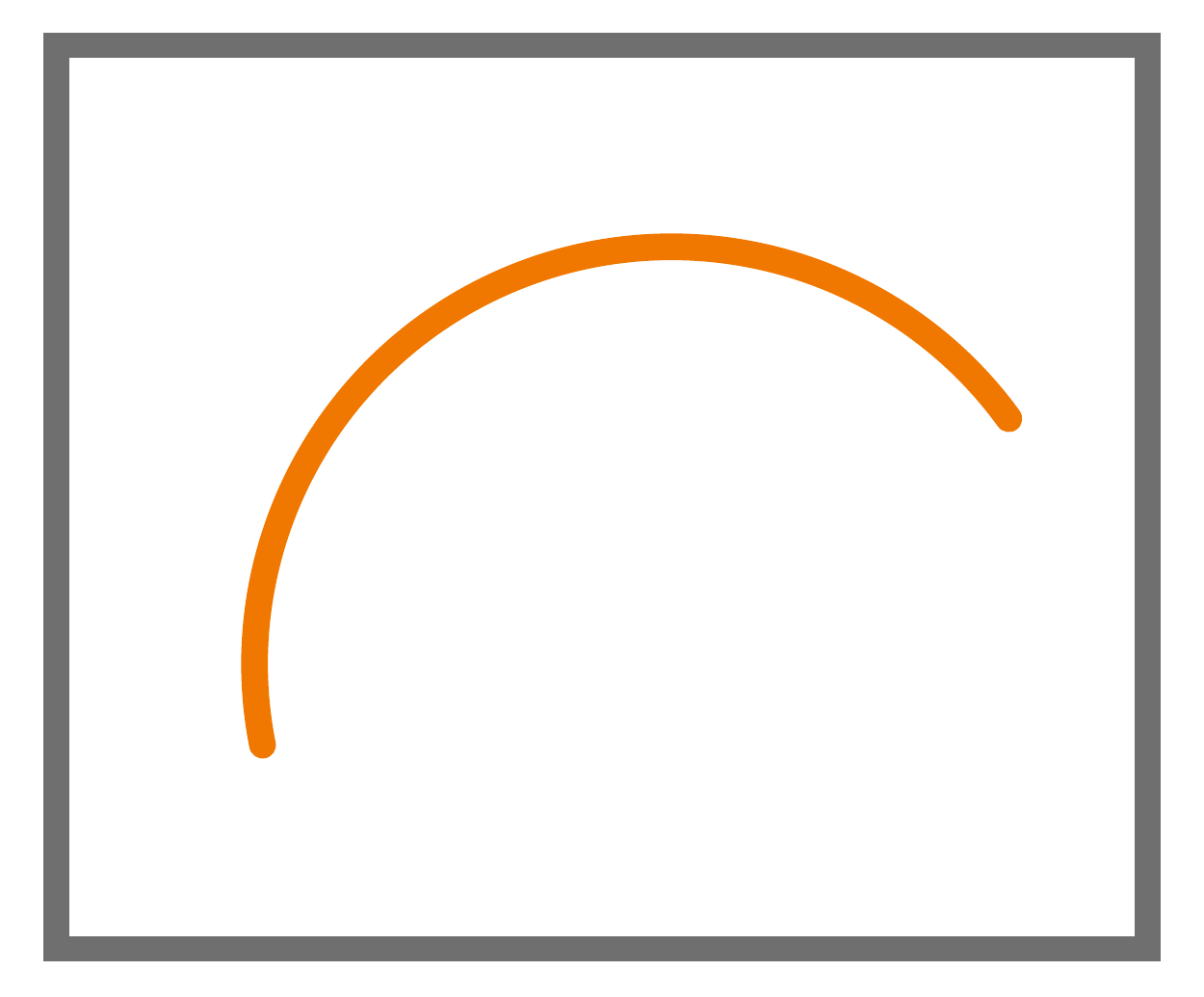}
\includegraphics[width=0.315\textwidth]{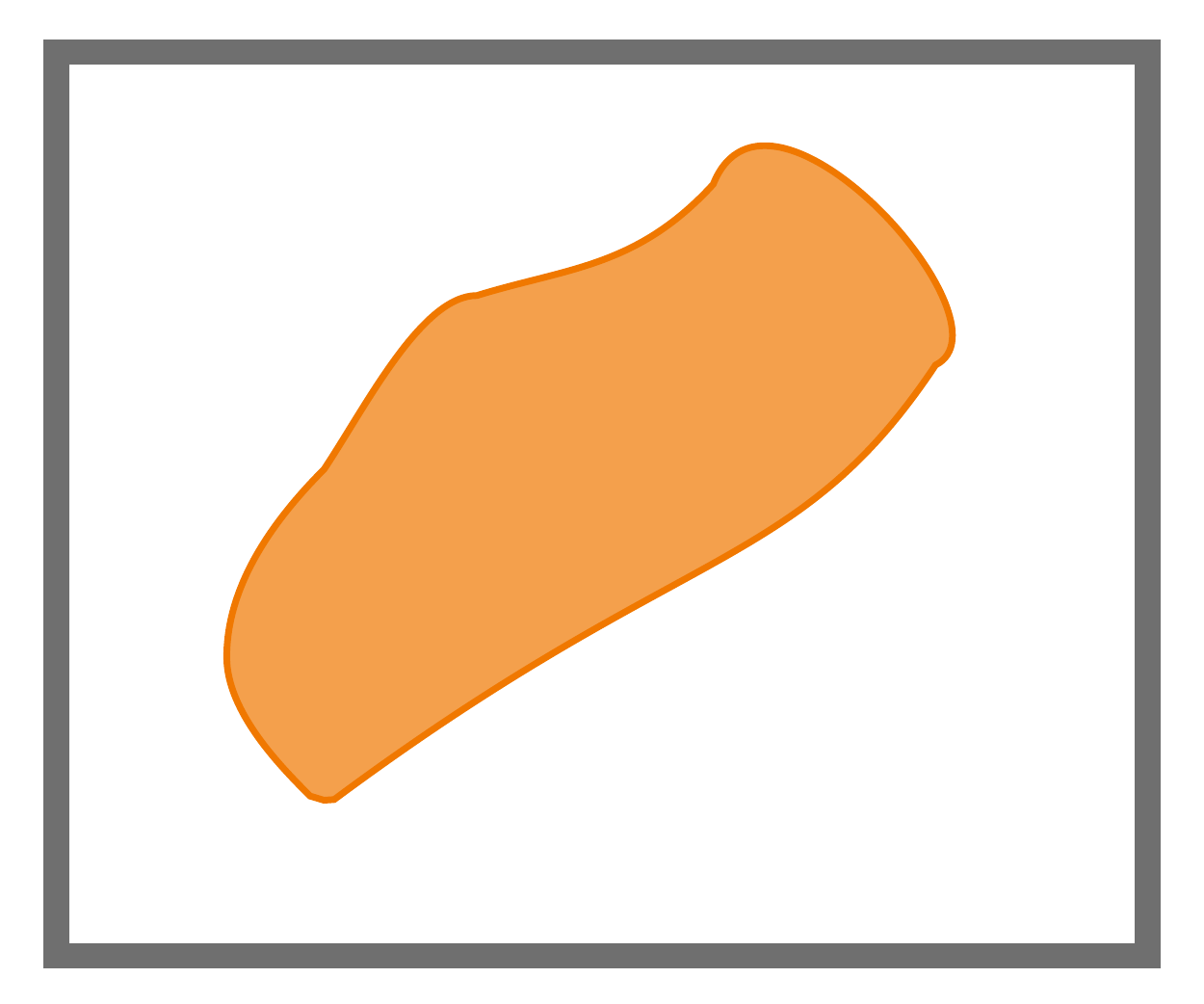}
\includegraphics[width=0.315\textwidth]{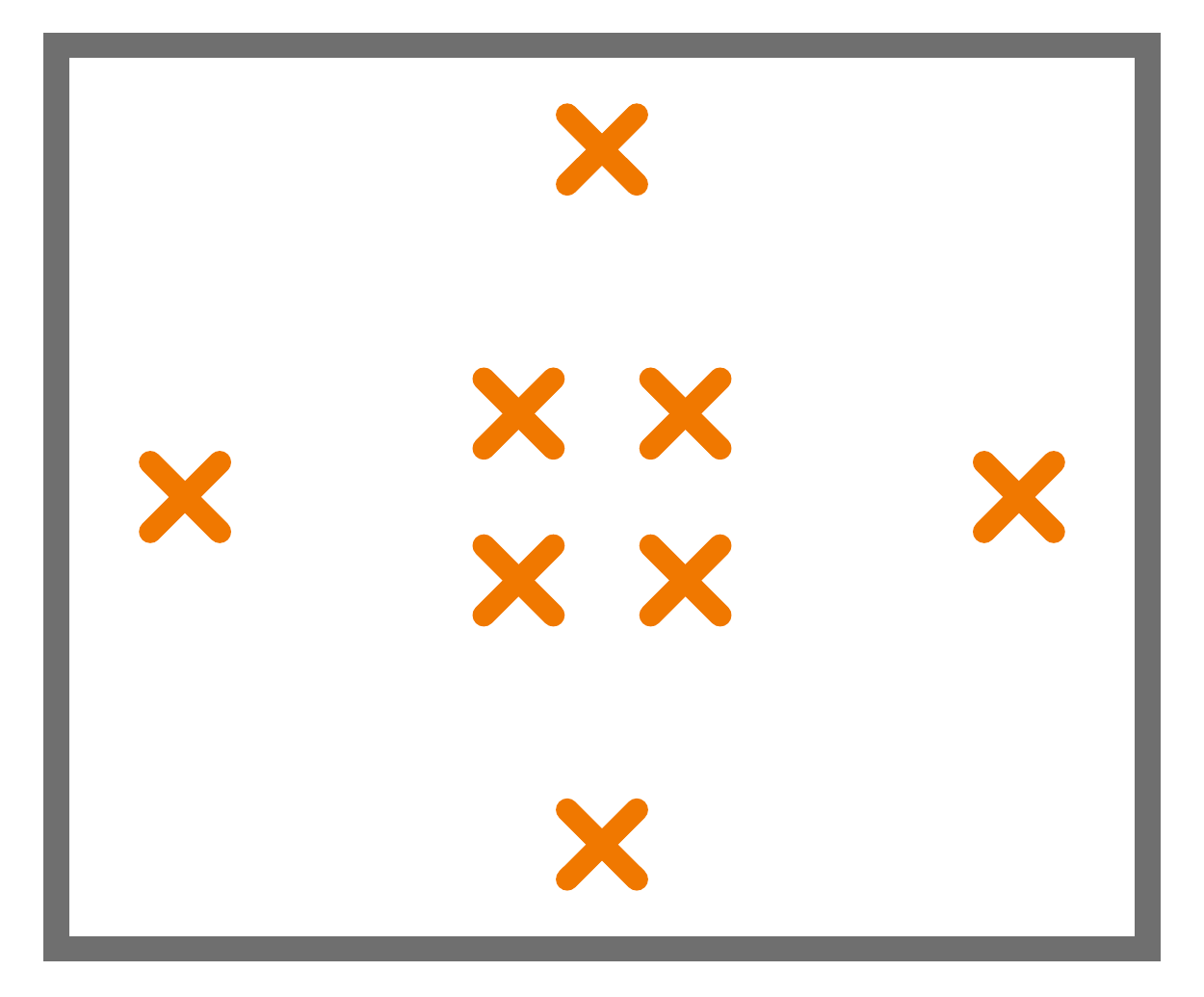}
\caption{Three schematics are shown in order to illustrate the meaning of the terms {\it discrete-} and {\it continuum} ambiguities. The grey colored box represents in each case the higher-dimensional space furnished by the partial wave amplitudes, be it for infinite partial wave models, or for truncated ones. \newline
Top: One-dimensional (for instance circular) arcs can be traced out by continuum ambiguity transformations, both for infinite and truncated models. \newline
Center: More general connected continua in amplitude space, containing an infinite number of points belonging to the same cross section, can be generated by use of the continuum ambiguity (\ref{eq:ContAmbTrafoDefinition}). However, this phenomenon is only present once the partial wave series goes to infinity. The connected patches are also referred to as {\it islands of ambiguity} \cite{BowcockBurkhardt, AtkinsonEtAlContAmb}. \newline
Bottom: Discrete ambiguities refer to cases where the cross section is the same for disconnected, discretely located points in amplitude space. These ambiguities are most prominent in TPWAs \cite{BowcockBurkhardt, Gersten} (see section \ref{sec:DiscrAmbs} below). However, two-fold discrete ambiguities can also appear for infinite partial wave models, where elastic unitarity is employed \cite{BowcockBurkhardt}.
}
\label{fig:AmbiguityConceptSchematics}
\end{figure}

A different kind of ambiguity arises whenever the amplitude $A(W,\theta)$ has a zero in the angular variable, for instance in $\cos \theta$ \cite{LPKokNote}. This can be seen by splitting the original amplitude $A(W,\theta)$ into the product of the linear factor belonging to the complex zero $\alpha$, times a reduced amplitude $\hat{A} (W,\theta)$:
\begin{equation}
 A (W, \theta) =  \hat{A} (W,\theta) \left( \cos \theta - \alpha \right) \mathrm{,} \label{eq:DiscreteAmbKoksAmplitudeDecomposition}
\end{equation}
Writing the differential cross section for this case, i.e.
\begin{equation}
 \sigma_{0} = \big| \hat{A} (W,\theta) \big|^{2} \left( \cos \theta - \alpha^{\ast} \right) \left( \cos \theta - \alpha \right) \mathrm{,} \label{eq:DiffCSGeneralDiscreteAmbiguity}
\end{equation}
it is evident that the complex conjugation of $\alpha$, i.e. $\alpha \rightarrow \alpha^{\ast}$, does not change this observable. Since $\alpha$ is an angular zero, it has to be connected to the partial wave amplitudes in some way. Thus, by leaping from one value of $\alpha$ to another one $\alpha^{\ast}$, one achieves the same effect in the amplitude space. This means, one transitions to a discretely diconnected point in this space, which yields the exact same cross section. In this way, these so-called {\it discrete ambiguities} acquire their name and they are a most prominent (but not fully exclusive) feature of truncated partial wave analyses (TPWAs). The latter term refers to any analysis that involves a truncation of the infinite series (\ref{eq:BasicInfinitePWExpansion}) at some angular momentum $L$. \newline
With this knowledge, also the name continuum ambiguity given to the general rotations (\ref{eq:ContAmbTrafoDefinition}) can be understood. As it turns out \cite{BowcockBurkhardt}, the vast size of this class of symmetry transformations, owing to the fact that they can be performed with in principle any function $\Phi(W,\theta)$, makes it possible to trace out connected arcs or even whole regions in amplitude space, which all have the same cross section. In fact, quite involved and sophisticated studies have been done in the past, in order to estimate and calculate such ambiguity-continua \cite{AtkinsonEtAlContAmb}. Figure \ref{fig:AmbiguityConceptSchematics} gives a schematic illustration of the different types of ambiguites in partial wave analyses. \newline
In this work, we investigate both continuum- and discrete ambiguities as purely mathematical phenomena, which occur once partial waves are to be extracted from the quadratic form defined by the cross section (\ref{eq:DiffCSDefinition}). We will compare the large class of symmetry transformations generated by the general rotations (\ref{eq:ContAmbTrafoDefinition}), to the smaller class of discrete symmetries caused by root-conjugation and elaborate how and under which circumstances traces of the latter class can be found in the former. \newline
The amount of ambiguity encountered may of course be reduced by introducing further physical constraints into the analysis, the most prominent one being the unitarity of the $\hat{S}$-matrix \cite{BowcockBurkhardt}. We do not further pursue ambiguities under unitarity-constraints here, but leave them as a further avenue of exploration. \newline
It should just be mentioned that TPWAs performed below the first inelastic channel, where elastic unitarity is a very powerful contraint, are known to have discrete ambiguites,

\clearpage

so-called Crichton-ambiguities \cite{Crichton}.
The explorations of continuum ambiguities by Atkinson et al \cite{AtkinsonEtAlContAmb} have also been performed under quite strict unitarity constraints. \newline

We focus on the scalar amplitudes in order to keep the discussion as simple and illustrative as possible.
However, it should be stated that the obtained results often carry over to analyses of more complicated reactions involving particles with spin ($\pi N$-scattering, photoproduction, $\ldots$), in many cases without large modifications.\footnote{Some possible complications for the generalization to spin reactions are hinted at in the conclusions of this work.} Therefore, what is discussed here may also turn out to be relevant in recently initiated programs on analyses of so-called {\it complete} sets of polarization-data, performed for the spin-reactions (see for instance \cite{RWorkmanEtAlCompExPhotoprod, YWEtAl2014, Grushin}). \newline
This work is complementary to the study of \v{S}varc et al \cite{Svarc2017}, which deals with related issues of ambiguities in partial wave analyses.

\section{Continuum ambiguities and the mixing formula} \label{sec:ContAmbs}

Here, we consider continuum ambiguities, i.e. new partial wave solutions generated by transforming the original amplitude $A(W,\theta)$ as in equation (\ref{eq:ContAmbTrafoDefinition}), using a general energy- and angle-dependent phase-rotation $e^{i \Phi (W, \theta)}$. We choose to write the latter as a Legendre-series:
\begin{equation}
 e^{i \Phi (W, \theta)} = \sum_{k = 0}^{\infty} L_{k} (W) P_{k} (\cos \theta) \mathrm{.} \label{eq:PhaseRotLegendreDecomposition}
\end{equation}
As mentioned in the introduction, quite a lot of work has been done in the past on the capability of such rotations, which themselves have infinitely many real degrees of freedom, to generate ambiguous partial wave solutions. Here, we want to focus only on one aspect of the problem, namely the transformation of the original partial waves $A_{\ell}$ into waves $\tilde{A}_{\ell}$ belonging to the rotated amplitude $\tilde{A} (W, \theta)$, caused by the rotation (\ref{eq:PhaseRotLegendreDecomposition}). In the following derivation, we employ the notation $x=\cos \theta$. The projection integral for the transformed waves becomes
\begin{align}
 \tilde{A}_{\ell} (W) &= \frac{1}{2} \int_{-1}^{1} d x \tilde{A} (W,x) P_{\ell} (x) \nonumber \\
   &= \frac{1}{2} \int_{-1}^{1} d x e^{i \phi (W, x)} A (W,x) P_{\ell} (x) \nonumber \\
   &= \frac{1}{2} \int_{-1}^{1} d x \sum_{k = 0}^{\infty} L_{k} (W) P_{k} (x) A (W,x) P_{\ell} (x) \nonumber \\
  &= \sum_{k = 0}^{\infty} L_{k} (W) \hspace*{2pt} \frac{1}{2} \int_{-1}^{1} d x A (W,x) P_{k} (x) P_{\ell} (x) \mathrm{,} \label{eq:MixingDerivation1}
\end{align}
\newline \\
%
%
where in the last step, the permutation of the integral and the infinite $k$-sum was just assumed to be valid. \newline
The product of Legendre polynomials under the integral in (\ref{eq:MixingDerivation1}) is again expandable into the basis of Legendre polynomials. The resulting formula is known from the theory of the rotation group and can be written using either the Wigner $3j$-symbols, or the well-known Glebsch-Gordan coefficients \cite{LegendreProductFormula}:
\begin{align}
 P_{k} (x) P_{\ell} (x) &= \sum_{m = \left| k - \ell \right|}^{ k + \ell } \left( \begin{array}{ccc} k & l & m \\ 0 & 0 & 0 \end{array} \right)^{2} (2m+1) P_{m} (x) \nonumber \\
 &= \sum_{m = \left| k - \ell \right|}^{ k + \ell } \left< k,0 ; \ell,0 | m , 0 \right>^{2} P_{m} (x) \mathrm{.} \label{eq:AdamsFormula}
\end{align}
For the remainder of this work, the Clebsch-Gordan coefficients $\left< k,0 ; \ell,0 | m , 0 \right>$ are utilized. Using this recoupling-formula, the partial wave projection (\ref{eq:MixingDerivation1}) becomes
\begin{widetext}
\begin{align}
 \tilde{A}_{\ell} (W) &= \sum_{k = 0}^{\infty} L_{k} (W) \hspace*{2pt} \frac{1}{2} \int_{-1}^{1} d x A (W,x) \sum_{m = \left| k - \ell \right|}^{ k + \ell } \left< k,0 ; \ell,0 | m , 0 \right>^{2} P_{m} (x) \nonumber \\
 &= \sum_{k = 0}^{\infty} L_{k} (W) \hspace*{2pt} \sum_{m = \left| k - \ell \right|}^{ k + \ell } \left< k,0 ; \ell,0 | m , 0 \right>^{2}  \hspace*{2pt}  \frac{1}{2} \int_{-1}^{1} d x A (W,x) P_{m} (x) \nonumber \\
 &= \sum_{k = 0}^{\infty} L_{k} (W) \hspace*{2pt} \sum_{m = \left| k - \ell \right|}^{ k + \ell } \left< k,0 ; \ell,0 | m , 0 \right>^{2} A_{m} (W) \mathrm{.} \label{eq:MixingDerivation2Final}
\end{align}
\end{widetext}
We see that the final result on the right hand side takes the form of a linear combination, or mixing, of the partial waves $A_{\ell} (W)$ from the original amplitude. The precise form of the mixing is of course dictated by the energy-dependent Legendre coefficients $L_{k} (W)$ that define the phase-rotation (\ref{eq:PhaseRotLegendreDecomposition}). Since this mixing formula is vital to the remainder of this work, we state it again in closed form
%
\begin{widetext}
\begin{equation}
 \boxed{\tilde{A}_{\ell} (W) = \sum_{k = 0}^{\infty} L_{k} (W) \hspace*{2pt} \sum_{m = \left| k - \ell \right|}^{ k + \ell } \left< k,0 ; \ell,0 | m , 0 \right>^{2} A_{m} (W) \mathrm{.} }  \label{eq:MixingFormulaBoxed}
\end{equation}
\end{widetext}
The general relation given in equation (\ref{eq:MixingFormulaBoxed}) has been derived using straightforward algebra and identities involving the Legendre polynomials. However, we have not found it reproduced, in this form, in the literature. \newline
For reactions involving particles with spin on the other hand, similar mixing-phenomena have been found either derived explicitly, or at least hinted at. Dean and Lee \cite{DeanLee} give a very detailed treatment of analogous relations for $\pi N$-scattering. Omelaenko \cite{Omelaenko} hints, near the end of his famous paper on discrete ambiguities in photoproduction, at similar circumstances for this particular reaction. Angle-dependent phase-rotations and their effects in photoproduction are also discussed by Keaton and Workman \cite{KeatonWorkmanII}. \newline
Some mathematical comments on the mixing formula (\ref{eq:MixingFormulaBoxed}) are in order. First of all, angle-independent phase-rotations are defined only by the lowest Legendre coefficient $L_{0} (W)$, with all higher one's vanishing (see equation (\ref{eq:PhaseRotLegendreDecomposition})). The mixing formula immediately tells that for these purely energy-dependent rotations, no mixing occurs at all and all partial waves are rotated by the same angle. \newline
However, once the continuum ambiguity phase $\Phi(W,\theta)$ has at least some angular dependence, the Legendre expansion (\ref{eq:PhaseRotLegendreDecomposition}) regains the full complexity of an infinite series. However, it is indeed feasible to construct phase-rotations whose Legendre series converges rather quickly. In fact, for most examples considered in this work, they do. However, the mixing formula (\ref{eq:MixingFormulaBoxed}) then implies that for any angle-dependence of the continuum ambiguity, mixing of partial waves necessarily occurs and furthermore is defined by an infinite tower of strictly speaking non-vanishing Legendre coefficients $L_{k} (W)$. \newline
Having discussed the effect of the general continuum ambiguity transformations on partial waves, we now introduce discrete ambiguities proper and outline the way in which they leave traces in the former, larger class of symmetry transformations.

\section{Discrete ambiguities in TPWAs and generating phase-rotations} \label{sec:DiscrAmbs}

Next we consider TPWAs, i.e. those based on the partial wave series (\ref{eq:BasicInfinitePWExpansion}) cut off at some maximal angular momentum $L$. Gersten \cite{Gersten} has first noted the usefulness of decomposing such polynomial amplitudes into the product over their linear factors, i.e. by writing
\begin{align}
A (W, \theta)  &= \sum_{\ell = 0}^{L} (2 \ell + 1) A_{\ell} (W) P_{\ell} (\cos \theta) \nonumber \\
 &\equiv \lambda \prod_{i = 1}^{L} \left( \cos \theta - \alpha_{i} \right) \mathrm{,} \label{eq:GerstenDecomposedScalarAmplitude}
\end{align}
where the $\alpha_{i}$ are a set of $L$ complex zeros defining the amplitude. The complex normalization factor $\lambda$ is, in the convention chosen above, proportional to the highest partial wave: $\lambda \propto A_{L}$. \newline
Furthermore, since the differential cross section (\ref{eq:DiffCSDefinition}) is a modulus squared, even in the truncated PWA one energy dependent overall phase has to be fixed prior to fitting the model (\ref{eq:GerstenDecomposedScalarAmplitude}) to data. One common choice could be to require the $S$-wave to be real and positive: $A_{0} = \mathrm{Re}\left[ A_{0} \right] > 0$. This is the convention we will adhere to later. However, one could also choose to fix the normalization $\lambda$ in (\ref{eq:GerstenDecomposedScalarAmplitude}) to be real, thereby also implying the same convention for the highest wave $A_{L}$. \newline
As mentioned in the introduction, the complex conjugation of a zero of any, either truncated or even infinite, partial wave expansion generates a discrete ambiguity. Since in the truncated case, the amplitude (\ref{eq:GerstenDecomposedScalarAmplitude}) is nothing but a product over linear factors, the cross section
\begin{equation}
 \sigma_{0} = \left| \lambda \right|^{2} \prod_{i = 1}^{L} \left( \cos \theta - \alpha_{i}^{\ast} \right)  \left( \cos \theta - \alpha_{i} \right) \mathrm{,} \label{eq:DiffCSTruncatedAmplitude}
\end{equation}
is unchanged by all possibilities of conjugating subsets of roots \cite{Gersten}. There exist in total $2^{L}$ such possibilities and we adhere to the formalization of all those possibilities introduced by Gersten \cite{Gersten}. Therefore, we define a set of $2^{L}$ maps
\begin{equation}\bm{\uppi}_{\hspace*{0.035cm}p} \left(\alpha_{i}\right) := \begin{cases}
                    \alpha_{i} &\mathrm{,} \hspace*{3pt} \mu_{i} \left(p\right) = 0 \\
                    \alpha_{i}^{\ast} &\mathrm{,} \hspace*{3pt} \mu_{i} \left(p\right) = 1
                   \end{cases}\mathrm{,} \label{eq:GerstenMapsScalarExample}
\end{equation}
where the binary representation of the number $p$,
\begin{equation}
 p = \sum_{i = 1}^{L} \mu_{i} \left( p \right) 2^{(i-1)}\mathrm{,} \label{eq:BinaryRepresentationForP}
\end{equation}
has been employed. The index $p$ just labels all combinatorically possible ambiguities acting on the roots $\alpha_{i}$, with $\bm{\uppi}_{\hspace*{0.035cm}0}$ being the identity. \newline
Now, it is easy to define ambiguity-transformed truncated amplitudes $A^{(p)} (W, \theta)$ which, since the number of factors in (\ref{eq:GerstenDecomposedScalarAmplitude}) is unchanged by any of the Gersten-ambiguities (\ref{eq:GerstenMapsScalarExample}), retains the same truncation order $L$ as the original amplitude
\begin{align}
A^{(p)} (W, \theta) &= \lambda \prod_{i = 1}^{L} \left( \cos \theta - \bm{\uppi}_{\hspace*{0.035cm}p} \left[\alpha_{i}\right] \right) \nonumber \\
&\equiv \sum_{\ell = 0}^{L} (2 \ell + 1) A^{(p)}_{\ell} (W) P_{\ell} (\cos \theta) \mathrm{.} \label{eq:GerstenTransformedScalarAmplitude}
\end{align}
The ambiguous amplitudes $A^{(p)}$ have the same cross section as the original model $A$. According to remarks made in the introduction, this means that they have to be connected to the original amplitude by rotations (see Figure \ref{fig:ContinuumAmbiguityGeometry}). These phase-rotations are, once the Gersten-formalism has been established, computed without effort:
\begin{align}
e^{i \varphi_{p} (W, \theta)} &= \frac{A^{(p)} (W, \theta)}{A(W,\theta)} \nonumber \\
&= \frac{\left( \cos \theta - \bm{\uppi}_{\hspace*{0.035cm}p} \left[\alpha_{1}\right] \right) \ldots \left( \cos \theta - \bm{\uppi}_{\hspace*{0.035cm}p} \left[\alpha_{L}\right] \right)}{\left( \cos \theta - \alpha_{1} \right) \ldots \left( \cos \theta - \alpha_{L} \right)} \mathrm{.} \label{eq:GerstenFormalismPhaseRotation}
\end{align}
Remembering the definition of the maps (\ref{eq:GerstenMapsScalarExample}), it can be seen quickly that the resulting expression has modulus $1$ for all $\cos \theta \in [-1,1]$, as it should. \newline
Some more remarks have to be made about the result (\ref{eq:GerstenFormalismPhaseRotation}). First of all, for all ambiguities except the identity $\bm{\uppi}_{\hspace*{0.035cm}0}$ (which leads to $e^{i \varphi_{0} (W, \theta)} = 1$), the phase-rotation is explicitly angle-dependent. As mentioned in section \ref{sec:ContAmbs}, a purely energy-dependent phase-rotation rotates all partial waves by the same angle. The discrete Gersten-ambiguities have a different nature, leading via the conjugations of the roots $\alpha_{i}$ to more intricate transformations on the level of partial waves $A_{\ell}$. Already for low truncation orders $L$, conjugations of single waves can be observed, or more generally rotations of different waves by different angles. In order to achieve this, the generating phase-rotations (\ref{eq:GerstenFormalismPhaseRotation}) have to have at least some angle-dependence. \newline
Secondly, in establishing the discrete Gersten-ambiguities to be generated by phase-rotations (\ref{eq:GerstenFormalismPhaseRotation}), a connection has been drawn between the discrete partial wave ambiguities discussed in this section and the more general continuum ambiguities treated in section \ref{sec:ContAmbs}. In particular, since the generating phases (\ref{eq:GerstenFormalismPhaseRotation}) are angle-dependent they have, by means of equation (\ref{eq:MixingFormulaBoxed}) above, to lead to partial wave mixing. In any case, an angle-dependent phase has an infinite Legendre-expansion. However, from their definition the phases (\ref{eq:GerstenFormalismPhaseRotation}) again lead to manifestly truncated amplitudes (\ref{eq:GerstenTransformedScalarAmplitude}). Therefore, these generating phase-rotations are finely tuned such that they lead to exact cancellations on the right hand side of the mixing formula (\ref{eq:MixingFormulaBoxed}), for all $\ell > L$. \newline

Gersten \cite{Gersten} stated, without proof, that the transformations (\ref{eq:GerstenMapsScalarExample}) exhaust all possibilities to form discrete ambiguities in a TPWA. To be more precise, he mentions a further discrete symmetry, namely
\begin{equation}
 A^{(p)} (W, \theta) \longrightarrow - \left[ A^{(p)} (W, \theta) \right]^{\ast} \mathrm{,} \label{eq:GerstensOverallSignAmbiguity}
\end{equation}
which has however been removed by fixing a suitable phase-convention in the analysis, requiring one specific partial wave (for instance $A_{0}$) to be real and positive. \newline
We have to state that we consider Gersten's claim to be true. There really are no more ways to transform to disconnected points in amplitude space where the truncated PWA-model is ambiguous. \newline
However, having reformulated the Gersten-ambiguities in a language that fits the general continuum ambiguities of section \ref{sec:ContAmbs}, we would like to reformulate the claim in a different guise: \newline
{\it The phase-rotations $e^{i \varphi_{p} (W, \theta)}$ form a discrete sub-class of the general continuum ambiguity phases $e^{i \Phi (W, \theta)}$, representing all possible phase-rotations capable of rotating the original truncated amplitude $A(W,\theta)$ again into a truncated one. \newline
Thus, all the remaining infinite rotations contained in the larger class of symmetries $e^{i \Phi (W, \theta)}$ produce rotated models which are no longer truncated at $L$. The generating phases $e^{i \varphi_{p} (W, \theta)}$ are fully exhaustive in their capability to produce truncated models out of continuum ambiguity transformations.} \newline
Like Gersten, we do not have a precise mathematical proof of this claim. However, in the next section a numerical method is introduced capable of substantiating what has been stated above.

\section{Functional minimization formalism and the \textit{exhaustiveness} of the Gersten-ambiguities} \label{sec:FunctionalFormalism}

In the following, we again employ the notation $x = \cos \theta$. Furthermore, since phase-rotations (such as those in equation (\ref{eq:GerstenFormalismPhaseRotation})) will have to be searched numerically in what is to come, we switch from working with the phases themselves directly to the complex rotation-functions
\begin{equation}
 F (W,x) := e^{i \Phi(W,x)} \mathrm{.} \label{eq:PhaseRotFunctDefinition}
\end{equation}
Using the rotations has several advantages. Mainly, equations such as (\ref{eq:GerstenFormalismPhaseRotation}) only fix the phases $\Phi(W,x)$ themselves up to the branch-point singularity of the logarithm, which has to be encountered once the exponential is inverted. One could fix a convention, such as choosing the principal branch of the logarithm for the phases. Usage of the rotation functions circumvents this problem altogether.
In the following, we will sometimes loosely refer to the concept of vector spaces of functions. However, observe that the functions $F(W,x)$ do not form a vector space, since they do not close under addition and scalar multiplication. The functions $\Phi(W,x)$ on the other hand, do. \newline
From now on, we consider the action of the rotation (\ref{eq:PhaseRotFunctDefinition}) in a general continuum ambiguity transformation (\ref{eq:ContAmbTrafoDefinition}), i.e. $A(W,x) \rightarrow \tilde{A} (W,x) = F(W,x) A(W,x)$. The amplitude $A(W,x)$ is truncated at $L$ and a known input. \newline
In order to look for Gersten-type ambiguities, or potential further symmetries with similar properties, we solve the following two constraints at a fixed energy $W$:
\begin{itemize}
 \item[(I)] The rotated amplitude $\tilde{A}$, coming out of an amplitude $A$ truncated at $L$, has to be truncated as well, i.e.
 \begin{equation}
 \tilde{A}_{L + k} (W) = 0 \mathrm{,} \hspace*{5pt} \forall k = 1,\ldots,\infty \mathrm{.} \label{eq:RotAmplTruncatedAsWellRequireThis}
 \end{equation}
 \item[(II)] The complex solution-function $F (W, x)$ has to have modulus 1 for each value of $x$.
 \begin{equation}
 \left| F (W, x) \right|^{2} = 1 \mathrm{,} \hspace*{5pt} \forall x \in \left[ -1, 1 \right] \mathrm{.} \label{eq:PhaseModulus1}
 \end{equation}
\end{itemize}
The problem proposed here is a problem from functional analysis (or {functional problem} for short), since one tries to scan a full vector space of functions $\Phi(W,x)$ (implied up to logarithmic singularities by our solutions $F(W,x)$), for solutions of the problem. The obtained complex function is a solution to the infinite set of functional equations
\begin{align}
 \tilde{A}_{L + k} (W) &= \frac{1}{2} \int_{-1}^{+1} d x  F (W, x)  A (W, x) P_{L + k} (x) \equiv 0 \mathrm{,} \nonumber \\
 &\hspace*{22.75pt} \forall \hspace*{2pt} k = 1, \ldots, \infty \mathrm{.} \label{eq:FunctProblemIntEquations}
\end{align}
This set of equations corresponds to the formal statements of the functional problem we are trying to solve. However, it has to be clear that for any practical numerical calculation, an equation system built out of infinitely many functionals can never be solved. Therefore, in all practical examples we impose a restriction on the index $k$, making it range up to some finite, but sufficiently large, value $K_{\mathrm{cut}}$:
\begin{equation}
 k = 1, \ldots, K_{\mathrm{cut}} \mathrm{.} \label{eq:NIndexRestriction}
\end{equation}
Now, we formally define a quantity which, through it's minimization, allows for the solution of conditions (I) and (II) above. Also, due to the length of some of the ensuing expressions, explicit energy-dependences are in most cases implicit. The quantity to be minimized reads
\begin{widetext}
\begin{align}
 &\bm{W} \left[ F(x) \right] := \sum_{x} \left(  \mathrm{Re}\left[F(x)\right]^{2}  + \mathrm{Im}\left[F(x)\right]^{2} - 1 \right)^{2}  + \mathrm{Im} \left[  \frac{1}{2} \int_{-1}^{+1} dx \hspace*{1pt} F(x) A(x) \right]^{2}  \nonumber \\
  & \hspace*{52.5pt} + \sum_{k \geq 1} \Bigg\{ \mathrm{Re} \left[ \frac{1}{2} \int_{-1}^{+1} dx \hspace*{1pt} F(x) A(x) P_{L + k} (x) \right]^{2} + \mathrm{Im} \left[ \frac{1}{2} \int_{-1}^{+1} dx \hspace*{1pt} F(x) A(x) P_{L + k} (x) \right]^{2}  \Bigg\} \mathrm{.} \label{eq:FunctProblem}
\end{align}
\end{widetext}
This $\bm{W} \left[ F(x) \right]$ maps any whole phase-rotation function $F(x)$ to a real number. Therefore, it is also formally a functional. \newline
The individual terms in the minimization functional (\ref{eq:FunctProblem}) implement all the required constraints on the rotation-function $F(x)$. Minimization of the first term in the first line makes the function unimodular, cf. constraint (II) above. The sum $\sum_{x}$ is written in order to indicate that in any practical example, this term is evaluated on a discrete grid of equidistant points $\left\{ x_{n} \right\} \in [-1,1]$ (more on this below). The second term in the first line invokes an overall phase convention for the partial waves, by making the $S$-wave $A_{0}$ real. However, note that it does not make the latter positive (as in the convention declared in section \ref{sec:DiscrAmbs}), such that additional sign-ambiguities may be expected for the solutions. \newline
Finally, the third term filling the entire second line of equation (\ref{eq:FunctProblem}) formally implements constraint (I), by setting all partial wave projections above $\tilde{A}_{L}$ to zero, once it adopts it's minimum. In any practical minimization, the sum over $k$ is truncated at some $K_{\mathrm{cut}}$ (see equation (\ref{eq:NIndexRestriction})). \newline
We now come to the central statement of this section. We claim that once any suitable scheme for the minimization of the functional $\bm{W} \left[ F(x) \right]$ is applied, then those minima consistent with zero up to a good numerical approximation  will yield as solutions only the discrete Gersten-ambiguities. This can be written in idealized form as
\begin{align}
 \bm{W} \left[ F (W,\theta) \right] &\longrightarrow \mathrm{min.} \equiv 0 \mathrm{,} \label{eq:MinimFunctionalDef} \\
 \mathrm{for} \hspace*{2.5pt}  F (W,\theta) &\longrightarrow F_{p} (W,\theta) = e^{i \varphi_{p} (W, \theta)} \mathrm{,} \nonumber \\
 &\hspace*{23.5pt} p = 0,\ldots,(2^{L} - 1) \mathrm{.} \label{eq:FunctionLimit}
\end{align}
Of course, as mentioned below equation (\ref{eq:FunctProblem}), an additional sign-ambiguity exists due to the fact that the $S$-wave is only fixed to be real, but not positive, in our definition of $\bm{W} \left[ F(x) \right]$. However, such sign-ambiguities can be resolved easily, once the minimization has been performed. \newline
Any numerical scheme used to find general minima, or solution-functions, from the minimization of (\ref{eq:FunctProblem}) needs to implement some method to parametrize the functions $F(x)$ as generally as possible. Here, we employ a Legendre-expansion
\begin{equation}
 F (\left\{y_{\ell^{\prime}},w_{\ell^{\prime}}\right\}) (x) := \sum_{\ell' = 0}^{\mathcal{L}_{\mathrm{cut}}} (y_{\ell^{\prime}} + i w_{\ell^{\prime}}) P_{\ell^{\prime}} (x) \equiv \sum_{\ell' = 0}^{\mathcal{L}_{\mathrm{cut}}} L_{\ell^{\prime}} P_{\ell^{\prime}} (x)  \mathrm{,} \label{eq:PhaseRotTrafoLegendreSeriesTruncated}
\end{equation}
with $y_{\ell^{\prime}} = \mathrm{Re} \left[ L_{\ell^{\prime}} \right]$ and $w_{\ell^{\prime}} = \mathrm{Im} \left[ L_{\ell^{\prime}} \right]$. The latter are the parameters for which to solve. The expansion (\ref{eq:PhaseRotTrafoLegendreSeriesTruncated}) becomes numerically tractable with a truncation at some, possibly large, expansion index $\mathcal{L}_{\mathrm{cut}}$.\footnote{Angle-dependent rotations are always, strictly speaking, infinite expansions in $x$ (see section \ref{sec:ContAmbs}). However, in practical cases it is clearly impossible to solve for infinitely many Legendre coefficients. \\ \\ \\ With the finite expansion, we want to simulate a convergent infinite series. For practical examples, $\mathcal{L}_{\mathrm{cut}}$ has to be chosen much larger than the order for which the calculable Gersten-phases already achieve a good convergence. Then, this Ansatz works numerically, as illustrated by the example below.} \newline
Since the partial waves $A_{\ell}$ of the non-rotated amplitude are assumed to be known, one can directly use the mixing-formula (\ref{eq:MixingFormulaBoxed}) in order to parametrize the rotated partial waves $\tilde{A}_{\ell}$ (above $L$) in terms of the minimization parameters $y_{\ell^{\prime}}$ and $w_{\ell^{\prime}}$. This has the advantage of avoiding the need to explicitly implement numerical integration into the minimization-routine. However, the mixing-formula as used here is slightly modified due to two facts: first of all the original amplitude is truncated at $L$, secondly the Legendre expansion of the phase-rotation is cut off at $\mathcal{L}_{\mathrm{cut}}$. The result reads
\begin{widetext}
\begin{equation}
 \tilde{A}_{L + k} (\left\{y_{\ell^{\prime}},w_{\ell^{\prime}}\right\}) = \sum_{\ell^{\prime} = k}^{\mathrm{min} \left(2 L + k, \mathcal{L}_{\mathrm{cut}}\right)} (y_{\ell^{\prime}} + i w_{\ell^{\prime}}) \hspace*{2pt} \sum_{m = \left| L + k - \ell^{\prime} \right|}^{ L } \left< \ell^{\prime},0 ; \ell,0 | m , 0 \right>^{2} A_{m} \mathrm{,} \hspace*{5pt} \forall \hspace*{2pt} k = 1, \ldots, K_{\mathrm{cut}} \mathrm{.} \label{eq:MixingFormulaTruncatedOriginalAmplitudeForFit}
\end{equation}
\end{widetext}
This expression implies that in the chosen Ansatz for the functional minimization, the maximal index $\mathcal{L}_{\mathrm{cut}}$ sets a limit on the parameter $K_{\mathrm{cut}}$. The maximal choice, which we always use in the following, is
\begin{equation}
 K_{\mathrm{cut}} = \mathcal{L}_{\mathrm{cut}} \mathrm{,} \label{eq:MaximalChoiceOfKcut}
\end{equation}
%
The minimization scheme based on the Legendre-parametrization (\ref{eq:PhaseRotTrafoLegendreSeriesTruncated}) and the mixing formula (\ref{eq:MixingFormulaTruncatedOriginalAmplitudeForFit}) has turned out to be quite well-behaved numerically. Another, less favorable, Ansatz for the parametrization of $F(x)$ consists of using a discretization of this function for a discrete set of values $\left\{ x_{n} \right\} \in [-1,1]$. We briefly summarize this alternative procedure in appendix \ref{sec:FunctProblemDiscretizationAnsatz}, but do not utilize it further in the main discussion. \newline
Since the first term in the functional (\ref{eq:FunctProblem}) features a summation over $x$ in any case, a discrete grid of points $\left\{ x_{n} \right\} \in [-1,1]$ is needed for the Legendre-Ansatz as well. We employ a total number of $N_{I}$ equidistant points with separation
\begin{equation}
 \Delta x = \frac{1 - (-1)}{N_{I}} = \frac{2}{N_{I}} \mathrm{.} \label{eq:DelatXDef1}
\end{equation}
To define this sequence of base-points, a simple prescription is used:
\begin{equation}
\hspace*{65pt} x_{n} := - 1 + \left( \frac{1 + 2 (n - 1)}{2} \right)  \Delta x \mathrm{.} \hspace*{15pt} \label{eq:BasePointsX1}
\end{equation}
\hspace*{-50pt} \newline \newline
The points therefore make up the set \newline
%
\begin{equation}
x_{n} \in \Bigg\{  \frac{\Delta x}{2} -1 , \ldots , \left( \frac{ 1 + 2( N_{I} - 1)  }{2} \right) \ast \Delta x -1  \Bigg\} \mathrm{.} \label{eq:BasePointsXSet1}
\end{equation}
%
\newline
Using the definitions (\ref{eq:PhaseRotTrafoLegendreSeriesTruncated}), (\ref{eq:MixingFormulaTruncatedOriginalAmplitudeForFit}) and (\ref{eq:BasePointsX1}), as well as the fact that the truncation order $L$ and partial waves $A_{\ell}$ of the original amplitude are known input, the functional $\bm{W} \left[ F(x) \right]$ can be written as an ordinary function depending on the parameters $\left\{ y_{\ell^{\prime}},w_{\ell^{\prime}} \right\}$. The result, which is then optimized in the Legendre-Ansatz, becomes
\begin{widetext}
 \begin{align}
 \bm{W}_{\mathcal{L}} \left( \left\{y_{\ell^{\prime}},w_{\ell^{\prime}}\right\} \right) &:= \sum_{\left\{x_{n}\right\}} \left(  \mathrm{Re} \left[F (\left\{y_{\ell^{\prime}},w_{\ell^{\prime}}\right\}) (x_{n})\right]^{2}  + \mathrm{Im}\left[F (\left\{y_{\ell^{\prime}},w_{\ell^{\prime}}\right\}) (x_{n})\right]^{2} - 1 \right)^{2} + \mathrm{Im}\left[ \tilde{A}_{0} (\left\{y_{\ell^{\prime}},w_{\ell^{\prime}}\right\}) \right]^{2} \nonumber \\
  & + \hspace*{1pt} \sum_{k = 1}^{K_{\mathrm{cut}}} \left( \mathrm{Re} \left[\tilde{A}_{L + k} (\left\{y_{\ell^{\prime}},w_{\ell^{\prime}}\right\})\right]^{2} + \mathrm{Im} \left[\tilde{A}_{L + k} (\left\{y_{\ell^{\prime}},w_{\ell^{\prime}}\right\})\right]^{2} \right) \mathrm{.} \label{eq:FunctProblemLegendreVersion1}
\end{align}
\end{widetext}
A useful feature of model-independent expansions into basis-functions such as (\ref{eq:PhaseRotTrafoLegendreSeriesTruncated}) is that, once they are employed, complicated functionals become just ordinary functions depending on the expansion-coefficients. The explicit mathematical form of the function (\ref{eq:FunctProblemLegendreVersion1}) is elaborated in more detail in appendix \ref{sec:MinimizationFunctionalAsOrdinaryFunction} but for the ensuing discussion, it is not really needed. \newline
An open remaining question is about which initial conditions for the $\left\{y_{\ell^{\prime}},w_{\ell^{\prime}}\right\}$ to choose for the minimization process. We employ an ensemble consisting of $N_{\mathrm{MonteCarlo}}$ sets of start-parameters. How many to choose depends on the order $L$ of the original truncated model. Mostly, we employed values around $N_{\mathrm{MonteCarlo}} = 50,\ldots,100$ for the treatment of simple toy-model examples, with generally satisfactory results. \newline
For the precise method to generate the $N_{\mathrm{MonteCarlo}}$ start-configurations, we have made good experiences by just drawing each parameter randomly from the interval $\left[ -1,1 \right]$, for example by using $\mathrm{RandomReal}\left[ \left\{ -1 , 1 \right\} \right]$ in MATHEMATICA. Also, all numerical minimizations shown in the following have been done with MATHEMATICA. \newline

\begin{figure*}[h]
\centering
\includegraphics[width=0.4315\textwidth]{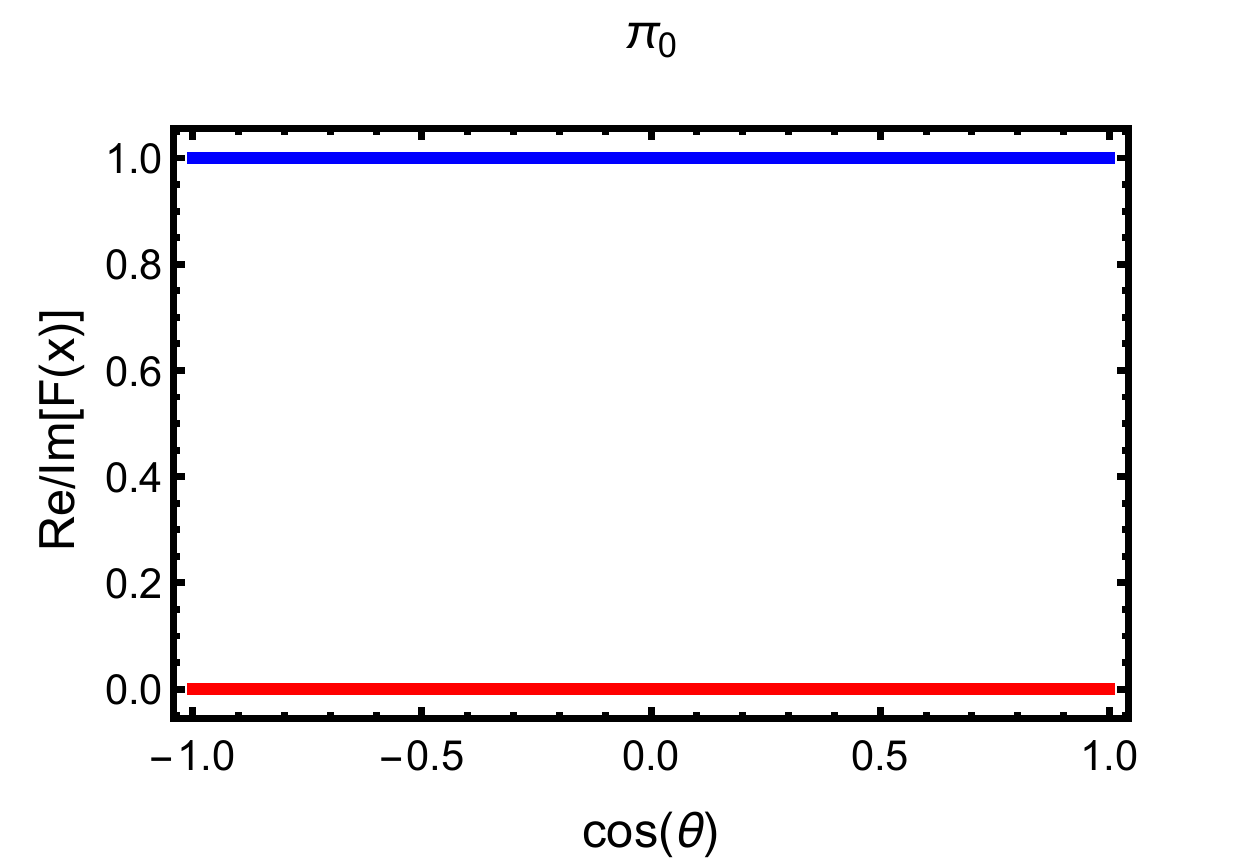}
\includegraphics[width=0.4315\textwidth]{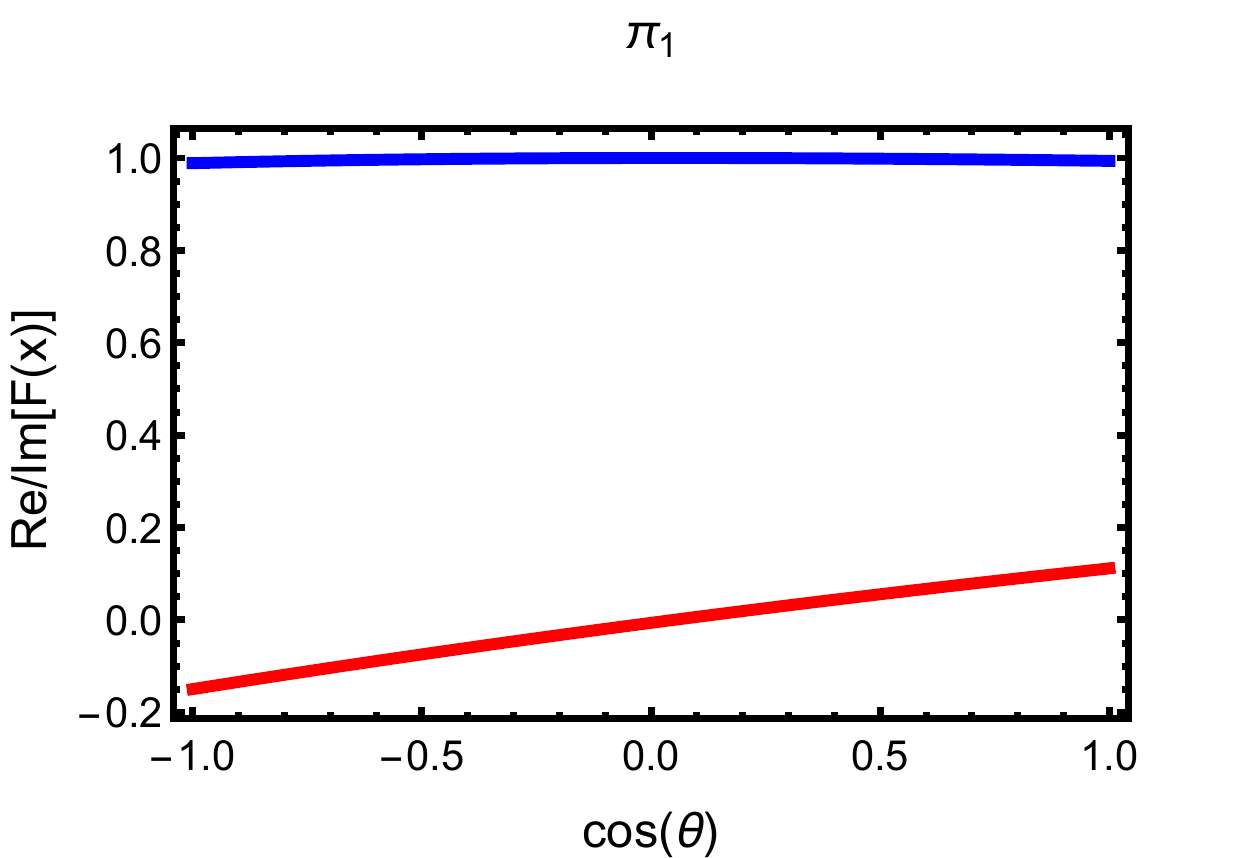} \\
\includegraphics[width=0.4315\textwidth]{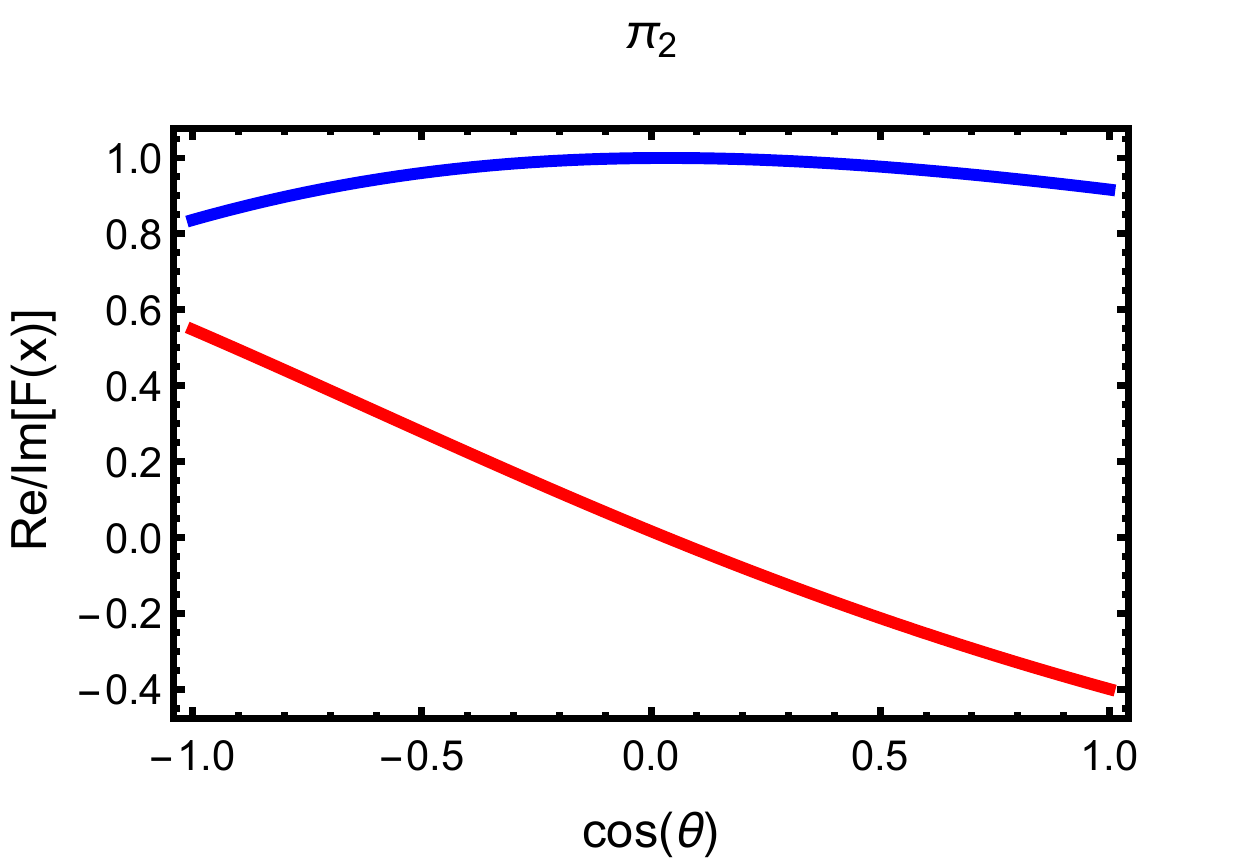}
\includegraphics[width=0.4315\textwidth]{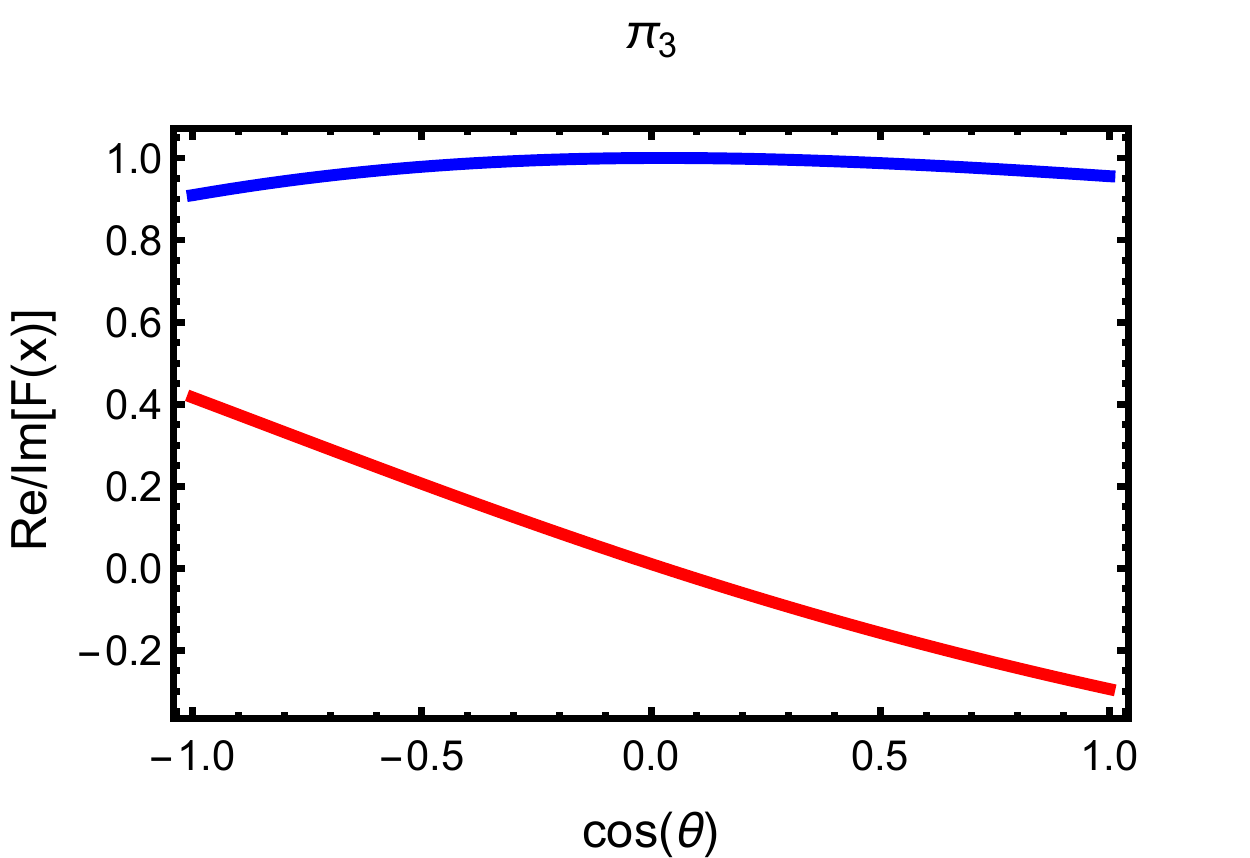}
\caption{These plots show the real- (blue) and imaginary-parts (red) of the phase-rotations (\ref{eq:GerstenPhasesToyModel}) extracted from the toy-model amplitude (\ref{eq:DefLmax2ToyModel}) defined in the main text. The individual figures are labelled via the respective ambiguity $\bm{\uppi}_{\hspace*{0.45pt}p}$ belonging to each phase $e^{i \varphi_{p} (x)}$. (color online)}
\label{fig:GerstenAmbiguityGeneratingPhases}
\end{figure*}

What remains to be done is to demonstrate the machinery presented in this section on a particular example. We consider a simple toy-model consisting of an amplitude truncated at $L=2$, with partial waves given in arbitrary units:

\begin{align}
 A(x) &= \sum_{\ell = 0}^{2} (2 \ell + 1) A_{\ell} P_{\ell} (x) \nonumber \\
  &= A_{0} + 3 A_{1} P_{1} (x) + 5 A_{2} P_{2} (x) \nonumber \\
  &= 5 + 3 (0.4 + 0.3 \hspace*{1pt} i) x + \frac{5}{2} (0.02 + 0.01 \hspace*{1pt} i) (3 x^{2} - 1) \mathrm{.} \label{eq:DefLmax2ToyModel}
\end{align}
Note that in addition to the truncation, this model is constructed in such a way that the non-vanishing partial waves show a soft convergence-behavior. Once the Gersten-decomposition (\ref{eq:GerstenDecomposedScalarAmplitude}) is computed for this example, the following values for the complex normalization-factor
\begin{equation}
 \lambda = 0.15 + 0.075 \hspace*{1pt} i \mathrm{,} \label{eq:LambdaFactorToyModel}
\end{equation}
as well as the two roots
\begin{align}
 \alpha_{1} &= - 7.05858 - 4.63163 \hspace*{1pt} i \mathrm{,} \label{eq:GerstenRoot1ToyModel} \\
 \alpha_{2} &= - 1.74142 + 3.03163 \hspace*{1pt} i \mathrm{,} \label{eq:GerstenRoot2ToyModel}
\end{align}
are obtained. Since the toy-model (\ref{eq:DefLmax2ToyModel}) is truncated at $L=2$, there exist $2^{2} = 4$ Gersten-ambiguities, which accoring to equations (\ref{eq:GerstenMapsScalarExample}) and (\ref{eq:BinaryRepresentationForP}) are enumerated as follows
\begin{align}
    \bm{\uppi}_{\hspace*{0.45pt}0} \left(\alpha_{1}, \alpha_{2}\right) &= \left( \alpha_{1}, \alpha_{2}  \right) &\mathrm{,} \quad \bm{\uppi}_{\hspace*{0.45pt}1} \left(\alpha_{1}, \alpha_{2}\right) &= \left( \alpha_{1}^{\ast}, \alpha_{2}  \right) \mathrm{,} \label{eq:GerstenLmax2Ambiguities1} \\
  \bm{\uppi}_{\hspace*{0.45pt}2} \left(\alpha_{1}, \alpha_{2}\right) &= \left( \alpha_{1}, \alpha_{2}^{\ast}  \right) &\mathrm{,} \quad \bm{\uppi}_{\hspace*{0.45pt}3} \left(\alpha_{1}, \alpha_{2}\right) &= \left( \alpha_{1}^{\ast}, \alpha_{2}^{\ast}  \right) \mathrm{.} \label{eq:GerstenLmax2Ambiguities2}
 \end{align}
The generating phases of the discrete ambiguities (\ref{eq:GerstenLmax2Ambiguities1}) and (\ref{eq:GerstenLmax2Ambiguities2}) can be evaluated using equation (\ref{eq:GerstenFormalismPhaseRotation}) from section \ref{sec:DiscrAmbs}. Four different rotations are obtained
\begin{equation}
 e^{i \varphi_{0} (x)} = 1 \mathrm{,} \hspace*{2.5pt} e^{i \varphi_{1} (x)} \mathrm{,} \hspace*{2.5pt} e^{i \varphi_{2} (x)} \mathrm{,} \hspace*{2.5pt} e^{i \varphi_{3} (x)} \mathrm{,} \label{eq:GerstenPhasesToyModel}
\end{equation}
with all of them, except for the phase of the identity $\bm{\uppi}_{\hspace*{0.45pt}0}$, depending on $x = \cos \theta$ (energy-dependencies supressed). \newline
The phase-rotations (\ref{eq:GerstenPhasesToyModel}) are plotted in Figure \ref{fig:GerstenAmbiguityGeneratingPhases} as complex functions of $x$. Their Legendre coefficients, up to and including $L_{8}$, are collected in Table \ref{tab:ToyModelLegCoeffs}. Apart from the trivial dependence of $e^{i \varphi_{0}(x)}$, the remaining phase-rotations $e^{i \varphi_{1} (x)}$, $e^{i \varphi_{2} (x)}$ and $e^{i \varphi_{3} (x)}$ show a relatively quick convergence. This makes the toy-model (\ref{eq:DefLmax2ToyModel}) a well-suited example for the demonstration of the functional minimization formalism, since the range $\mathcal{L}_{\mathrm{cut}}$ of the Legendre-parametrization (\ref{eq:PhaseRotTrafoLegendreSeriesTruncated}) can be chosen comparatively low, making the calculations numerically tractable. \newline
With the toy-model (\ref{eq:DefLmax2ToyModel}) as input, we performed a numerical minimization of the abstract functional (\ref{eq:FunctProblem}). The Legendre-parametrization (\ref{eq:PhaseRotTrafoLegendreSeriesTruncated}) for the phase-rotations was utilized, such that the procedure reduced to the optimization of the ordinary function (\ref{eq:FunctProblemLegendreVersion1}), with the Legendre-coefficients $\left\{y_{\ell^{\prime}}, w_{\ell^{\prime}}\right\}$ as free parameters of the problem. The truncation-orders
\begin{equation}
 \mathcal{L}_{\mathrm{cut}} = K_{\mathrm{cut}} = 20 \mathrm{,} \label{eq:MaximalChoiceOfKcutExampleValues}
\end{equation}
were employed. Minimizations started from an ensemble of $N_{\mathrm{MonteCarlo}} = 50$ different initial parameter-configurations. The angular interval $x \in [-1,1]$ has been divided into $N_{I} = 400$ equidistant points $\left\{ x_{n} \right\}$. \newline
As a result of the functional minimization, we report that the anticipated exhaustiveness of the Gersten-ambiguities, formulated generally in equations (\ref{eq:MinimFunctionalDef}) and (\ref{eq:FunctionLimit}) above, has been fully confirmed. In the case at hand, this fact may be briefly expressed as
\begin{align}
 \bm{W}_{\mathcal{L}} \left( \left\{ y_{\ell^{\prime}}, w_{\ell^{\prime}} \right\} \right) &\longrightarrow \mathrm{min.} \equiv 0 \mathrm{,} \label{eq:MinimFunctionalDefToyModel} \\
 \mathrm{for} \hspace*{2.5pt}  F (x) &\longrightarrow e^{i \varphi_{p} (x)} \mathrm{,} \hspace*{2pt} p = 0,\ldots,3 \mathrm{.} \label{eq:FunctionLimitToyModel}
\end{align}
The consistency of the functional-minimum (\ref{eq:MinimFunctionalDefToyModel}) with zero means in this practical numerical case that the values of the adopted minima range around $10^{-29}, \ldots, 10^{-30}$. Local minima are found as well, but they are typically separated from the global (mathematical) minima by many orders of magnitude. They typically correspond to values of the order $1$ for the functional. These results have not been modified by raising $N_{\mathrm{MonteCarlo}}$. \newline
A graphical representation of the convergence-process for the functional minimizations is provided in Figures \ref{tab:FunctMinConvergencePlots1} and \ref{tab:FunctMinConvergencePlots2} below. There, four different randomly chosen initial functions have been picked, each of them leading to a different Gersten-ambiguity in the process of minimization. Then, numerical minimizations have been performed for eight different ascending values of the maximal number of iterations $N_{\mathrm{max}}$. For the maximal value $N_{\mathrm{max}} = 500$, the minimizations have converged to the precise Gerten-ambiguity in any case. However, apart from that, differences can be observed in the speed of convergence. 


%
\begin{table*}[h]
\centering
 \begin{tabular}{r|c|c|c|c}
  $L_{k}$ & $e^{i \varphi_{0} (x)}$ & $e^{i \varphi_{1} (x)}$ & $e^{i \varphi_{2} (x)}$ & $e^{i \varphi_{3} (x)}$ \\
  \hline
 $L_{0}$ &  $1$ & $0.997-0.01049 i$ & $0.95864+0.03697 i$ & $0.97741+0.02781 i$ \\
 $L_{1}$ & $0$ & $0.00182+0.13038\hspace*{1pt} i$ & $0.02769-0.48277\hspace*{1pt} i$ & $0.01563-0.35988\hspace*{1pt} i$ \\
 $L_{2}$ & $0$ & $-0.00581-0.00852\hspace*{1pt} i$ & $-0.08227+0.03939\hspace*{1pt} i$ & $-0.04507+0.03429\hspace*{1pt} i$ \\
 $L_{3}$ & $0$ & $0.00068+0.00028\hspace*{1pt} i$ & $0.0126+0.009\hspace*{1pt} i$ & $0.00769+0.00427\hspace*{1pt} i$ \\
 $L_{4}$ & $0$ & $-0.00005+9.6\ast 10^{-6} i$ & $0.00029-0.00249\hspace*{1pt} i$ & $0.00005-0.00148\hspace*{1pt} i$ \\
$L_{5}$ & $0$ & $2.3\ast 10^{-6} -2.3\ast 10^{-6} i$ & $-0.00037+0.00015\hspace*{1pt} i$ & $-0.00021+0.00011\hspace*{1pt} i$ \\
$L_{6}$ & $0$ & $-4.4\ast 10^{-8} +2.1\ast 10^{-7} i$ & $0.00005+0.00004\hspace*{1pt} i$ & $0.00003+0.00002\hspace*{1pt} i$ \\
$L_{7}$ & $0$ & $-4.98\ast 10^{-9} -1.3\ast 10^{-8} i$ & $1.6\ast 10^{-6} -9.1\ast 10^{-6} i$ & $4.0\ast 10^{-7} -5.5\ast 10^{-6} i$ \\
$L_{8}$ & $0$ & $7.0\ast 10^{-10} +4.9\ast 10^{-10} i$ & $-1.3\ast 10^{-6} +4.5\ast 10^{-7} i$ & $-7.6\ast 10^{-7} +3.5\ast 10^{-7} i$
 \end{tabular}
\caption{This Table collects the Legendre coefficients of the phase rotations (\ref{eq:GerstenPhasesToyModel}) corresponding to the toy-model amplitude (\ref{eq:DefLmax2ToyModel}) defined in the main text. All coefficients up to $L_{8}$ are shown. All numbers are printed to $5$ significant digits, in order to illustrate the quick convergence of these examples.}
\label{tab:ToyModelLegCoeffs}
\end{table*}
%


%
\begin{table*}[h]
\centering
\underline{\begin{Large}$e^{i \varphi_{0} (x)}$\end{Large}} \vspace*{5pt} \\
\includegraphics[width=0.245\textwidth]{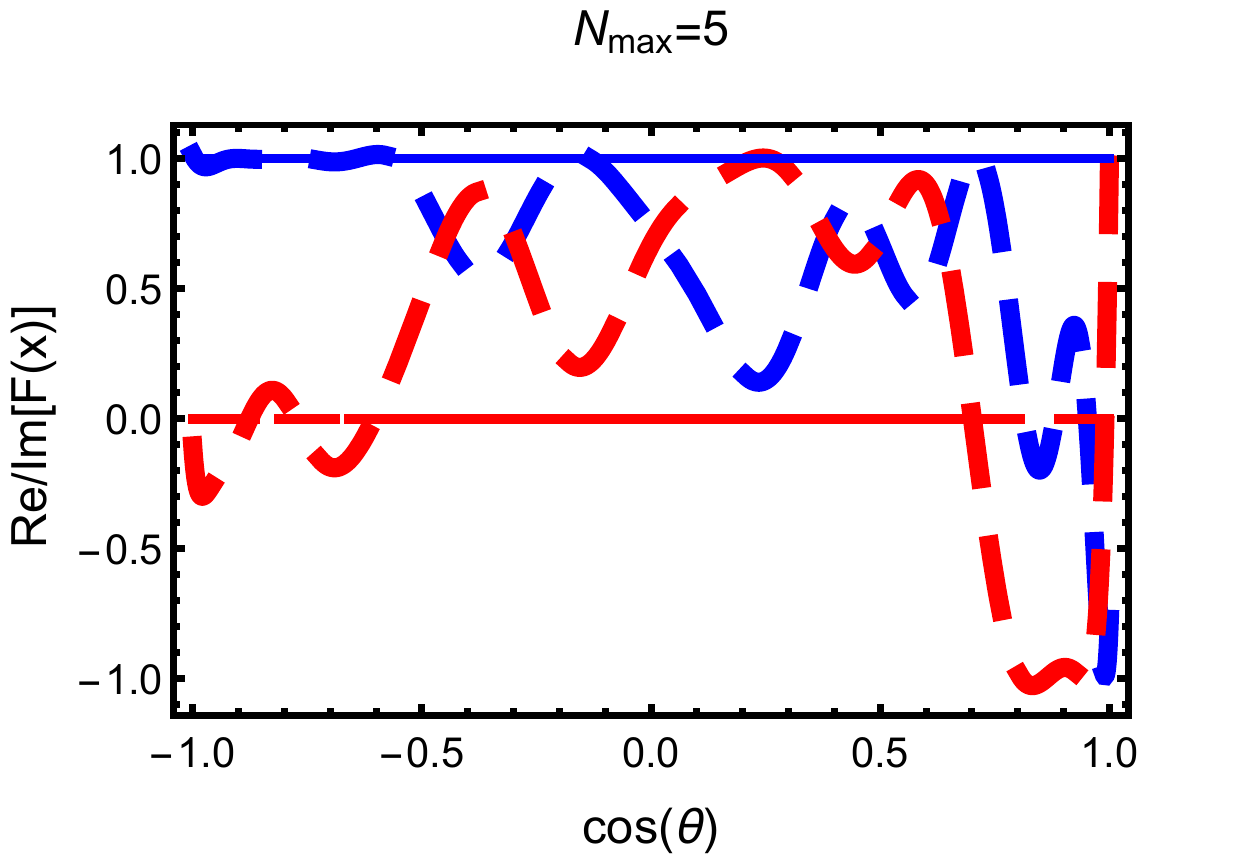}
\includegraphics[width=0.245\textwidth]{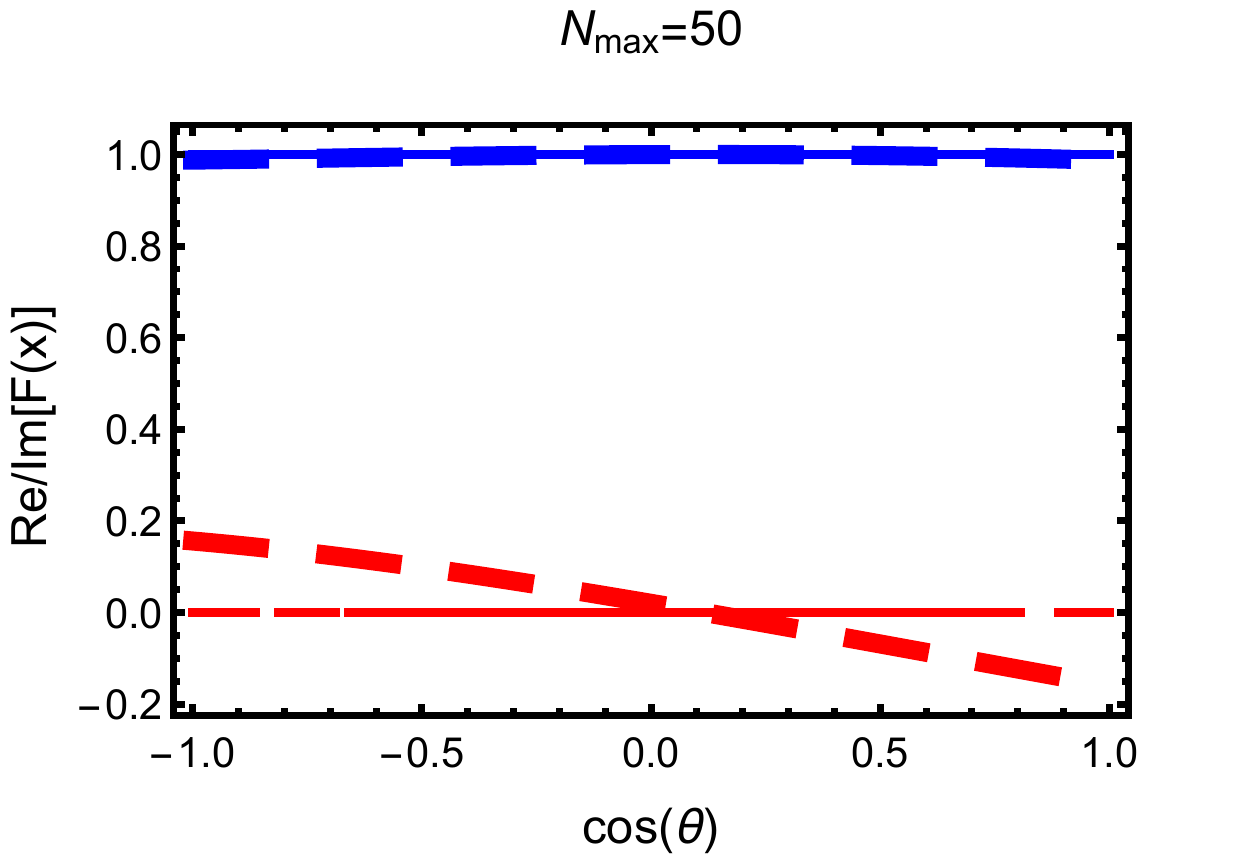}
\includegraphics[width=0.245\textwidth]{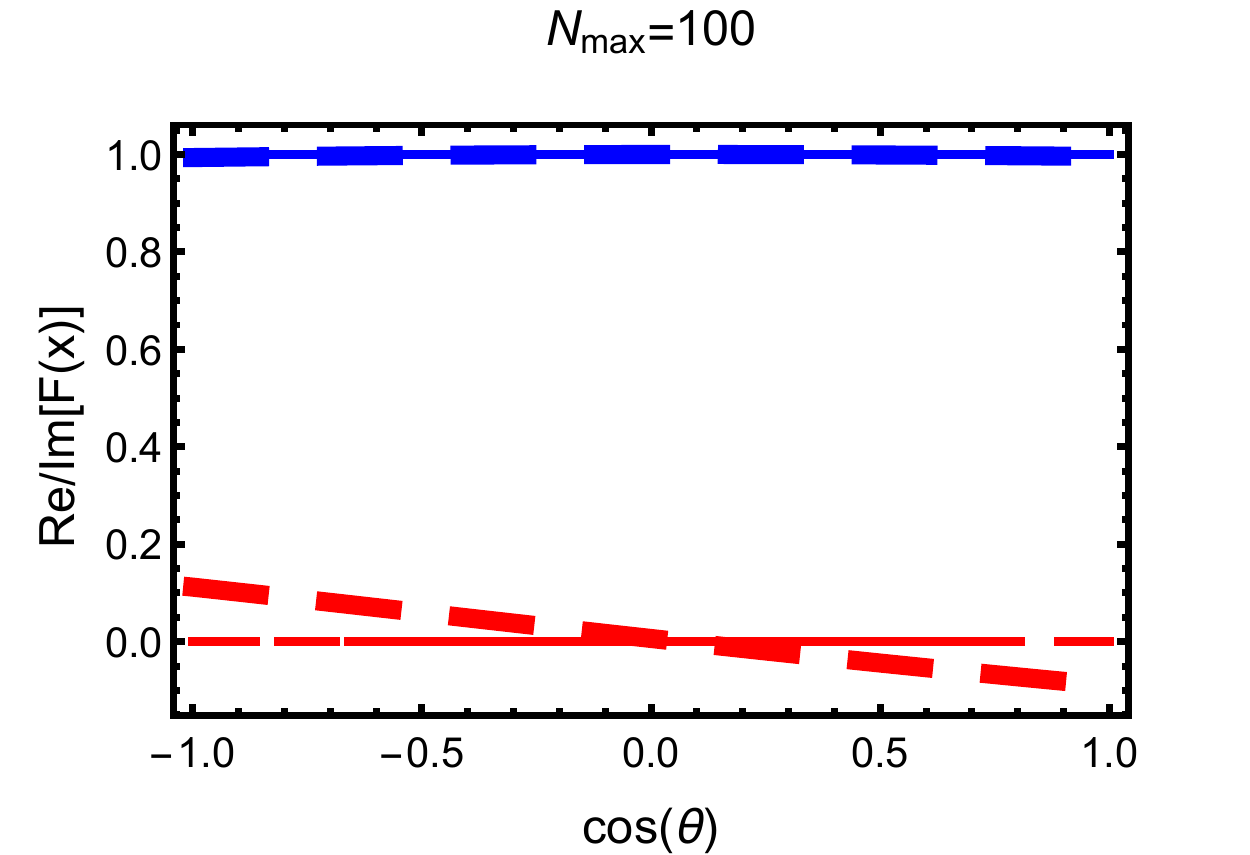}
\includegraphics[width=0.245\textwidth]{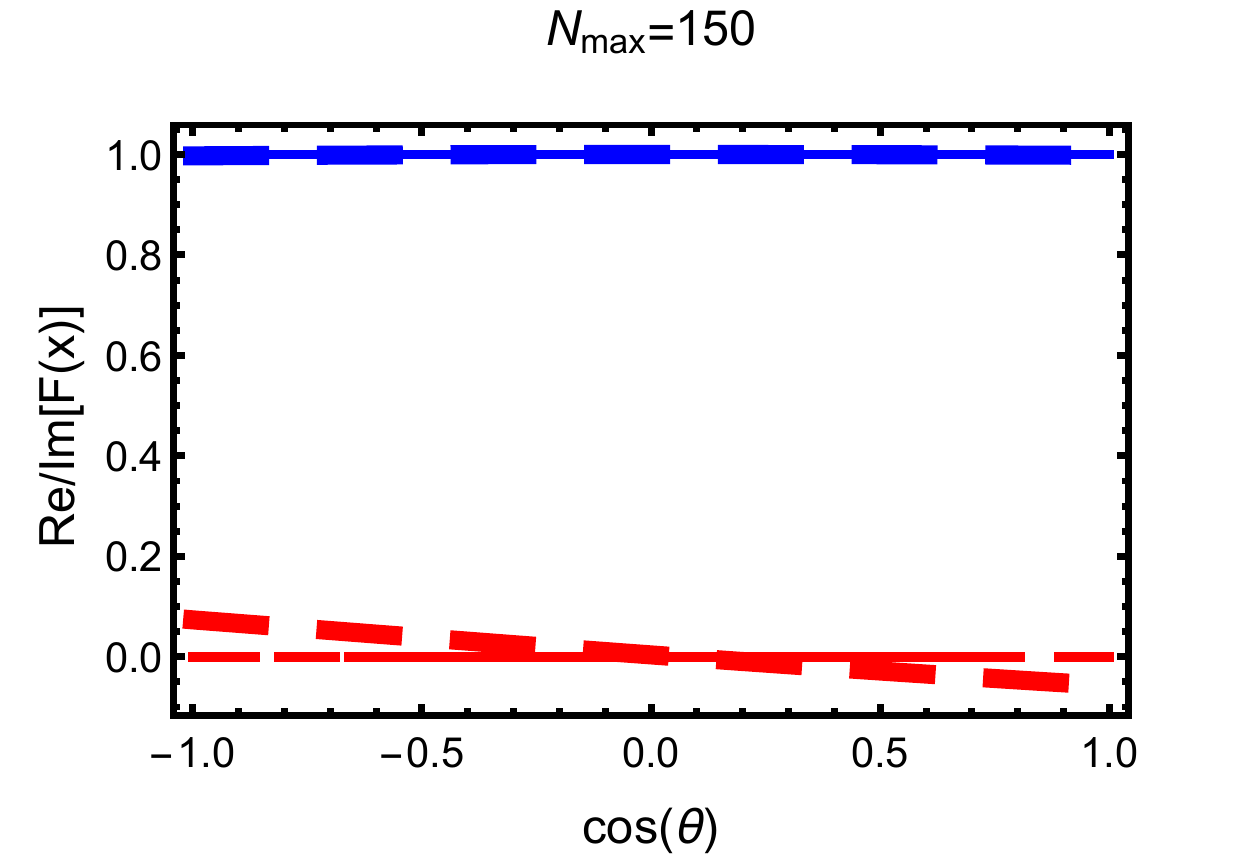} \\
\includegraphics[width=0.245\textwidth]{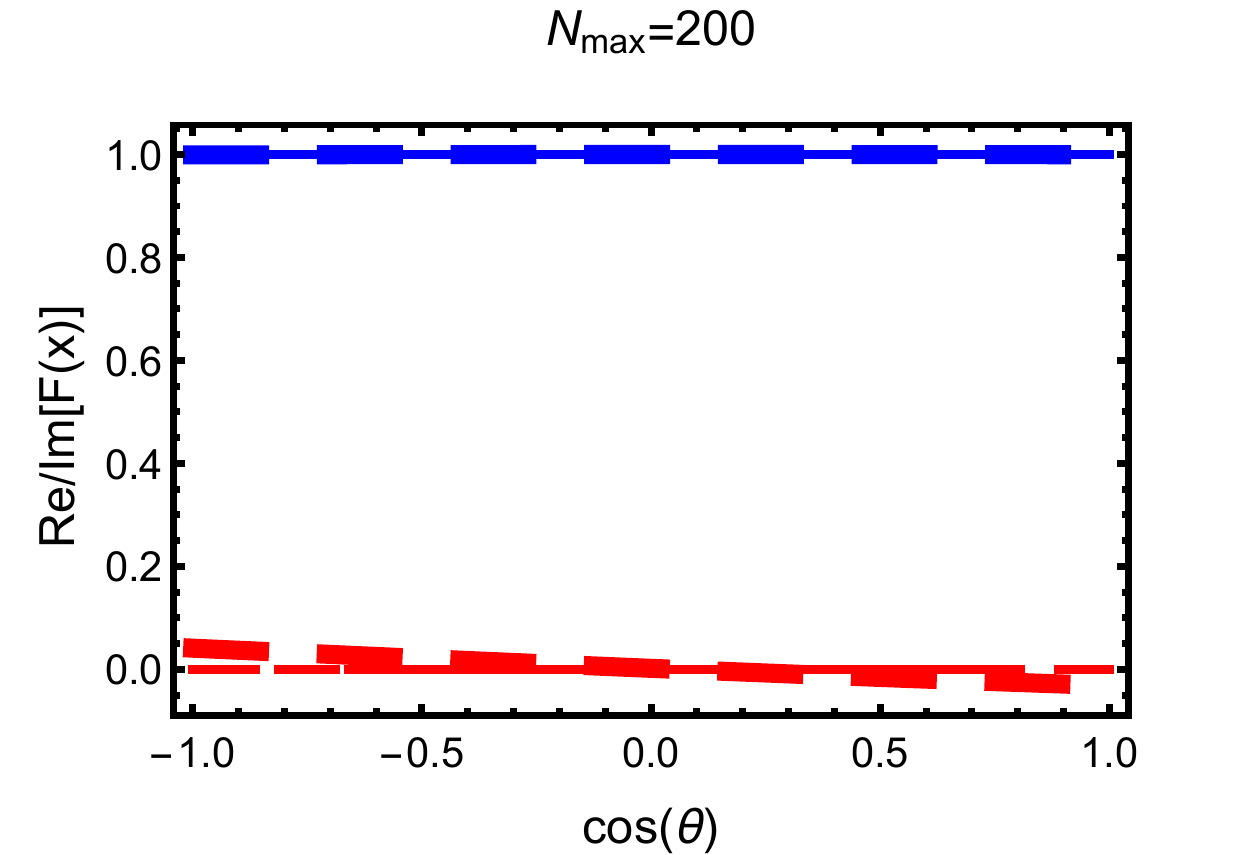}
\includegraphics[width=0.245\textwidth]{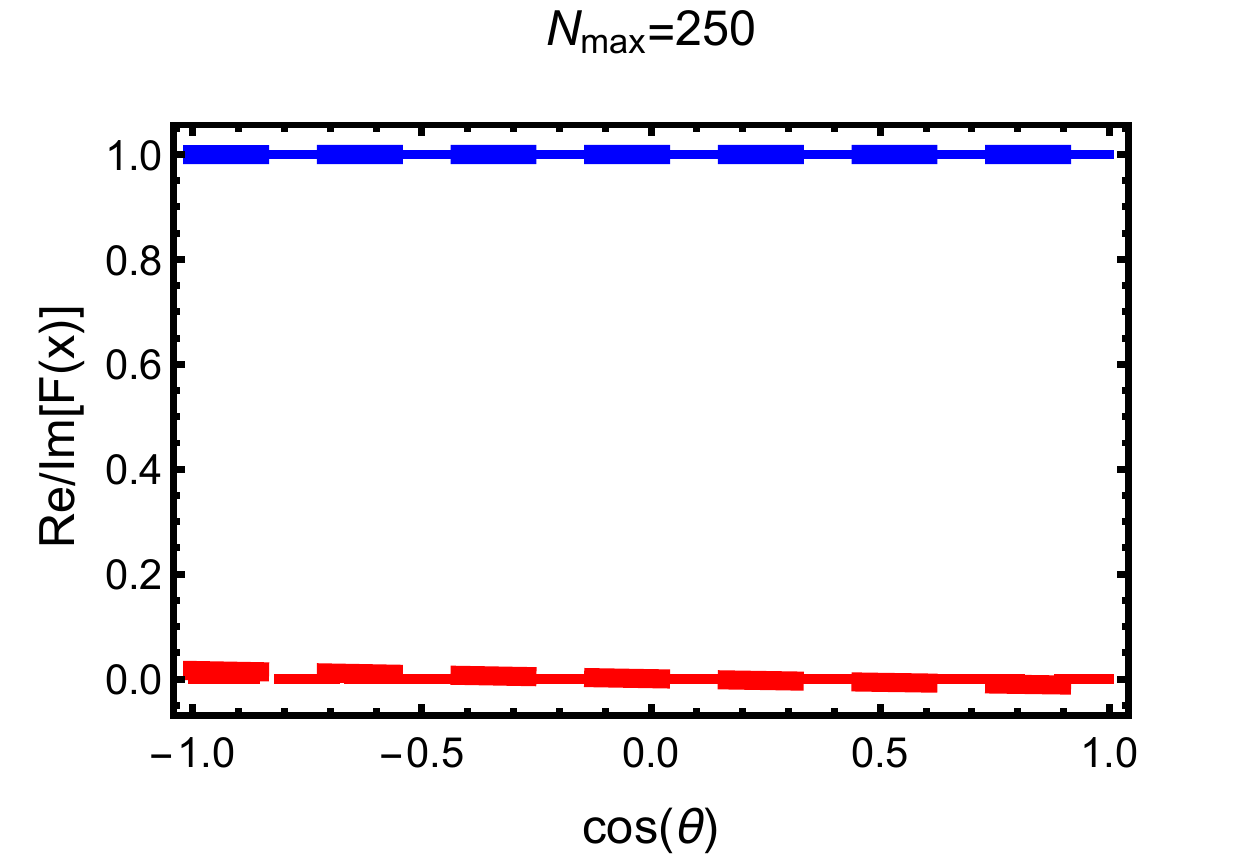}
\includegraphics[width=0.245\textwidth]{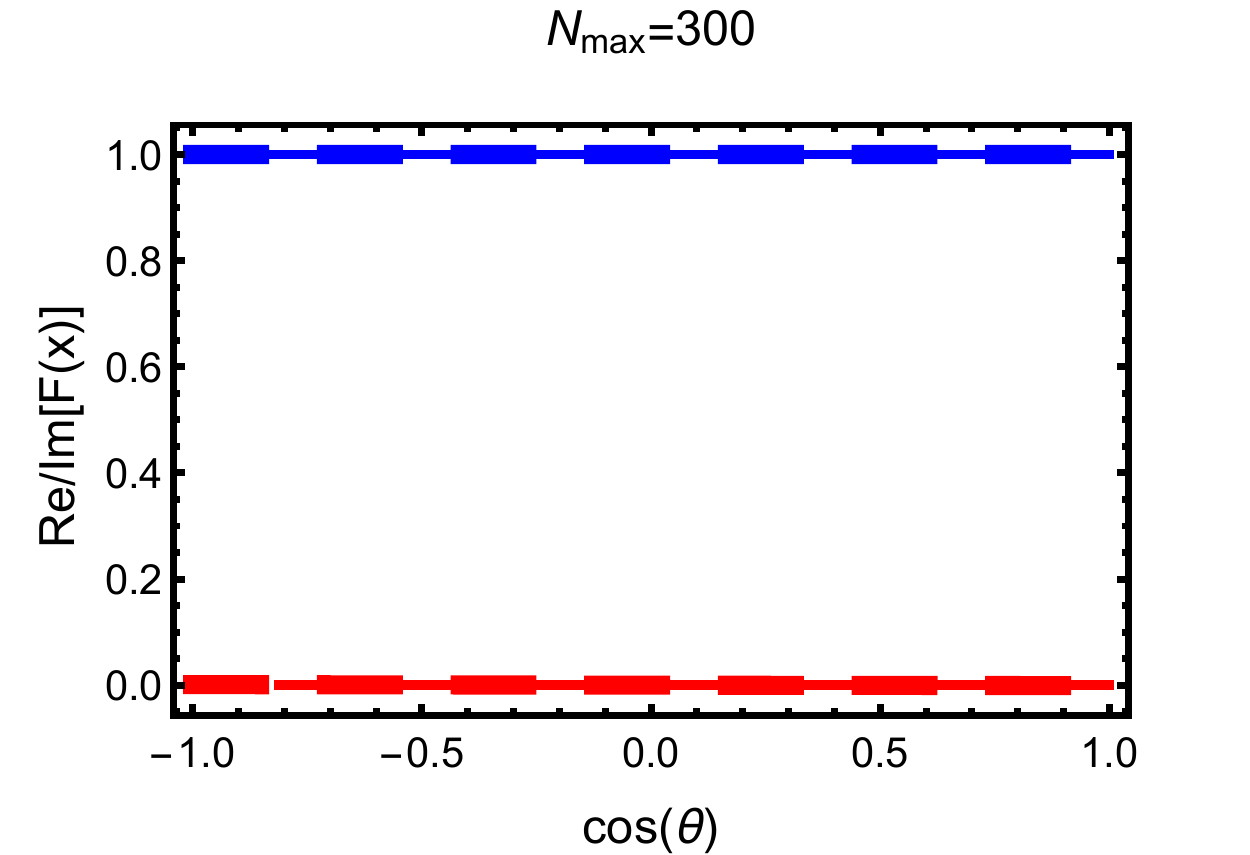}
\includegraphics[width=0.245\textwidth]{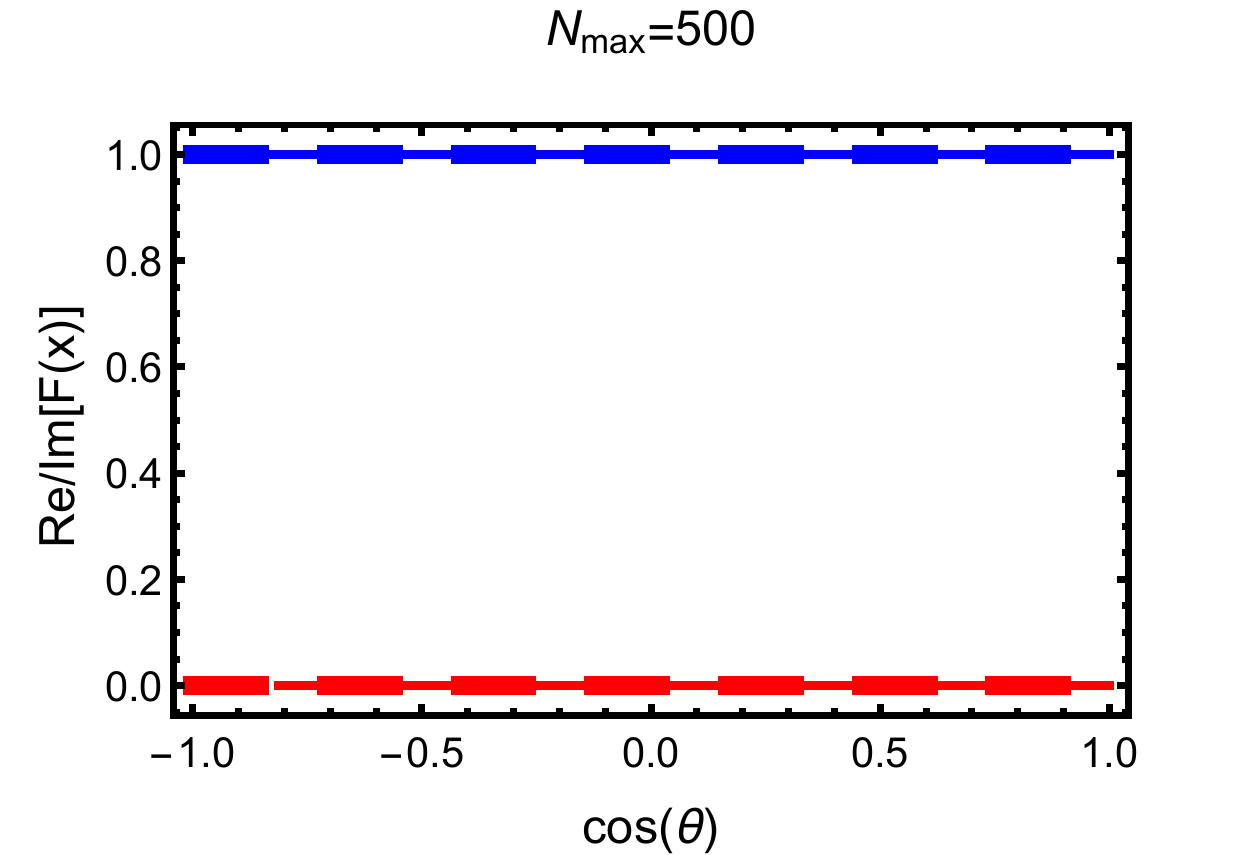} \\
\line(1,0){500} \\
\vspace*{5pt} \underline{\begin{Large}$e^{i \varphi_{1} (x)}$\end{Large}} \vspace*{5pt} \\
\includegraphics[width=0.245\textwidth]{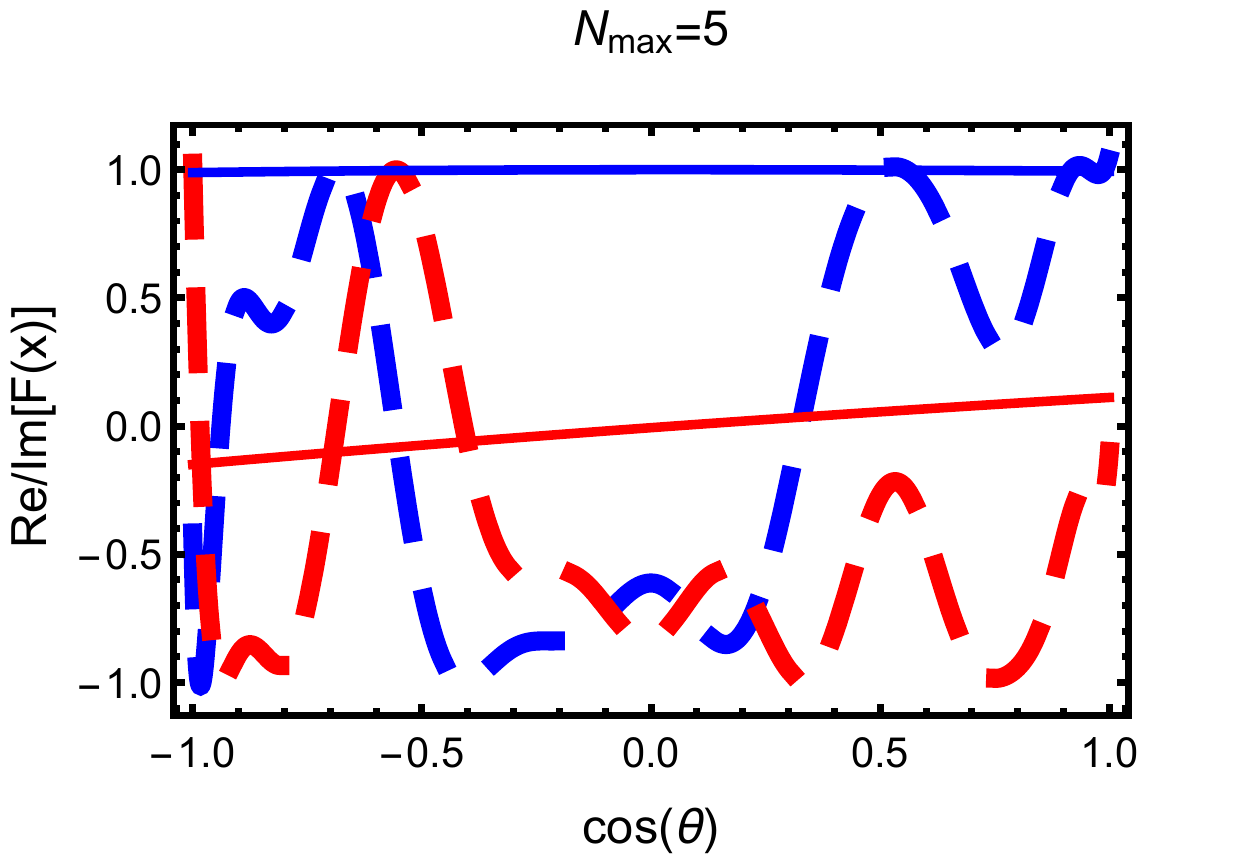}
\includegraphics[width=0.245\textwidth]{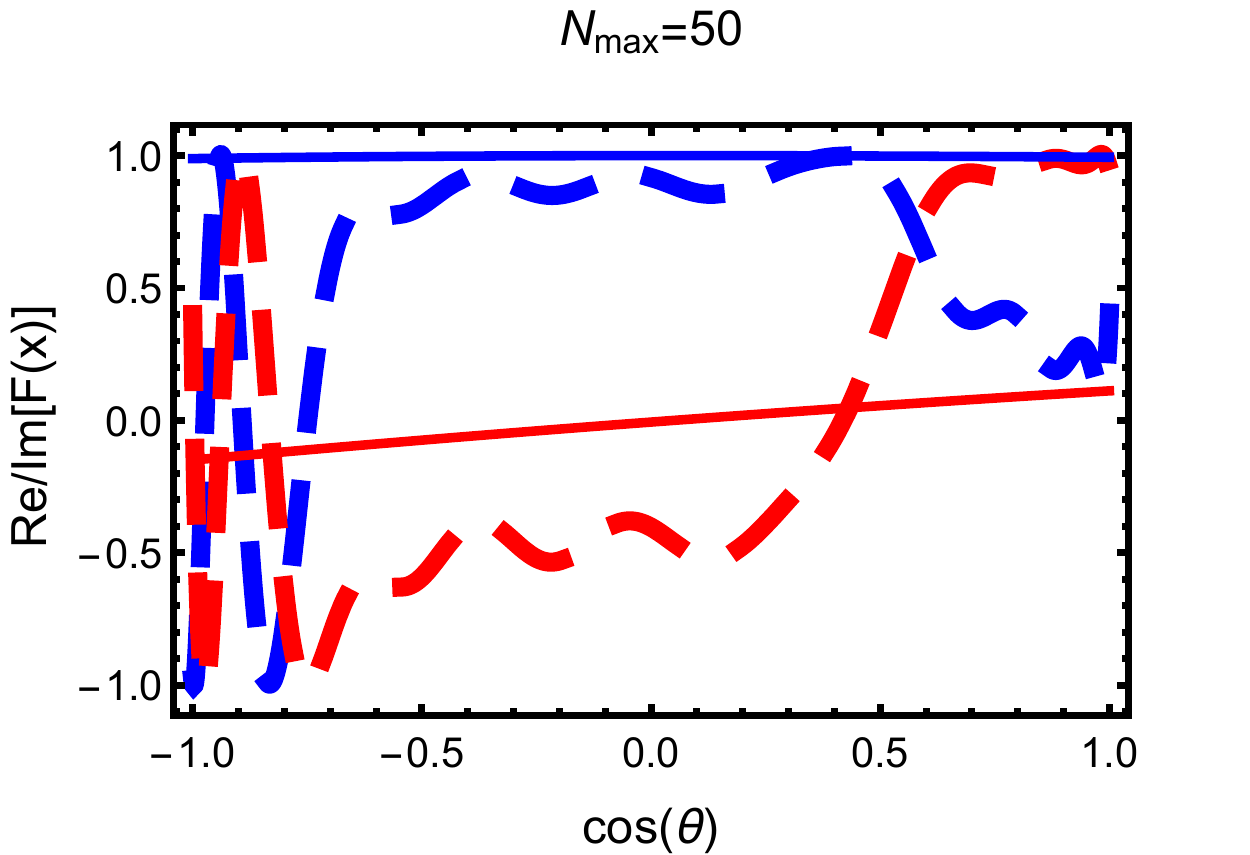}
\includegraphics[width=0.245\textwidth]{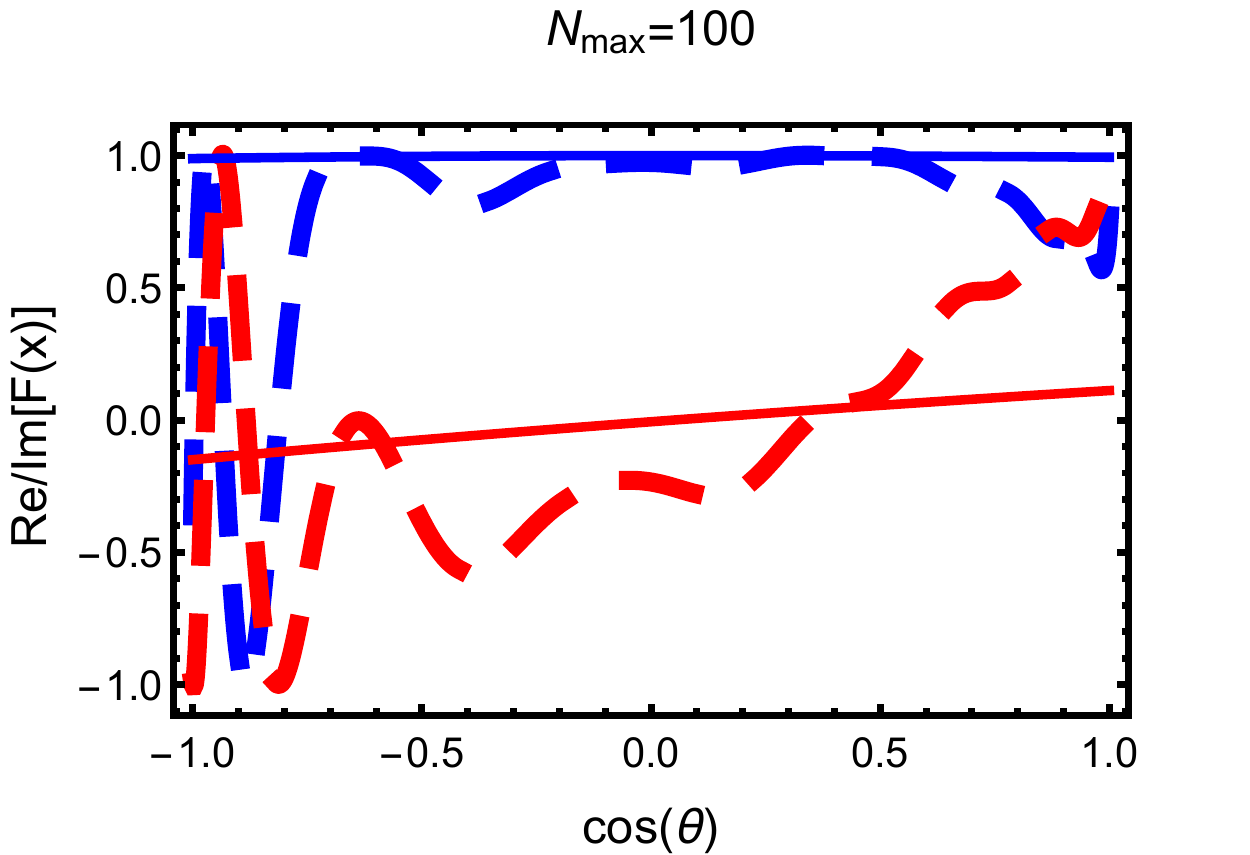}
\includegraphics[width=0.245\textwidth]{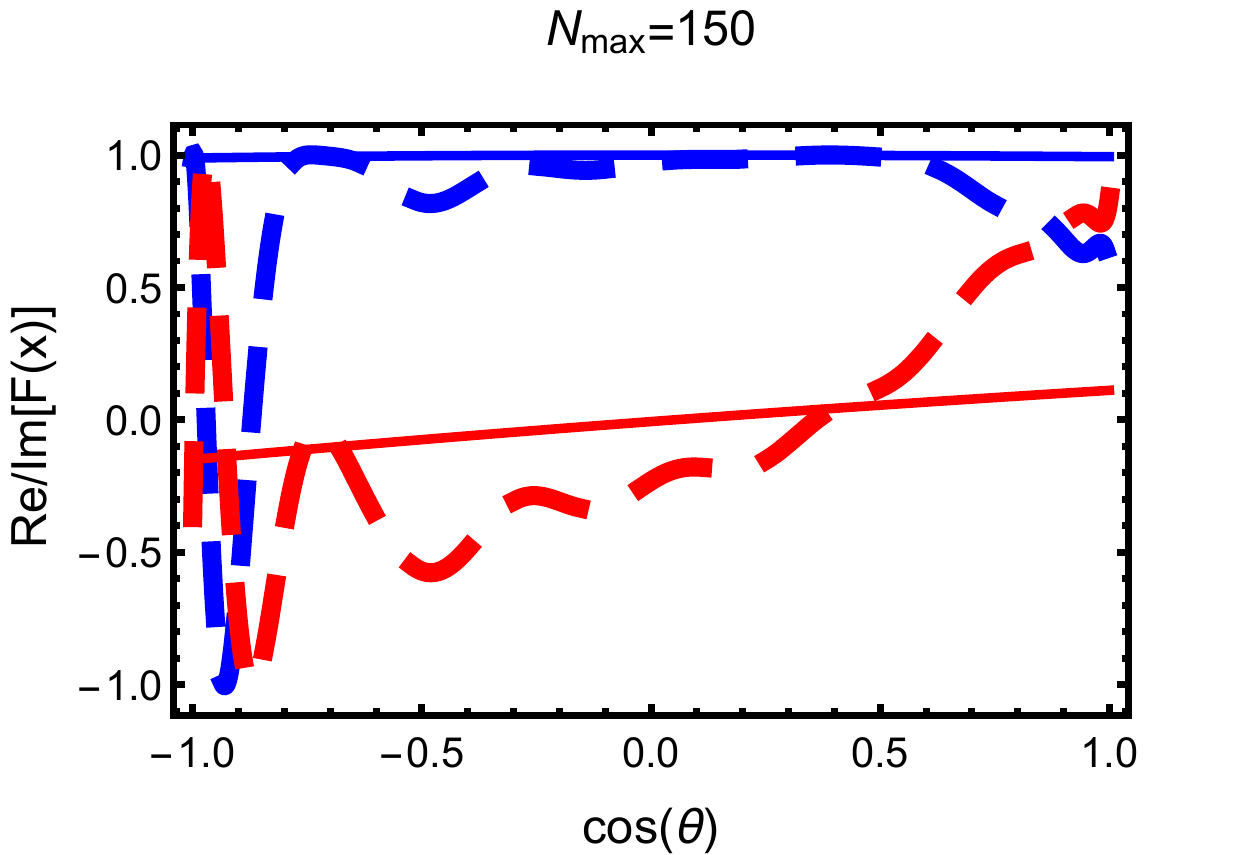} \\
\includegraphics[width=0.245\textwidth]{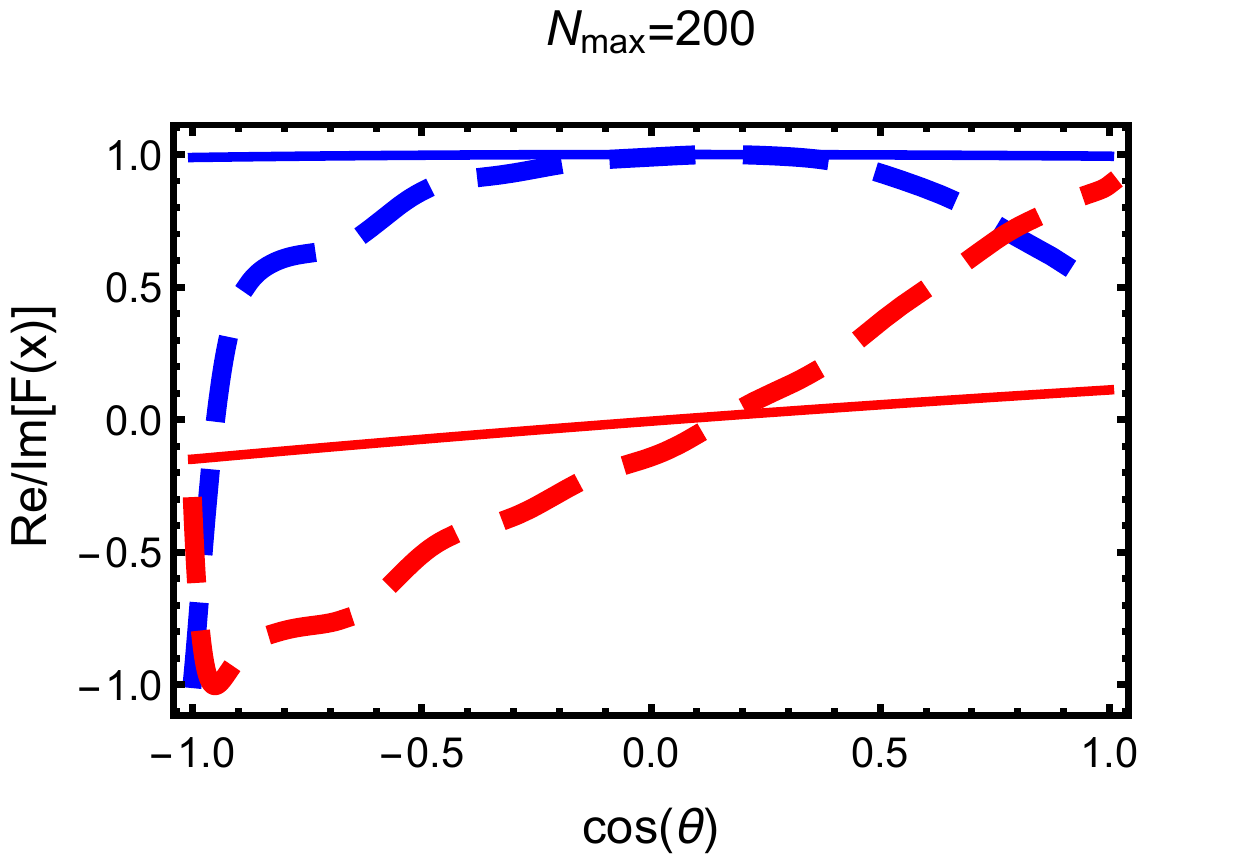}
\includegraphics[width=0.245\textwidth]{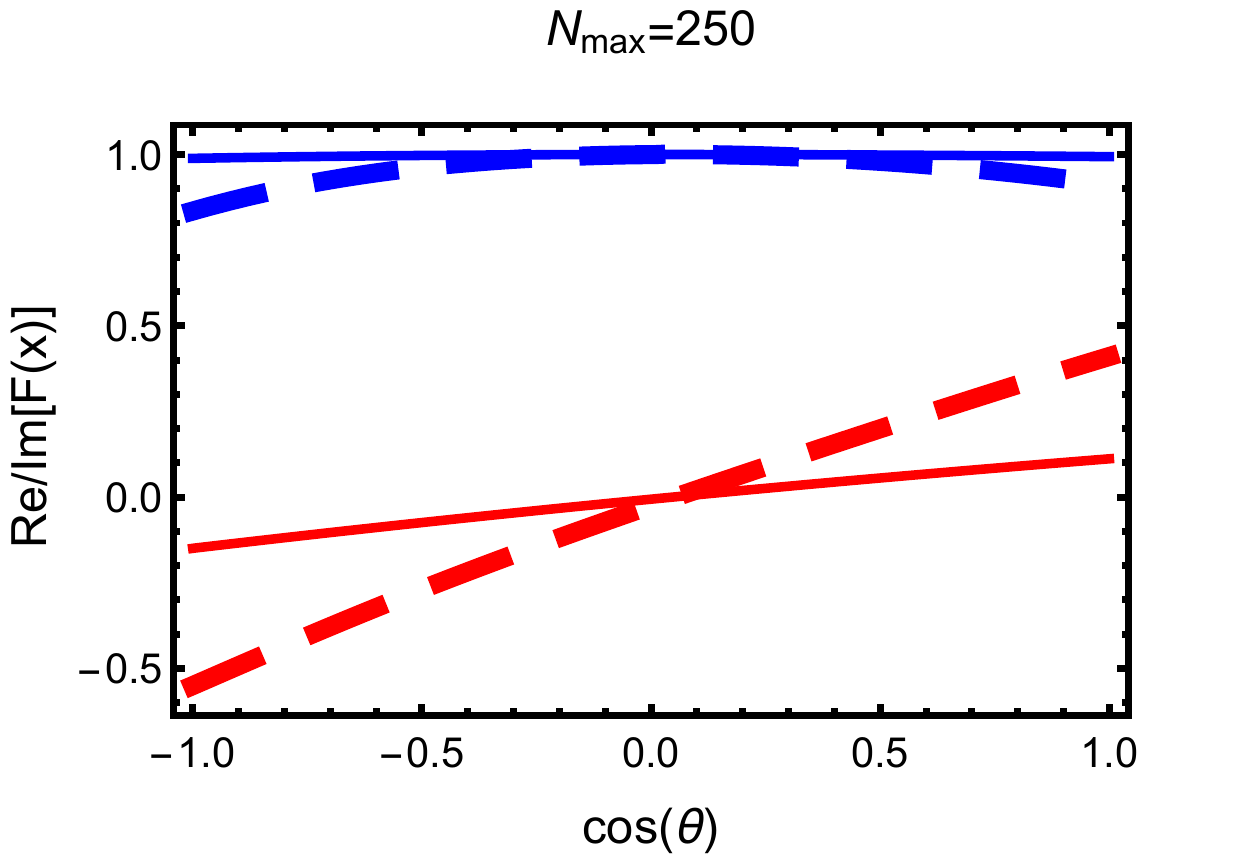}
\includegraphics[width=0.245\textwidth]{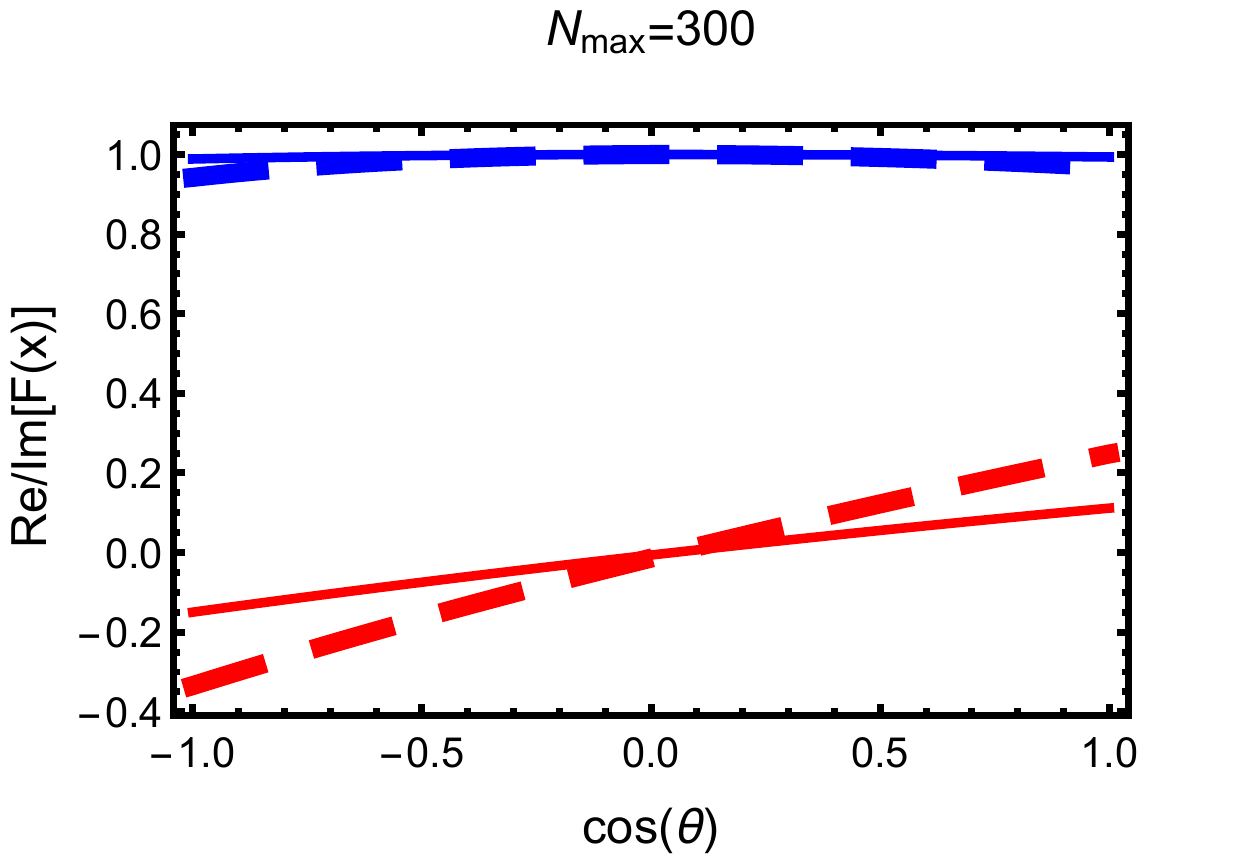}
\includegraphics[width=0.245\textwidth]{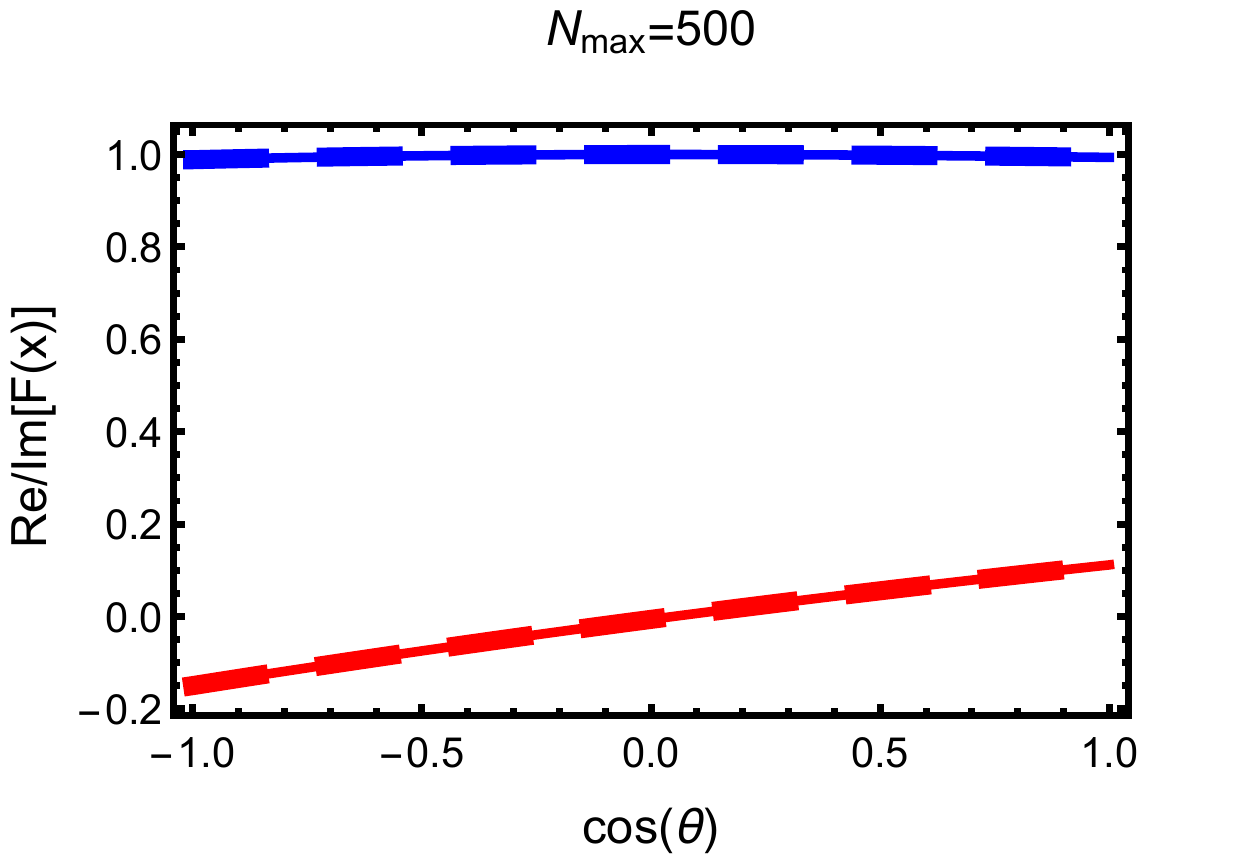} \\
\captionof{figure}[fig]{The convergence-process of the functional minimization procedure as described in the main text is demonstrated here. For the phase-rotations $e^{i \varphi_{0} (x)}$ and $e^{i \varphi_{1} (x)}$, generating the discrete ambiguities $\bm{\uppi}_{\hspace*{0.45pt}0}$ and $\bm{\uppi}_{\hspace*{0.45pt}1}$ of the toy-model (\ref{eq:DefLmax2ToyModel}), two randomly drawn initial functions have been picked from the applied ensemble. These initial conditions have, in the process of minimization, converged to these two respective phases. \newline
Minimizations have been performed by starting always at the same initial function, but applying different numbers for the maximal number of iterations $N_{\mathrm{max}}$ of the minimizer, as indicated in the headers of the plots. Values range from $N_{\mathrm{max}} = 5$ (minimizer has barely changed the initial function) up to $N_{\mathrm{max}}=500$ (convergence-condition fulfilled for any of the minimizations). \newline
In all plots, the real- and imaginary parts of the precise Gersten-ambiguity are drawn as blue and red solid lines. The results of the functional minimizations up to $N_{\mathrm{max}}$ are drawn as thick dashed lines, having the same color-coding for real- and imaginary parts. (color online)}
\label{tab:FunctMinConvergencePlots1}
\end{table*}
%


%
\begin{table*}[h]
\centering
\underline{\begin{Large}$e^{i \varphi_{2} (x)}$\end{Large}} \vspace*{5pt} \\
\includegraphics[width=0.245\textwidth]{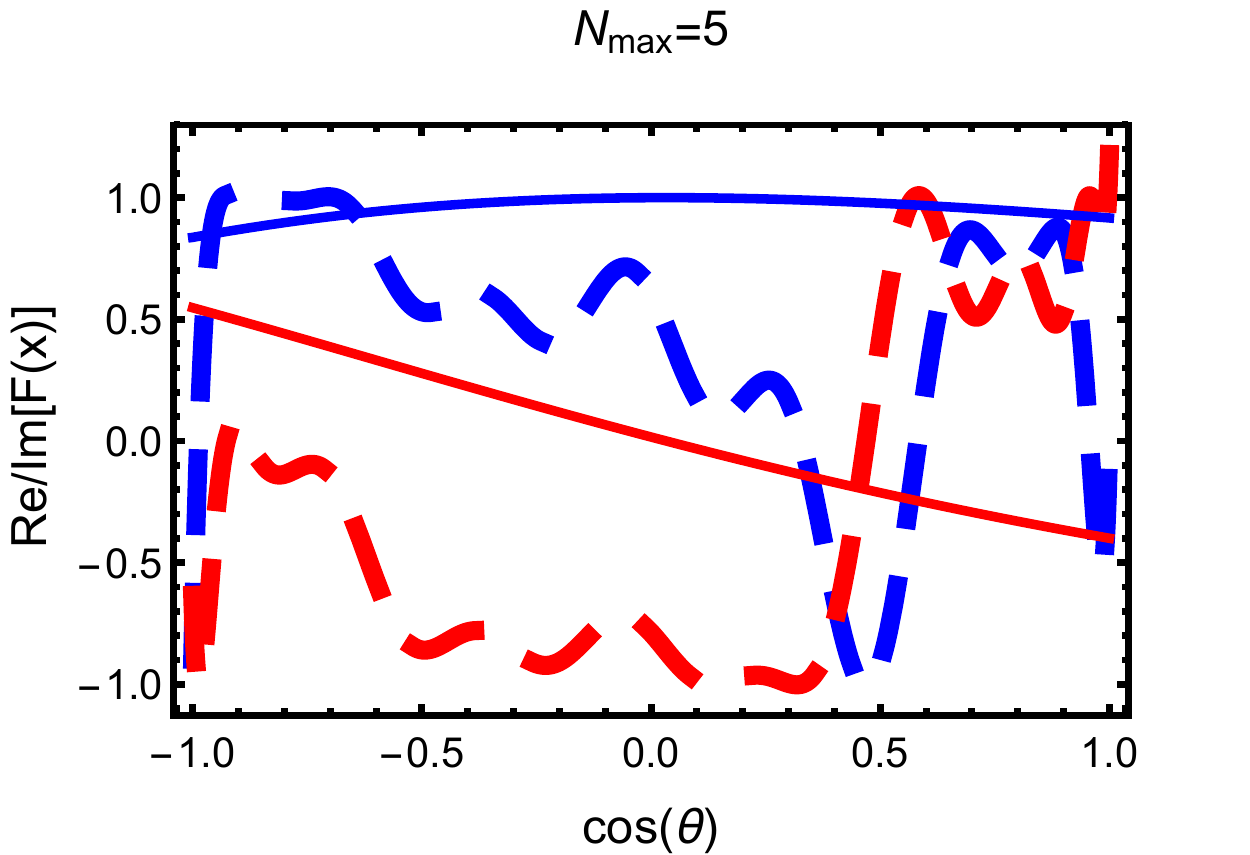}
\includegraphics[width=0.245\textwidth]{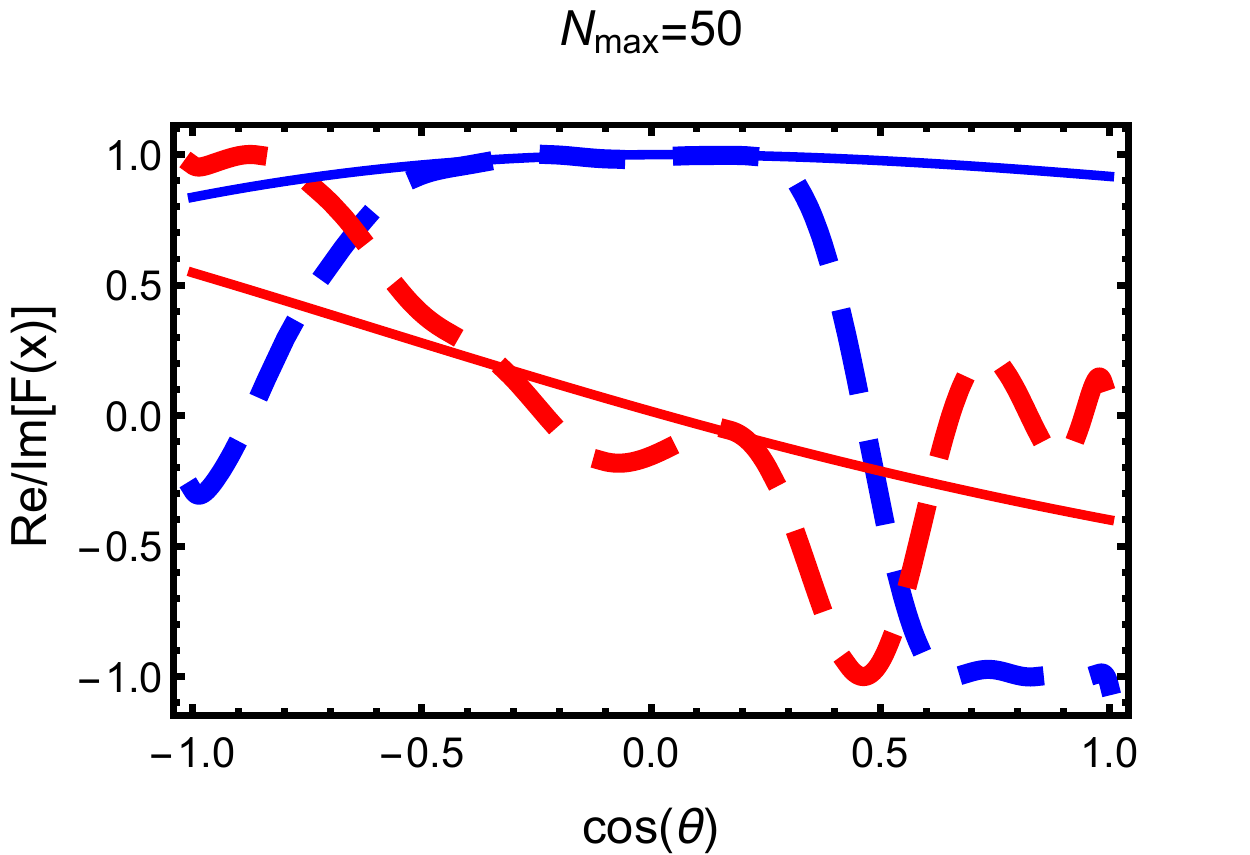}
\includegraphics[width=0.245\textwidth]{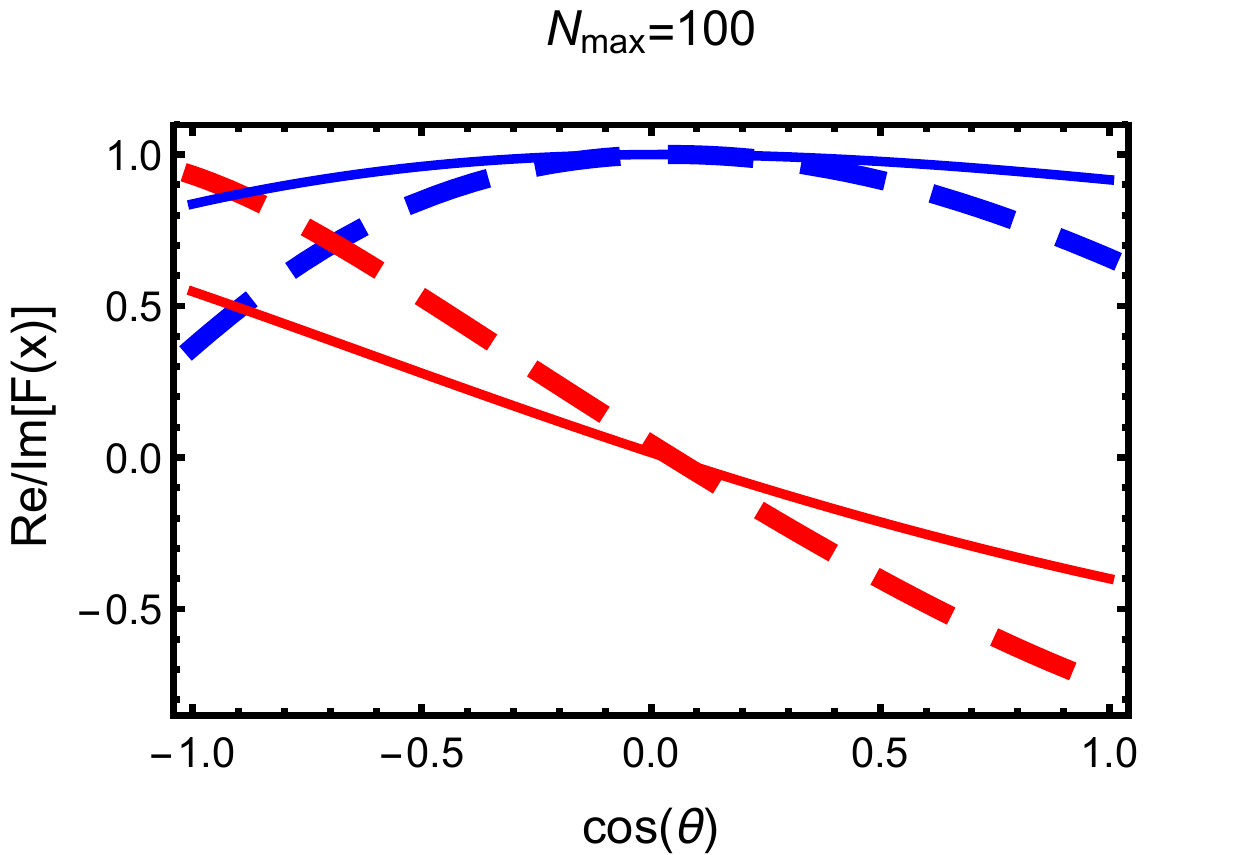}
\includegraphics[width=0.245\textwidth]{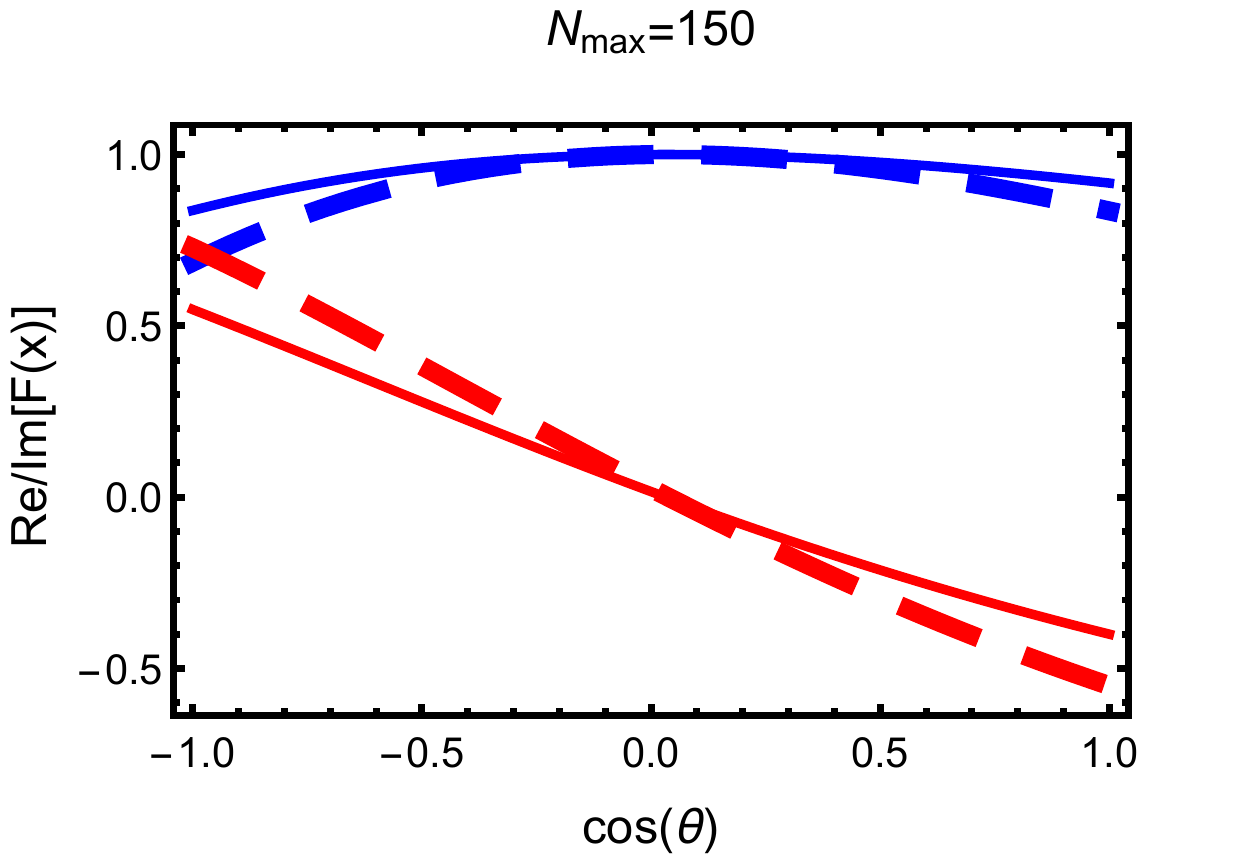} \\
\includegraphics[width=0.245\textwidth]{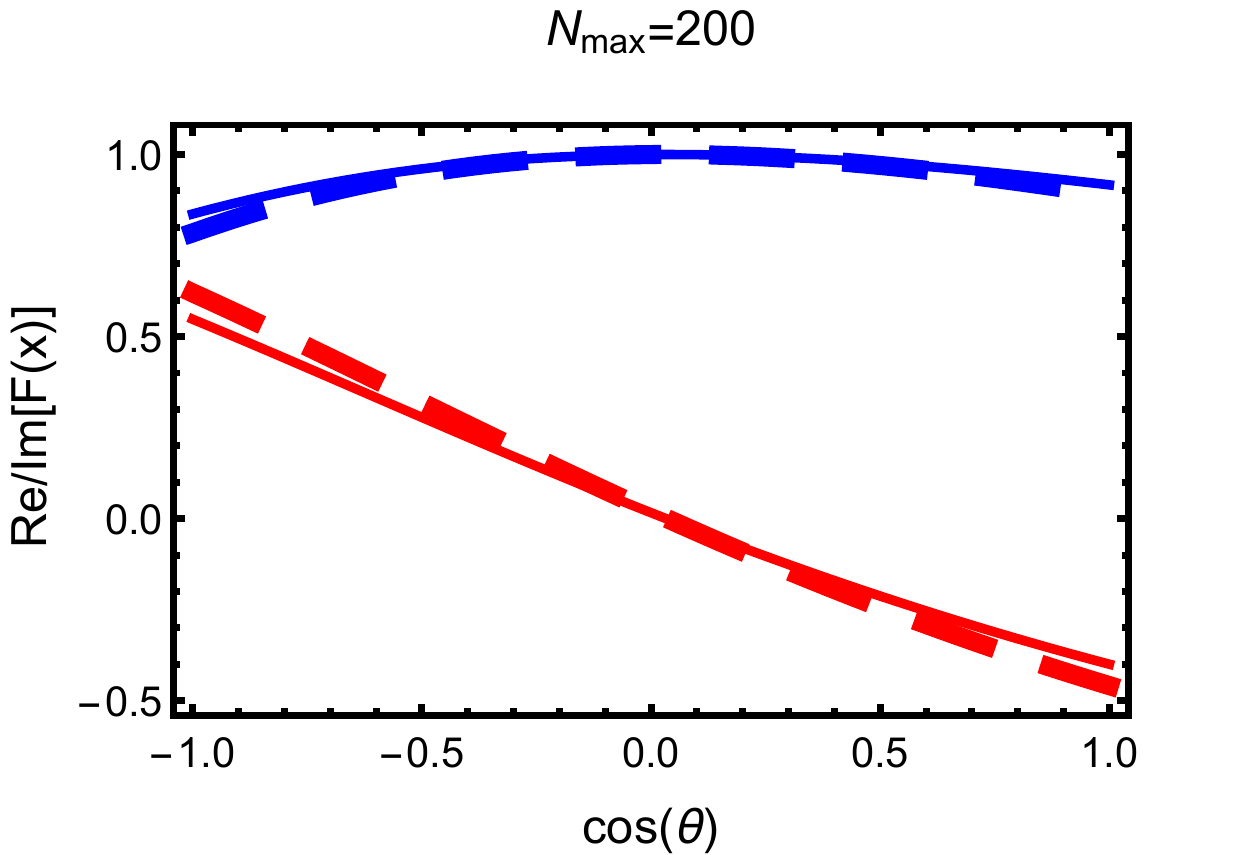}
\includegraphics[width=0.245\textwidth]{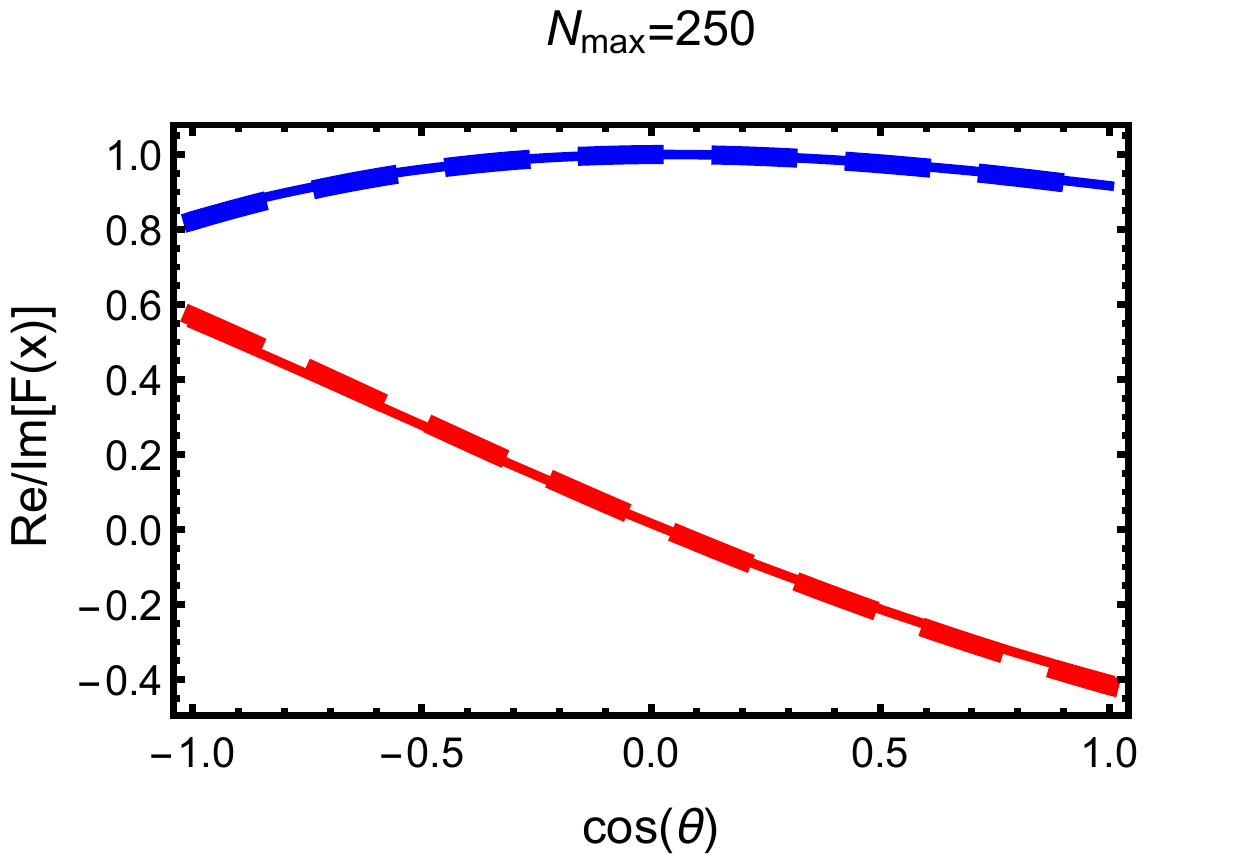}
\includegraphics[width=0.245\textwidth]{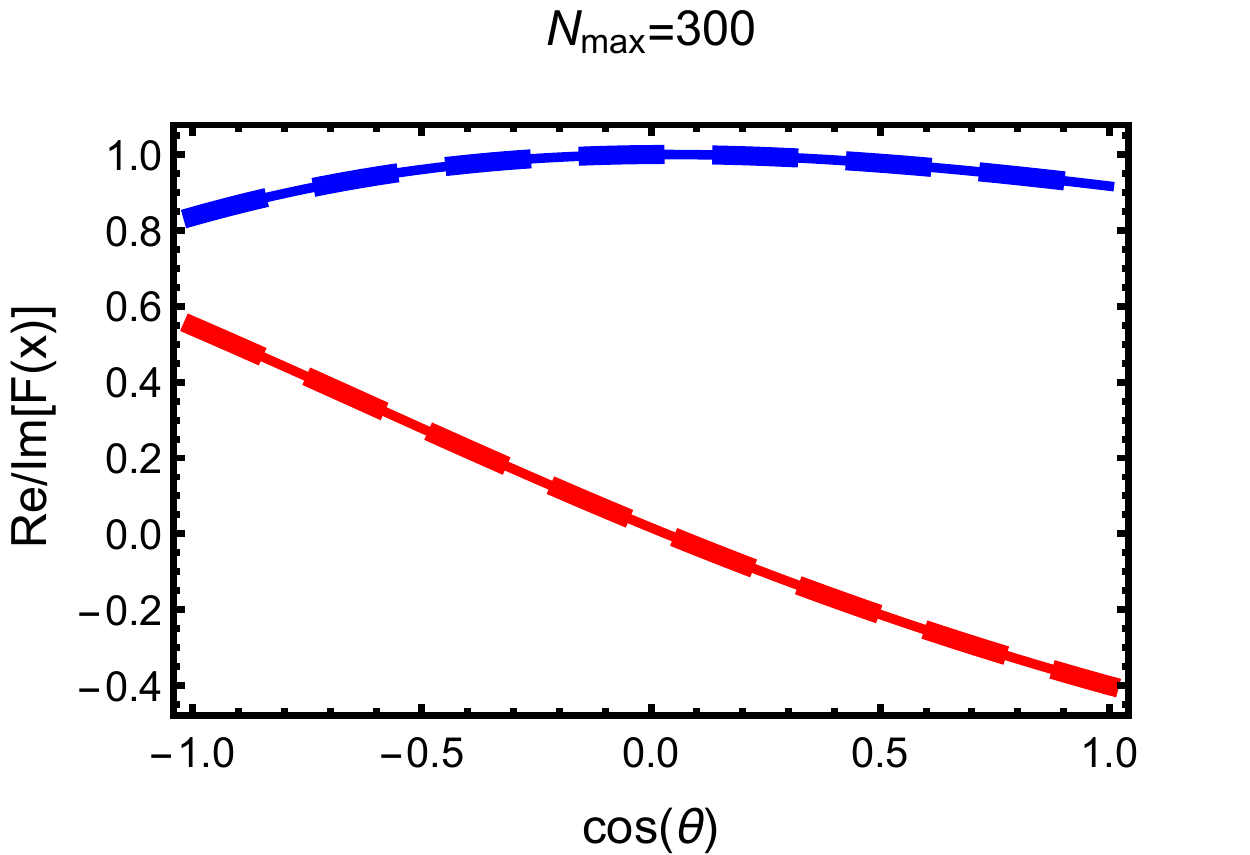}
\includegraphics[width=0.245\textwidth]{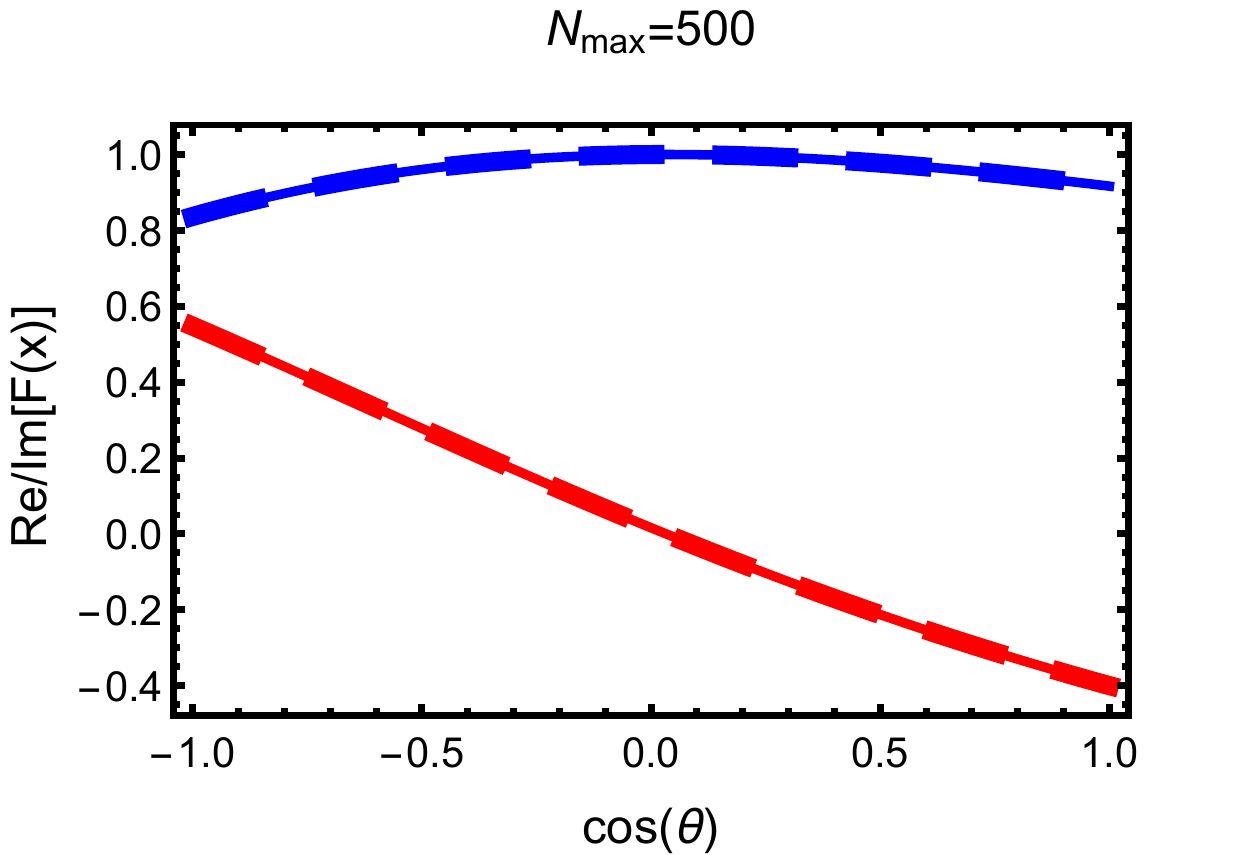} \\
\line(1,0){500} \\
\vspace*{5pt} \underline{\begin{Large}$e^{i \varphi_{3} (x)}$\end{Large}} \vspace*{5pt} \\
\includegraphics[width=0.245\textwidth]{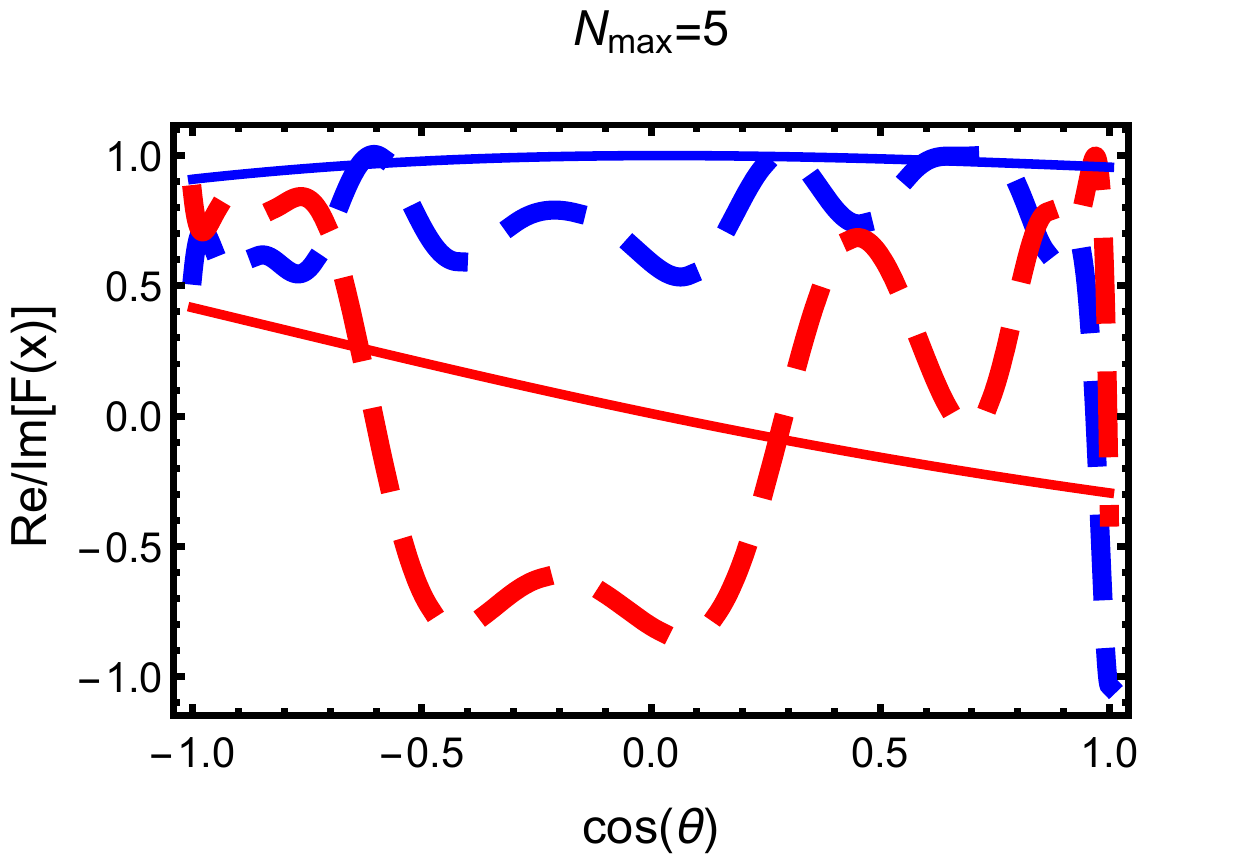}
\includegraphics[width=0.245\textwidth]{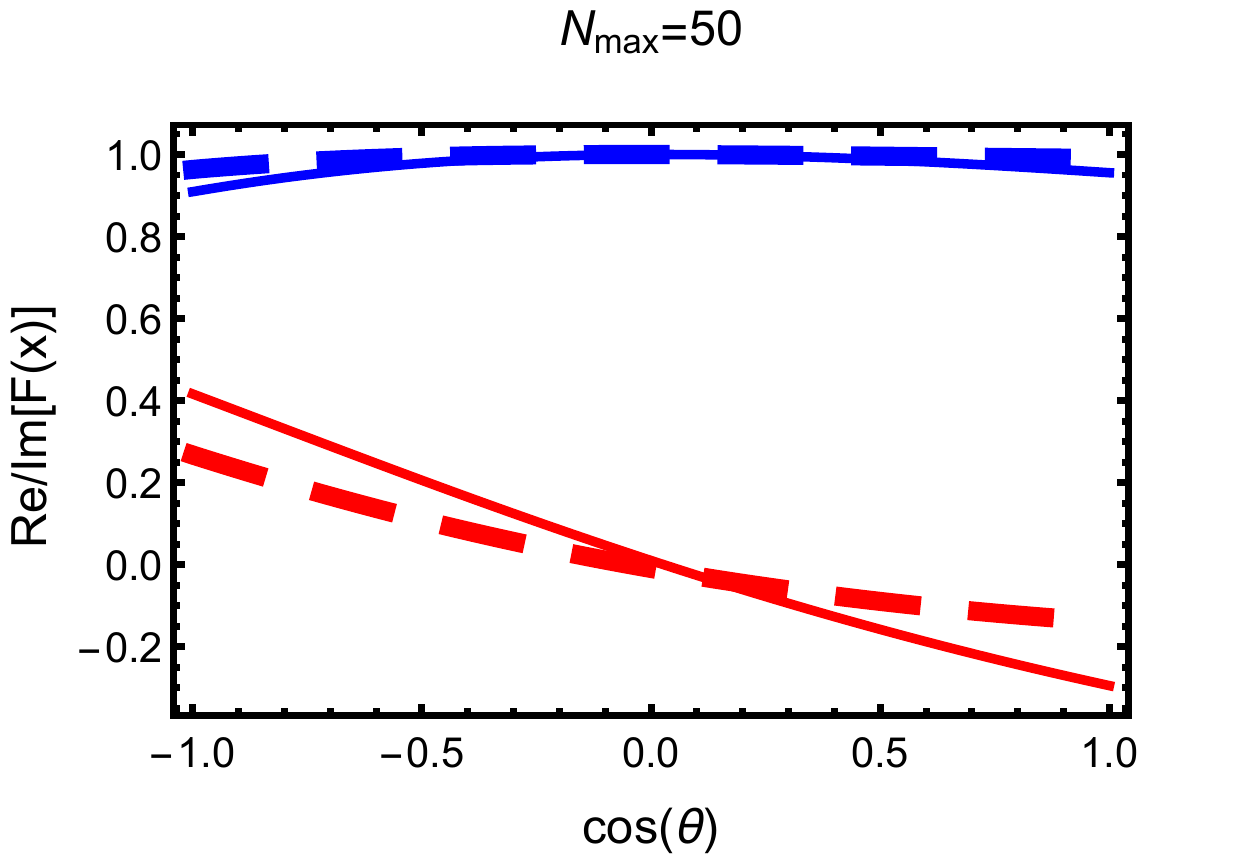}
\includegraphics[width=0.245\textwidth]{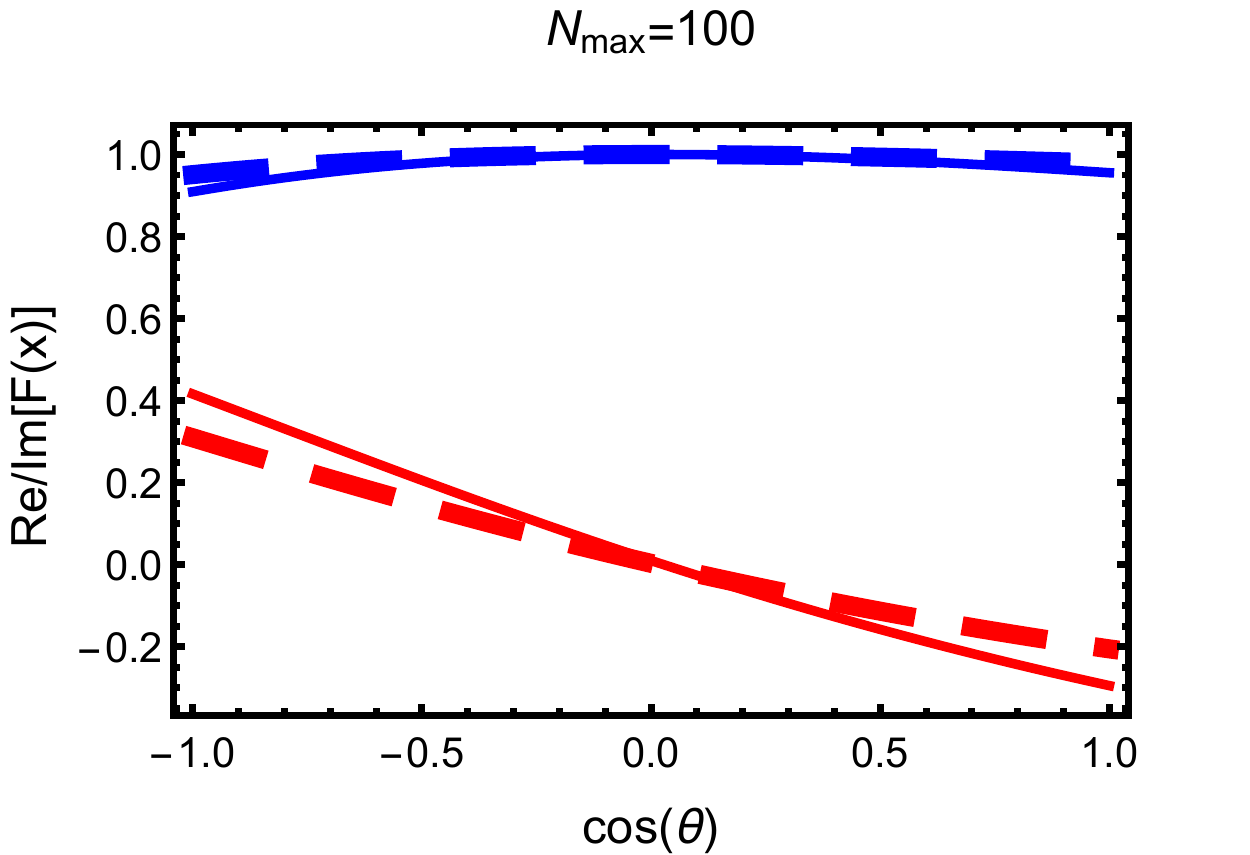}
\includegraphics[width=0.245\textwidth]{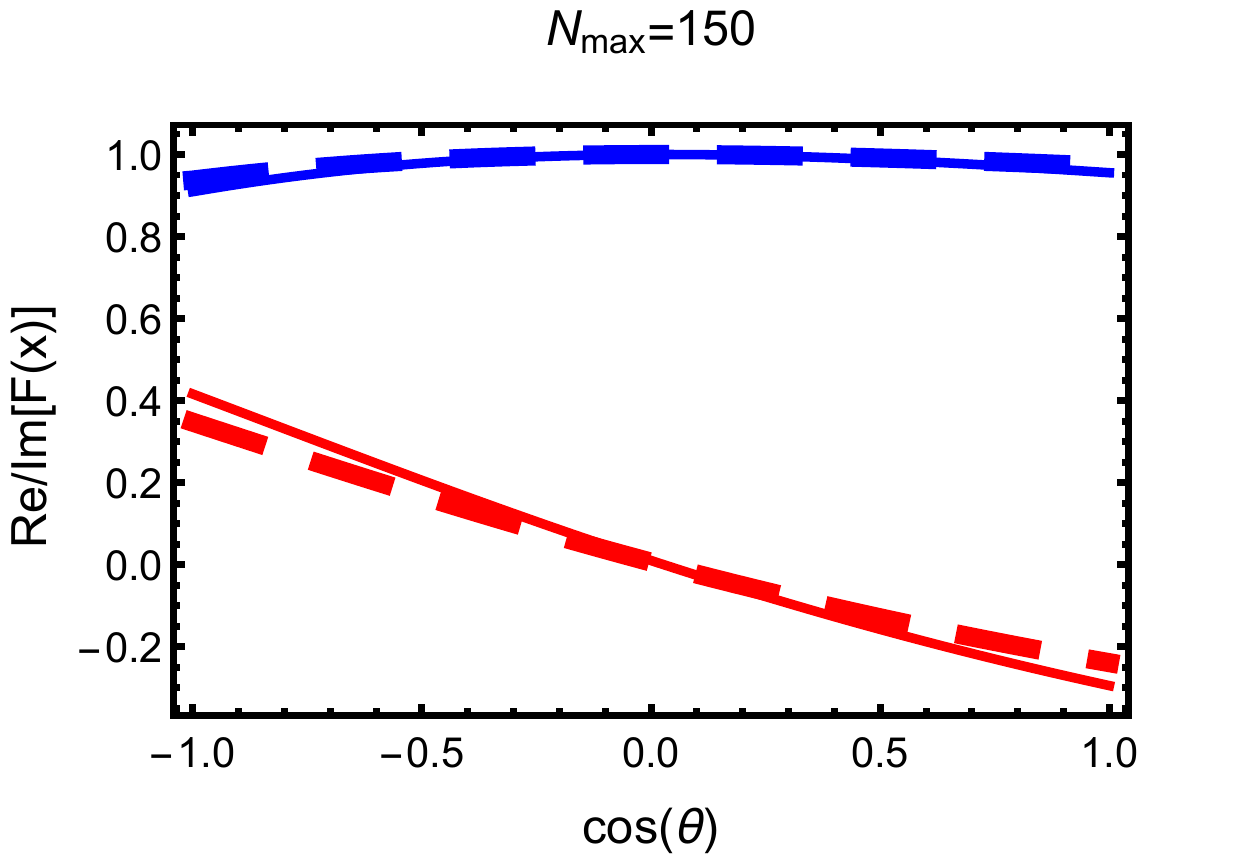} \\
\includegraphics[width=0.245\textwidth]{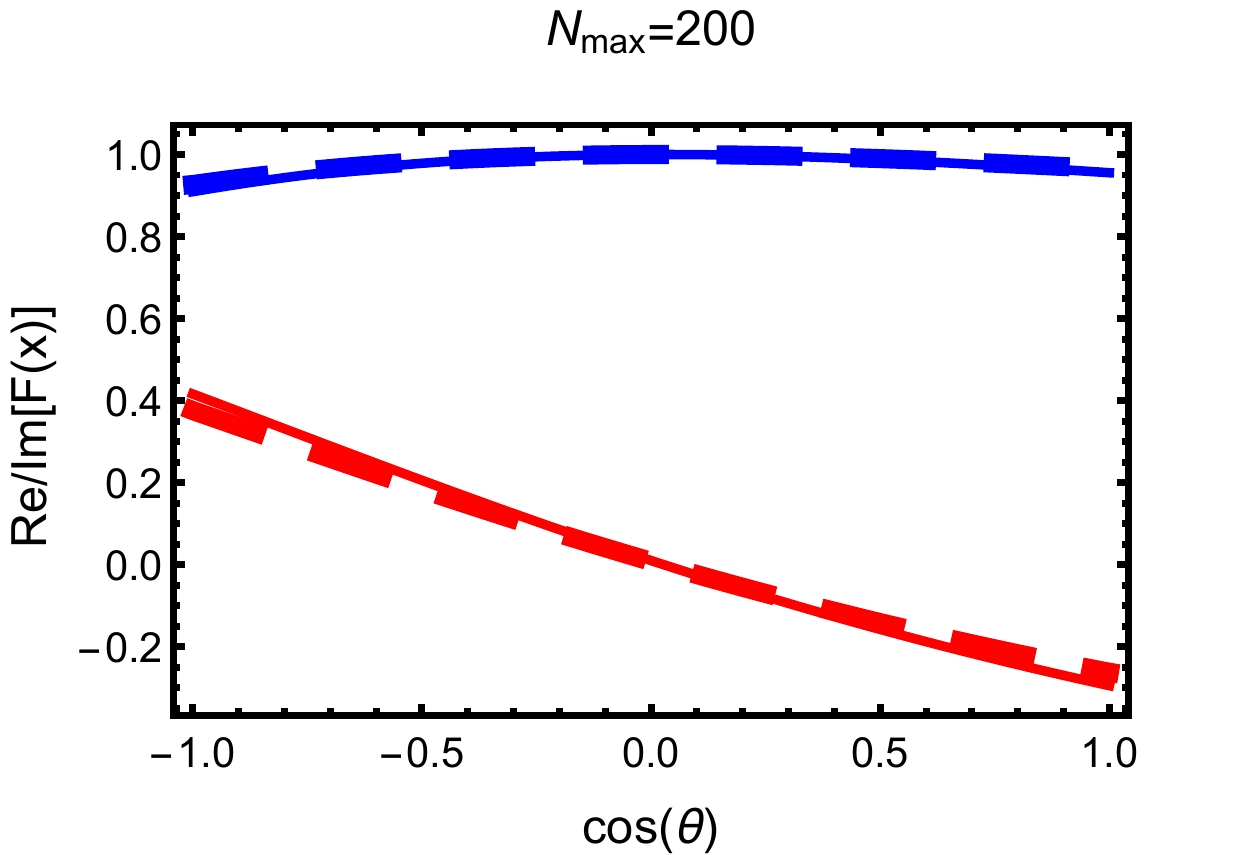}
\includegraphics[width=0.245\textwidth]{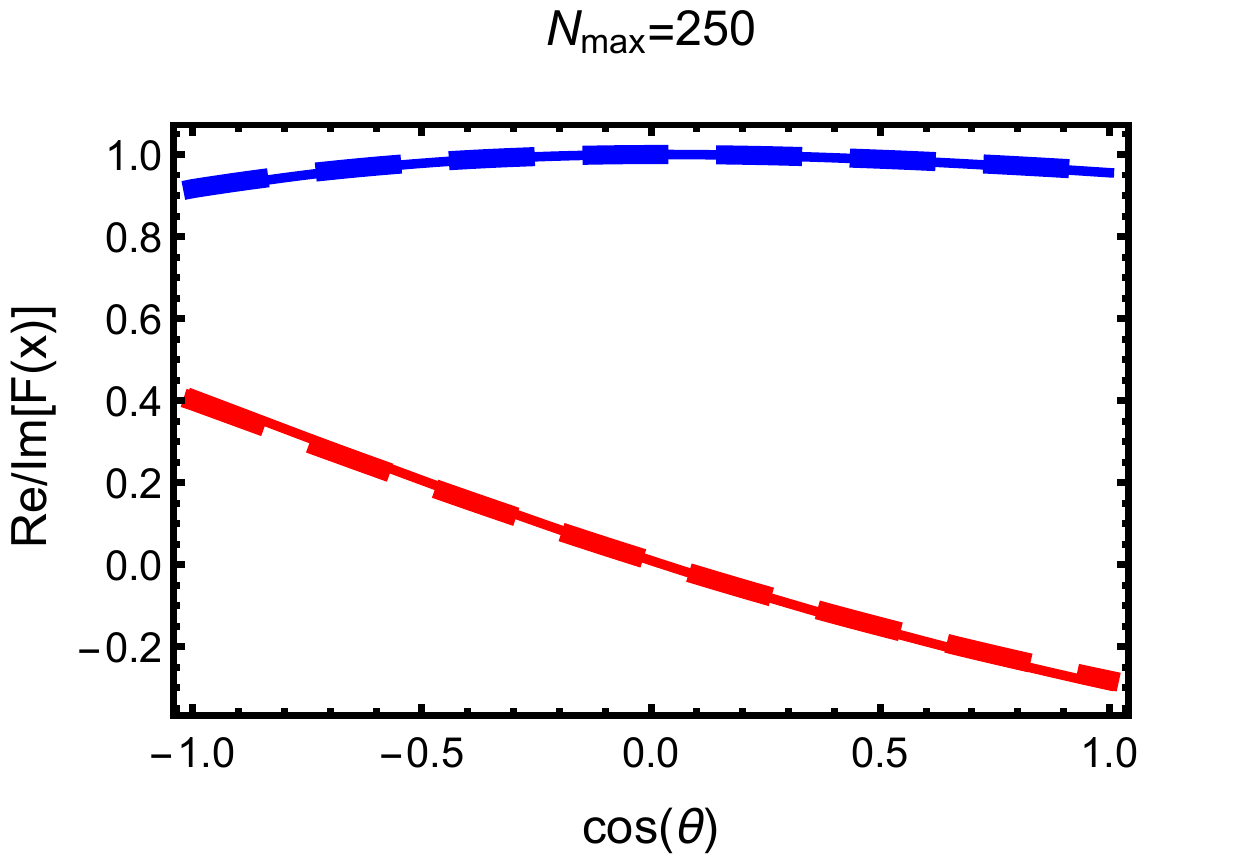}
\includegraphics[width=0.245\textwidth]{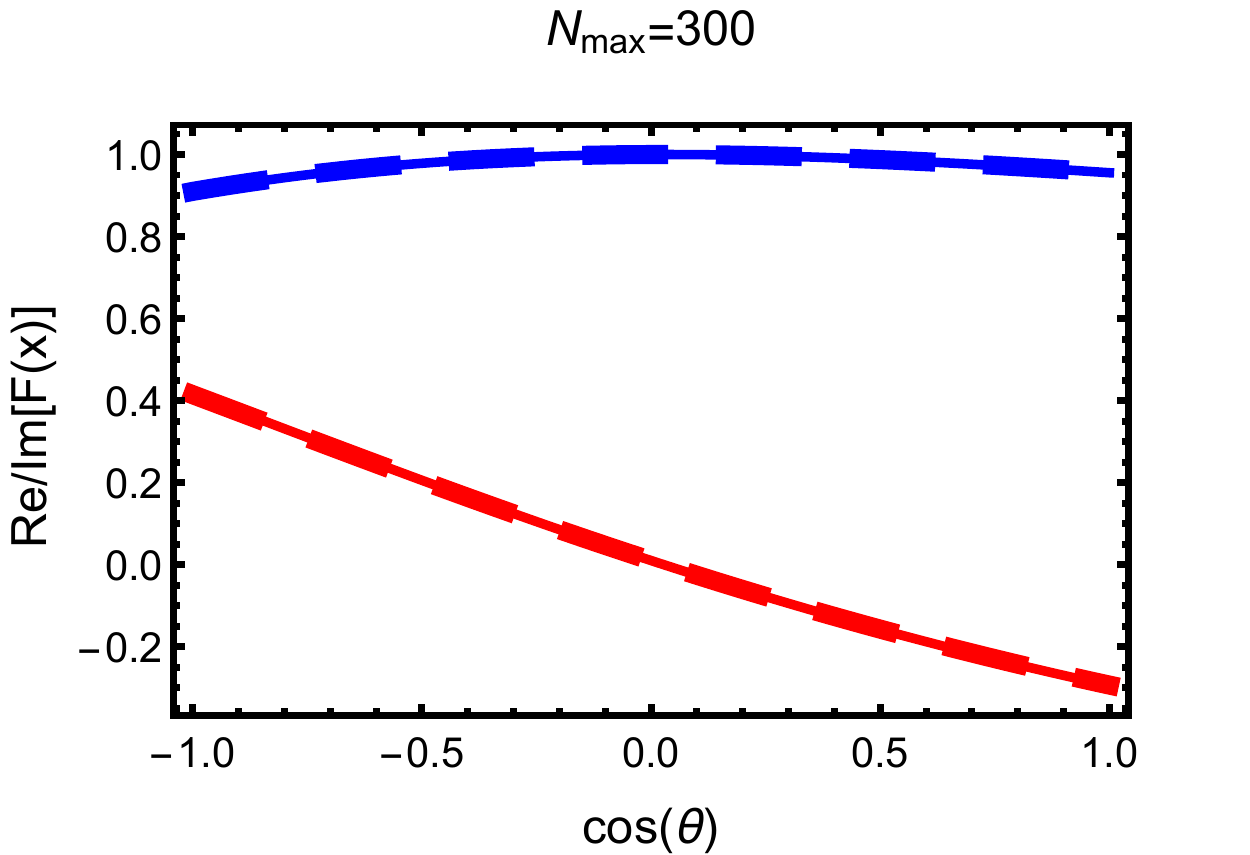}
\includegraphics[width=0.245\textwidth]{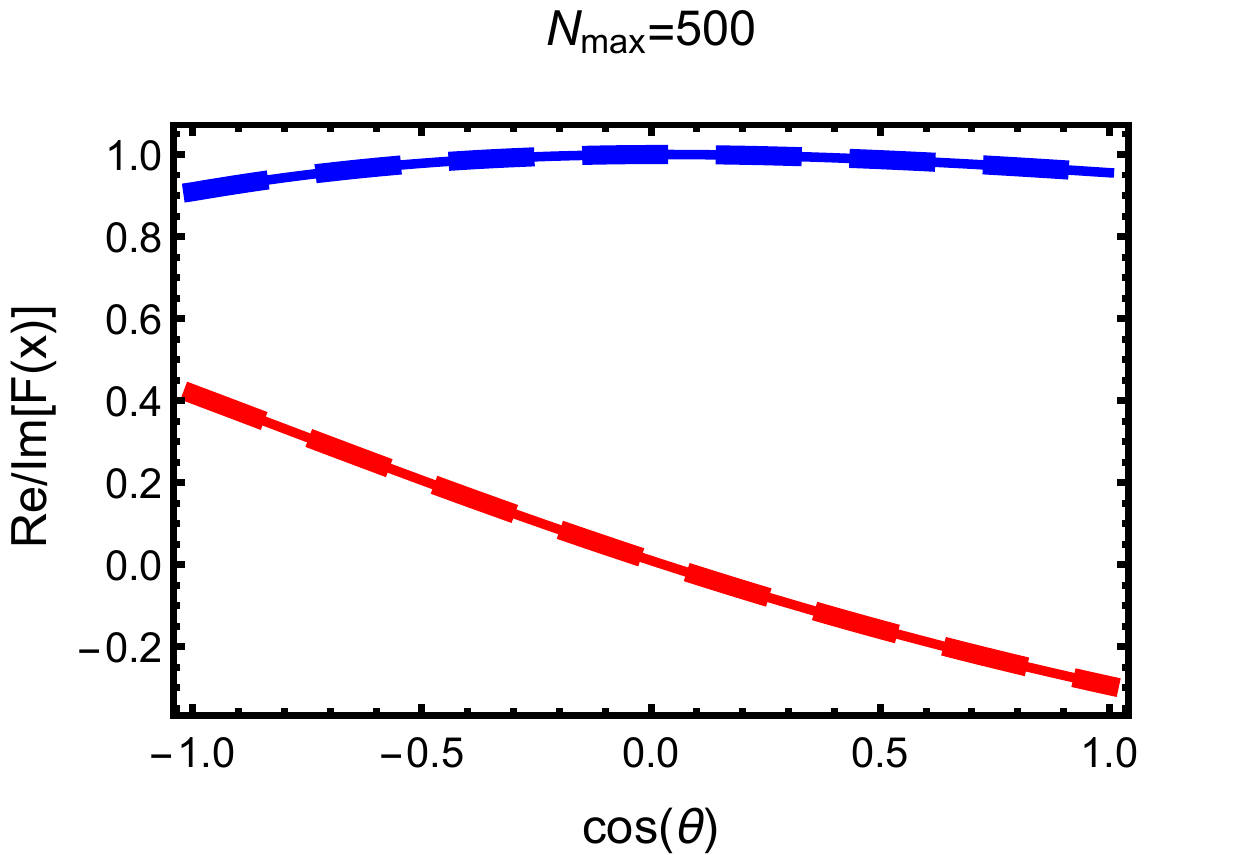} \\
\captionof{figure}[fig]{This is the continuation of Figure \ref{tab:FunctMinConvergencePlots1}. Convergence of the minimization of the functional (\ref{eq:FunctProblem}) is illustrated for the phases $e^{i \varphi_{2} (x)}$ and $e^{i \varphi_{3} (x)}$, which generate discrete symmetries for the toy-model (\ref{eq:DefLmax2ToyModel}).}
\label{tab:FunctMinConvergencePlots2}
\end{table*}

\clearpage

The identity $e^{i \varphi_{0} (x)}$ and full conjugation ambiguity $e^{i \varphi_{3} (x)}$ are found most quickly via the optimization, while the ambiguities $e^{i \varphi_{1} (x)}$ and $e^{i \varphi_{2} (x)}$ require more iterations.

\section{Conclusions and outlook} \label{sec:ConclusionsAndOutlook}

Ambiguites in the extraction of partial waves for the scalar case have been the main focus of this work. Continuum ambiguities caused by general energy- and angle-dependent phase-rotations, as well as discrete ambiguities stemming from the conjugation of zeros, have been formalized and compared. The discrete symmetries first defined by Gersten have been found to be a specific sub-class of the larger symmetry group of continuum ambiguities, with the property that they fully exhaust all possibilities to rotate an original truncated amplitude again into a truncated one. This sub-class is unique in the sense that no further transformations exist which can lead back to truncated models. \newline
Furthermore, the partial waves of the transformed amplitude have in all cases, i.e. for the full continuum ambiguity group as well as for the discrete symmetries, turned out to be mixings of the partial waves from the original amplitude. Since the discrete symmetries lead to truncated models again, they are finely tuned in such a way that exact cancellation occurs in all partial waves from the rotated model above the truncation order $L$. \newline

In order to substantiate the above mentioned exhaustiveness statement from a perspective which is orthogonal to the Gersten-formalism, a straightforward and, as far as we know, new method has been introduced based on the numerical minimization of functionals. Such functionals allow for a flexible way to scan the infinitely many possible phase-rotation functions $F(W,\theta) = e^{i \Phi(W,\theta)}$ for those obeying the implemented constraints, which in this case consisted of the vanishing of all transformed partial waves above $L$. First numerical tests for simple toy-models yielded consistent results and have in all cases confirmed the exhaustiveness statement on the discrete ambiguities. \newline

The present study is certainly just a beginning of further formal studies on partial wave ambiguities. We list in the following a few interesting open questions as well as further avenues of investigation:

\begin{itemize}
 \item[(i)] The exhaustiveness-, or uniqueness-, property of the Gersten-ambiguities has not been supported by a formal proof in this work. To perform this task, almost certainly a more sophisticated application of algebra or functional analysis will be needed. Still, a better mathematical understanding of the discrete ambiguites and why they appear, may also lead to a better grasp of the process of partial wave fitting and the quadratic equation-systems involved.
 \item[(ii)] A discrete class of angle-dependent phase-rotations has been formulated capable of rotating all models with the same truncation order $L$ into each other, i.e. of rotating $L \rightarrow L$. One may ask whether it is formally possible to raise the truncation order using angle-dependent phase-rotations, i.e. to rotate truncations
 \begin{equation}
  L \rightarrow L + \mathbb{N} \mathrm{.} \label{eq:RaiseTruncationOrderByPhaseQuestion}
 \end{equation}
 In the present case, the answer appears negative. However, it this not quite certain. Possibilities of changing truncation orders by phase-rotations would in any case be interesting. \newline
 The search for such phases may, for instance, be performed using ideas similar to the functional methods outlined in this work.
 \item[(iii)] The study of ambiguities in this work did not impose unitarity-constraints on the amplitude. It would certainly be interesting to see how to impose strict unitarity-requirements using ideas similar to the functional minimization, or how to link the findings of this work to the residual Crichton-ambiguities appearing below the first inelastic threshold.
 \item[(iv)] Finally, the formal treatment of ambiguities presented here may be extended to reactions with spin, or even with multi-particle final states. $\pi N$-scattering has been treated in some detail in the past \cite{DeanLee}. For photoproduction, no formal treatment of partial wave mixing and continuum- vs. discrete ambiguites as presented in this work, has been found. The functional methods developed here may also be extended to spin reactions.
\end{itemize}
However, a \textit{word of warning} should be said about reactions with spin. The following statements stem from preliminary considerations done for $\pi N$-scattering and for photoproduction of single pseudoscalar mesons, but may turn out to be more general, at least in the context of $2$-body reactions. \newline
It is well-known that for such reactions, the overall reaction amplitude can be parametrized in a model-independent way using $N$ invariant amplitudes
\cite{ChTab}, where the integer $N$ depends on the spins of the par\-ti\-ci\-pa\-ting particles. Upon converting to the CMS-frame, different schemes of spin-quantization can be used to obtain $N$ so-called spin-amplitudes. It is often convenient to use the basis of $N$ transversity-amplitudes $\left\{ b_{j} (W,\theta), j=1,\ldots,N \right\}$. The latter shall be chosen in the following. In case of $\pi N$-scattering, for instance, there exist $N=2$ amplitudes. Pseudoscalar meson photoproduction is described by $N=4$ amplitudes. \newline
Once more than one amplitude is in the game, it is important to distinguish different types of continuum ambiguity transformations, or in other words, rotations. The first, most general, kind of transformation rotates every transversity amplitude $b_{j}$ by a \textit{different} phase $\phi_{j}$ and is thus referred to as an $N$-fold continuum ambiguity\footnote{We use here the language of H\"{o}hler \cite{HoehlerBible}, who discusses $2$-fold continuum ambiguities in the context of $\pi N$-scattering.}
\begin{equation}
 \hspace*{10pt} b_{j} (W,\theta) \rightarrow e^{i \phi_{j}(W,\theta)} b_{j} (W,\theta) \mathrm{,} \hspace*{2.5pt} j=1,\ldots,N \mathrm{.} \label{eq:GeneralNFoldRotation}
\end{equation}
This is a much larger class of symmetry transformations than the rotation of all amplitudes by the \textit{same} phase $\Phi$, from now on referred to as a $1$-fold continuum ambiguity
\begin{equation}
 \hspace*{10pt} b_{j} (W,\theta) \rightarrow e^{i \Phi (W,\theta)} b_{j} (W,\theta) \mathrm{,} \hspace*{2.5pt} j=1,\ldots,N \mathrm{.} \label{eq:General1FoldRotation}
\end{equation}
From inspection of the well-known linear factor decompositions of the $\pi N$- and photoproduction amplitudes \cite{Omelaenko, Gersten}, we have been able to infer that at least in these two cases, the Gersten-type ambiguities (i.e. those stemming from root-conjugation) are in general generated by $N$-fold rotations (\ref{eq:GeneralNFoldRotation}) and \textit{not} by the $1$-fold ones (\ref{eq:General1FoldRotation}).\footnote{This fact has been observed only for the two example-reactions. We just assume that it carries over to more general spin reactions.} Thus, the fact that discrete Gersten-type ambiguities fall into the general $1$-fold rotations is a special feature only present for the scalar reactions, caused by the fact that there exists only one amplitude. For the more general cases with spin, one has to carefully distinguish which kind of symmetry is generated from which kind of rotation, and the generalization of the scalar results obtained in this work is by no means trivial. \newline

The distinction of $N$- and $1$-fold continuum ambiguities made here becomes interesting once one considers the observables measurable in a spin-reaction. It is again well-known that for $2$-body reactions, $N^{2}$ polarization observables $\mathcal{O}^{a}$ can be accessed, at least in principle \cite{ChTab}. When written in the transversity basis, there exists a subset of $N$ observables given just by a sum of moduli-squared of the amplitudes (proportionality in the following equation up to phase-space factors) 
\begin{equation}
 \hspace*{17.5pt} \mathcal{O}_{\bm{d}}^{a} (W,\theta) \propto \pm \left| b_{1} (W,\theta) \right|^{2} \pm \ldots \pm \left| b_{N} (W,\theta) \right|^{2} \mathrm{,} \label{eq:DiagObsGeneralForm}
 \end{equation}
for $a=1,\ldots,N$. The signs in front of each squared amplitude depend on the observable and conventions used. When considered as a bilinear form of the amplitudes, the $N$ observables $\mathcal{O}_{\bm{d}}^{a}$ are defined by diagonal matrices (thus the subscript $\bm{d}$). For $\pi N$-scattering, the diagonal observables would be the unpolarized cross section $\sigma_{0}$ and the target-polarization asymmetry $\hat{P}$ \cite{DeanLee, HoehlerBible}. In case of photoproduction, it is well-known that the single-spin observables $\sigma_{0}$, $\hat{\Sigma}$, $\hat{T}$ and $\hat{P}$ are diagonal \cite{Omelaenko, ChTab} in the transversity-basis. \newline
The remaining $N^{2} - N = N(N-1)$ observables are non-diagonal and thus composed of interference-terms
\begin{equation}
 \hspace*{17.5pt} \mathcal{O}_{\bm{nd}}^{a} (W,\theta) \propto \sum_{j,k} \bm{c}_{jk}^{a} b_{j}^{\ast} (W,\theta) b_{k} (W,\theta) \mathrm{,} \label{eq:NonDiagObsGeneralForm}
\end{equation}
in this case for $a=1,\ldots,N(N-1)$. The hermitean matrices $\bm{c}_{jk}^{a}$ always render these observables to be either the real- or imaginary-part of a particular linear combination of interference terms. For $\pi N$-scattering, the spin-rotation parameters $\hat{R}$ and $\hat{A}$ are non-diagonal in the transversity basis \cite{DeanLee, HoehlerBible}, while for photoproduction the same is true for all double-polarization observables of type beam-target, beam-recoil and target-recoil \cite{Omelaenko, ChTab}. \newline

Comparing the forms of the diagonal (\ref{eq:DiagObsGeneralForm}) and non-diagonal (\ref{eq:NonDiagObsGeneralForm}) observables, it is seen quickly that the former are generally always invariant under the $N$-fold rotations (\ref{eq:GeneralNFoldRotation}), while the latter are not. On the other hand, the $1$-fold rotations (\ref{eq:General1FoldRotation}) leave both kinds of observables invariant. \newline
Therefore, for the spin-reactions the interesting possibility emerges to obtain unique solutions in a TPWA once the energy-dependent overall phase has been fixed. The problem of such \textit{complete experiments} in TPWAs for photoproduction has been explored before \cite{RWorkmanEtAlCompExPhotoprod, YWEtAl2014, Grushin, Omelaenko}. A very recent publication \cite{TiatorEtAlElpro} treats the even more involved problem of electroproduction of pseudoscalar mesons. However, in these references the problem has not been formulated explicitly in the language involving rotations, which has been used in this work.

\begin{acknowledgments}
The work of Y.W. and R.B. was supported in part by the Deutsche Forschungsgemeinschaft (SFB/TR16)
and the European Community-Research Infrastructure Activity (FP7). \newline
A.\v{S}. and L.T. are supported by the Deutsche Forschungsgemeinschaft (SFB 1044).
The work of R.L.W. was supported through the US Department of Energy Grant DE-SC0016582.
\end{acknowledgments}


\appendix

\section{Ansatz for the minimization of $\bm{W} \left[ F(x) \right]$ using a discretization of the function $F(x)$} \label{sec:FunctProblemDiscretizationAnsatz}

In the following, we outline briefly a numerical alternative for the minimization of the functional (\ref{eq:FunctProblem}) which, contrary to the method of Legendre-expansions utilized in the main text, parametrizes the sought after phase-rotation functions $F(x)$ by discretization on the interval $x \in [-1,1]$. Therefore, we introduce a set of equidistant points $\left\{ x_{n} | n = 1,\ldots,N_{I} \right\}$ according to the prescription (\ref{eq:BasePointsX1}) used in the main text. The set of variables to be determined in the minimization procedure is given by the real- and imaginary parts of the function $F(x)$ on this grid, i.e.
\begin{equation}
 r_{n} := \mathrm{Re}\left[F(x_{n})\right], \hspace*{1pt} q_{n} := \mathrm{Im}\left[F(x_{n})\right], \hspace*{1pt} n = 1,\ldots,N_{I} \mathrm{.} \label{eq:DefOfDiscretizedFunctionValues}
\end{equation}
These variables fulfill a similar purpose as the real- and imaginary part of the Legendre coefficients used in the method described in the main text, cf. equation (\ref{eq:PhaseRotTrafoLegendreSeriesTruncated}). \newline
In order to minimize the quantity (\ref{eq:FunctProblem}), the latter has first of all to be evaluated. Therefore, numerical integration has to be defined. We choose here the simplest possible way to do so and use the same grid employed in the discretization (\ref{eq:DefOfDiscretizedFunctionValues}). Therefore, any integral can be calculated using the form
\begin{equation}
 \int_{-1}^{1} d x f (x) = \sum_{n = 1}^{N_{I}} \Delta x \hspace*{1.5pt} f \left( x_{n} \right) \mathrm{,} \hspace*{2pt} \mathrm{using} \hspace*{2pt} \Delta x = \frac{2}{N_{I}} \mathrm{.} \label{eq:IntOfFBasicDefinition}
\end{equation}
It is seen that in order to obtain a precise knowledge of the solution function, as well as a small error in the numerical integration (\ref{eq:IntOfFBasicDefinition}), the number of grid points $N_{I}$ has to be chosen as large as possible. In specific examples, we have had satisfactory results with numbers in the range $N_{I} = 250, \ldots, 500$. \newline
Now, all ingredients necessary to formulate the functional (\ref{eq:FunctProblem}) in the case of a minimization using the function-discretization (\ref{eq:DefOfDiscretizedFunctionValues}) have been assembled. Again, the truncated non-rotated amplitude $A(x)$ is a known input. Numerical initial conditions for the parameters $\left\{q_{n},r_{n}\right\}$ have to be drawn prior to fitting, for instance from the interval $[-1,1]$. An ensemble of initial parameter-configurations should then be used, performing a functional minimization for each of them. One should employ ensembles of at least $N_{MC} = 200$ configurations. \newline
Omitting further intermediate steps, we quote the final result for the functional:
\begin{widetext}
\begin{align}
 &\bm{W}_{\mathrm{discr.}} \Big( \left\{r_{n} , q_{n}\right\} \Big) := \sum_{n = 1}^{N_{I}} \left(  r_{n}^{2}  + q_{n}^{2} - 1 \right)^{2} + \left[ \sum_{n=1}^{N_{I}} \Delta x  \Big( r_{n} \mathrm{Im} \left[ A(x_{n}) \right] + q_{n} \mathrm{Re} \left[ A(x_{n}) \right]  \Big) \right]^{2}  \nonumber \\
  & \hspace*{1pt} + \sum_{k = 1}^{K_{\mathrm{cut}}} \Bigg\{ \left[ \sum_{n=1}^{N_{I}} \Delta x  \Big( r_{n} \mathrm{Re} \left[ A(x_{n}) \right] - q_{n} \mathrm{Im} \left[ A(x_{n}) \right]  \Big) P_{L+k} (x_{n}) \right]^{2} + \left[ \sum_{n=1}^{N_{I}} \Delta x  \Big( r_{n} \mathrm{Im} \left[ A(x_{n}) \right] + q_{n} \mathrm{Re} \left[ A(x_{n}) \right]  \Big) P_{L+k} (x_{n}) \right]^{2}  \Bigg\} \mathrm{.} \label{eq:ReImFitFunctional2}
\end{align}
\end{widetext}
Note that in this case, the parameters $N_{I}$ and $K_{\mathrm{cut}}$ can be tuned independently from each other. This is different from the minimization scheme using Legendre expansions described in the main text, where the parameters $\mathcal{L}_{\mathrm{cut}}$ and $K_{\mathrm{cut}}$ have been connected. \newline
Using the minimization with the discretization-functional (\ref{eq:ReImFitFunctional2}), we have obtained the same solutions as with the Legendre-parametrization (\ref{eq:FunctProblemLegendreVersion1}) for specific toy-model examples. However, the discretization method has proven to be the numerically more demanding and less stable of the two.

\section{The functional $\bm{W} \left[ F(x) \right]$ as an ordinary function $\bm{W} (\left\{L_{k}\right\})$, depending on Legendre coefficients} \label{sec:MinimizationFunctionalAsOrdinaryFunction}

As hinted at in the main text, the minimization functional (\ref{eq:FunctProblem}) becomes, once the phase-rotation function $F(x)$ is parametrized as a Legendre series, an ordinary function depending on the Legendre coefficients. Here, we derive a formal expression of the resulting functional, which is then for finite Legendre expansions equal to the form (\ref{eq:FunctProblemLegendreVersion1}) used in our numerical minimizations. \newline
We assume here an initial non-rotated amplitude $A(W, \theta)$ truncated at $L$ (i.e. the truncated version of equation (\ref{eq:BasicInfinitePWExpansion})) and for the most general formal case, an infinite Legendre series for the phase-rotation function
\begin{equation}
 F(x) = \sum_{k = 0}^{\infty} L_{k} P_{k} (x) \mathrm{.} \label{eq:InfiniteLegExpansionAppendices}
\end{equation}
Now, all three terms appearing in the formal definition (\ref{eq:FunctProblem}) of the minimization functional $\bm{W} \left[ F(x) \right]$ are investigated with regard to their dependence on the complex Legendre coefficients $\left\{ L_{k} \right\}$. The first term, i.e. the sum over $x$, in equation (\ref{eq:FunctProblem}) imposes the unimodularity of $F(x)$. The squared modulus of the latter is appearing here, which under the present assumptions can be rewritten as follows
\begin{align}
 \left| F(x) \right|^{2} &= F^{\ast}(x) F(x) \nonumber \\
 & = \left( \sum_{k^{\prime} = 0}^{\infty} L^{\ast}_{k^{\prime}} P_{k^{\prime}} (x) \right) \left( \sum_{k = 0}^{\infty} L_{k} P_{k} (x) \right) \nonumber \\
 &= \sum_{k, k^{\prime} = 0}^{\infty} L^{\ast}_{k^{\prime}} L_{k} P_{k^{\prime}} (x) P_{k} (x) \nonumber \\
 &= \sum_{k, k^{\prime} = 0}^{\infty} \sum_{m = \left| k^{\prime} - k \right|}^{k^{\prime} + k} L^{\ast}_{k^{\prime}} L_{k} \left< k^{\prime},0 ; k,0 | m , 0 \right>^{2} P_{m} (x) \mathrm{.} \label{eq:FirstTermDerivation}
\end{align}
The second term in the first line of equation (\ref{eq:FunctProblem}) restricts the $S$-wave to an overall phase constraint. The imaginary part which is squared there becomes
\begin{align}
 &\mathrm{Im} \left[ \int_{-1}^{1} d x F(x) A(x) \right] \nonumber \\
 &= \mathrm{Im} \left[ \int_{-1}^{1} d x \left( \sum_{k^{\prime} = 0}^{\infty} L_{k^{\prime}} P_{k^{\prime}} (x) \right) \left( \sum_{\ell = 0}^{L} (2 \ell + 1) A_{\ell} P_{\ell} (x) \right) \right] \nonumber \\
 &= \mathrm{Im} \left[ \sum_{k^{\prime} = 0}^{\infty} \sum_{\ell = 0}^{L} L_{k^{\prime}} ( 2 \ell + 1) A_{\ell} \int_{-1}^{1} d x P_{k^{\prime}} (x) P_{\ell} (x) \right] \nonumber \\
 &= 2 \hspace*{2pt} \mathrm{Im} \left[ \sum_{k^{\prime} = 0}^{\infty} \sum_{\ell = 0}^{L} L_{k^{\prime}} A_{\ell} \delta_{k^{\prime} \ell} \right] =  2 \hspace*{2pt} \mathrm{Im} \left[ \sum_{\ell = 0}^{L} L_{\ell} A_{\ell} \right] \mathrm{,} \label{eq:SecondTermDerivation}
\end{align}
using just the basic orthogonality relation for the Legendre polynomials. \newline
Finally, the infinite sum over $k$ in the second line of the definition (\ref{eq:FunctProblem}) sets all the higher partial wave projections above $L$ to zero. Every summand in this infinite series consist of the modulus squared of a complex projection integral. This integral can again be formulated explicitly as a function of $\left\{ L_{k} \right\}$:
\begin{widetext}
 \begin{align}
  \int_{-1}^{1} d x F(x) A(x) P_{L+k} (x) &= \int_{-1}^{1} d x \left( \sum_{r = 0}^{\infty} L_{r} P_{r} (x) \right) \left( \sum_{\ell = 0}^{L} (2 \ell + 1) A_{\ell} P_{\ell} (x) \right) P_{L+k} (x) \nonumber \\
  &= \sum_{r=0}^{\infty} \sum_{\ell =0}^{L} (2 \ell + 1) A_{\ell} L_{r} \int _{-1}^{1} d x P_{r} (x) P_{\ell} (x) P_{L + k} (x) \nonumber \\
  &= \sum_{r=0}^{\infty} \sum_{\ell =0}^{L} \sum_{m = \left| \ell - L - k \right|}^{\ell + L + k} (2 \ell + 1) A_{\ell} L_{r}
   \left< \ell,0 ; L + k,0 | m , 0 \right>^{2} \int _{-1}^{1} d x P_{r} (x) P_{m} (x) \nonumber \\
  &= 2 \hspace*{2pt} \sum_{\ell =0}^{L} \sum_{m = \left| \ell - L - k \right|}^{\ell + L + k} \frac{(2 \ell + 1)}{(2 m + 1)} A_{\ell} L_{m} \left< \ell, 0 ; L + k , 0 | m, 0 \right>^{2} \mathrm{.} \label{eq:ThirdTermDerivation}
 \end{align}
\end{widetext}
Combining the intermediate results (\ref{eq:FirstTermDerivation}), (\ref{eq:SecondTermDerivation}) and (\ref{eq:ThirdTermDerivation}), we arrive at the final expression for the minimization functional $\bm{W}$ as a function of the Legendre coefficients
\begin{widetext}
\begin{align}
 \bm{W} (\left\{L_{k}\right\}) &= \sum_{x} \left( \sum_{k, k^{\prime} = 0}^{\infty} \sum_{m = \left| k^{\prime} - k \right|}^{k^{\prime} + k} L_{k^{\prime}}^{\ast} L_{k} \left< k^{\prime}, 0 ; k , 0 | m, 0 \right>^{2} P_{m} (x) - 1 \right)^{2} + \left( \sum_{\ell = 0}^{L} \mathrm{Im} \left[ L_{\ell} A_{\ell} \right]  \right)^{2} \nonumber \\
& \hspace*{-25pt} + \sum_{k = 1}^{\infty} \sum_{\ell, \tilde{\ell} = 0}^{L} \sum_{m = \left| \ell - L - k \right|}^{\ell + L + k} \sum_{\tilde{m} = \left| \tilde{\ell} - L - k \right|}^{\tilde{\ell} + L + k} \frac{(2 \ell + 1)}{(2 m + 1)} \frac{(2 \tilde{\ell} + 1)}{(2 \tilde{m} + 1)} A_{\ell}^{\ast} L_{m}^{\ast} A_{\tilde{\ell}} L_{\tilde{m}} \left< \ell, 0 ; L + k , 0 | m, 0 \right>^{2} \left< \tilde{\ell}, 0 ; L + k , 0 | \tilde{m}, 0 \right>^{2} \mathrm{.} \label{eq:FunctProblemAnalyticFunctionForm}
\end{align}
\end{widetext}
We note that this function is defined purely in terms of the information on the input amplitude, i.e. it's truncation order $L$ and partial waves $A_{\ell}$. Clebsch-Gordan coefficients and the Legendre polynomials appearing here are known. Therefore, the only free parameters here are just the real- and imaginary parts of the Legendre coefficients, as it should be. \newline
Furthermore, it should be noted that the expression (\ref{eq:FunctProblemAnalyticFunctionForm}) is still quite formal, especially since it still contains an infinite sum over $k$. For practical numerical purposes, the infinite sum over the partial wave projections, as well as the Legendre expansion (\ref{eq:InfiniteLegExpansionAppendices}), would have to be truncated.


\clearpage

\begin{thebibliography}{AA}


\bibitem{BowcockBurkhardt} J. E. Bowcock and H. Burkhardt, Rep. Prog. Phys. {\bf38}, 1099 (1975).

\bibitem{LPKokNote} L.P. Kok., \textit{Ambiguities in Phase Shift Analysis}, \\ In $\ast$Delhi 1976, Conference On Few Body Dynamics$\ast$, Amsterdam 1976, 43-46.

\bibitem{AtkinsonEtAlContAmb} D. Atkinson, L. P. Kok, M. de Roo and P. W. Johnson, Nucl. Phys. B {\bf 77}, 109 (1974).

\bibitem{Crichton} J. H. Crichton, {\it Nuovo Cimento}, A {\bf 45}, 256 (1966).

\bibitem{RWorkmanEtAlCompExPhotoprod} R.L. Workman, L. Tiator, Y. Wunderlich, M. D\"{o}ring, H. Haberzettl, Phys. Rev. C {\bf95} no.1, 015206 (2017).

\bibitem{YWEtAl2014}
Y. Wunderlich, R. Beck and L. Tiator, Phys. Rev. C {\bf 89}, no. 5, 055203 (2014).

\bibitem{Grushin} V.~F. Grushin, in {\it Photoproduction of Pions on Nucleons and Nuclei}, edited
by A.~A. Komar (Nova Science, New York, 1989), p. 1ff.

\bibitem{Svarc2017} A. \v{S}varc, Y. Wunderlich, H. Osmanovi\'{c}, M. Had\v{z}imehmedovi\'{c}, R. Omerovi\'{c}, J. Stahov,
V. Kashevarov, K. Nikonov, M. Ostrick, L. Tiator, and R. Workman, 	arXiv:1706.03211 [nucl-th].

\bibitem{LegendreProductFormula} For the first identity, see for instance: \newline
W.~J. Thompson, {\it Angular Momentum}, John Wiley \& Sons (2008). \newline
The second identity is a well-known relation between $3j$-symbols and Clebsch-Gordan coefficients.

\bibitem{DeanLee}  N.~W. Dean and P. Lee, Phys. Rev. D {\bf 5}, 2741 (1972).

\bibitem{Omelaenko}  A.~S. Omelaenko, Sov. J. Nucl. Phys. {\bf 34}, 406 (1981).

\bibitem{KeatonWorkmanII} G. Keaton and R. Workman, Phys. Rev. C {\bf 54}, 1437 (1996).

\bibitem{Gersten}  A. Gersten, Nucl. Phys. B {\bf 12}, p. 537 (1969).

\bibitem{ChTab}
W.~T. Chiang and F. Tabakin, Phys. Rev. C {\bf 55}, 2054 (1997).

\bibitem{HoehlerBible}
G. H\"{o}hler, \emph{Pion Nucleon Scattering}, Part 2, Landolt-B\"{o}rnstein:
Elastic and Charge Exchange Scattering of Elementary Particles, Vol.
9b (Springer-Verlag, Berlin, 1983). 

\bibitem{TiatorEtAlElpro}
L.~Tiator, R.L.~Workman, Y.~Wunderlich and H.~Haberzettl,
  arXiv:1702.08375 [nucl-th].


\end{thebibliography}
\end{document}